\documentclass[referee]{aa}

\usepackage{graphicx}
\usepackage{txfonts}

\bibpunct{(}{)}{;}{a}{}{,} 

\usepackage{subfig}
\usepackage{float}
\usepackage{comment}
\usepackage{algorithm,algorithmicx,algpseudocode}
\usepackage{amsmath}
\usepackage{bm}
\usepackage{amsfonts}
\usepackage{amssymb}
\usepackage{array}
\usepackage{url}
\usepackage{epstopdf}
\usepackage{color}
\usepackage{multirow}

\def\RR{\mathbb{R}}
\newcommand{\R}{\mathbb{R}}

\def\CC{\mathbb{C}}

\def\ff{\mathbf{f}}

\def\vv{\mathbf{v}}

\def\HH{\mathbf{H}}
\def\FF{\mathbf{F}}
\def\WW{\mathbf{W}}

\newcommand{\prt}[1]{\left(#1\right)}
\newcommand{\llv}[1]{\left\{#1\right\}}

\begin{document}

\title{Solar hard X-ray imaging by means of Compressed Sensing and \\ Finite Isotropic Wavelet Transform}

\author{M. A. Duval-Poo\inst{1}  \and M. Piana\inst{1,2} \and A. M. Massone\inst{2}}

\institute{Dipartimento di Matematica,  Universit\`{a} degli Studi di Genova, Via   Dodecaneso 35, 16146 Genova, Italy\\
\email{duvalpoo@dima.unige.it}, \email{piana@dima.unige.it}
\and
CNR-SPIN, Via Dodecaneso 33, 16146 Genova, Italy\\
\email{annamaria.massone@cnr.it}}


\abstract
{}
{This paper shows that compressed sensing realized by means of regularized deconvolution and the Finite Isotropic Wavelet Transform is effective and reliable in hard X-ray solar imaging.}
{The method utilizes the Finite Isotropic Wavelet Transform with Meyer function as the mother wavelet. Further, compressed sensing is realized by optimizing a sparsity-promoting regularized objective function by means of the Fast Iterative Shrinkage-Thresholding Algorithm. Eventually, the regularization parameter is selected by means of the Miller criterion.}
{The method is applied against both synthetic data mimicking the Spectrometer/Telescope Imaging X-rays ({\em{STIX}}) measurements and experimental observations provided by the Reuven Ramaty High Energy Solar Spectroscopic Imager ({\em{RHESSI}}). The performances of the method are compared with the results provided by standard visibility-based reconstruction methods.}
{The results show that the application of the sparsity constraint and the use of a continuous, isotropic framework for the wavelet transform provide a notable spatial accuracy and significantly reduce the ringing effects due to the instrument point spread functions.}

\keywords{Sun: flares -- Sun: X-rays, gamma rays -- Techniques: image processing}

\titlerunning{Hard X-ray imaging and compressed sensing}
\authorrunning{Duval Poo et al}

\maketitle

\section{Introduction}

Imaging spectroscopy is a powerful tool for exploring the physics underlying particle acceleration and transport in solar flares. In order to realize imaging spectroscopy in the hard X-ray range, in 2002 NASA launched the {\em{Reuven Ramaty High Energy Solar Spectroscopic Imager (RHESSI)}} \citep{lin2002reuven}, whose data have resulted in hard X-ray images of unprecedented angular and energy resolution. In a nutshell, {\em{RHESSI}} rotating collimators modulate the X-ray flux coming from the Sun, providing as a result sparse samples of its Fourier transform, named visibilities, picked up at specific points of the Fourier plane, named $(u,{\it{v}})$ plane in this context. 

The {\em{Spectrometer/Telescope for Imaging X-rays (STIX)}} \citep{benz2012spectrometer} is one of the ten instruments in the payload of {\em{Solar Orbiter}}, which will be launched by ESA close to the Sun in 2018. Analogously to {\em{RHESSI}} the main goal of {\em{STIX}} is to measure hard X-ray photons emitted during solar flares in order to determine the intensity, spectrum, timing and location of accelerated electrons near the Sun. The imaging system that characterizes this device relies on the Moir\'e pattern concept \citep{oster1964theoretical} and, similarly to {\em{RHESSI}}, it provides as well a sampling of the Fourier transform of the photon flux in the $(u,v)$ plane \citep{giordano2015process}. 

For both {\em{RHESSI}} and {\em{STIX}}, image reconstruction is needed to determine the actual spatial photon flux distribution from the few Fourier components acquired by the hard X-ray collimators and several methods have been realized to this goal \citep{hogbom1974aperture,cornwell1985simple,as02,boetal06,massone2009hard}, but none of them exploits a methodology that has been widely applied in astronomical imaging in the last decade, i.e. compressed sensing \citep{do06,cawa08,boetal08}. The present paper describes a possible use of compressed sensing  for regularized deconvolution in {\em{RHESSI}} and {\em{STIX}} imaging. In order to work, compressed sensing requires data incoherency and a sparse representation of the solution of the image reconstruction problem. Both {\em{RHESSI}} and {\em{STIX}}  sample the Fourier domain in a way characterized by a notable level of incoherency; on the other hand, a typical hard X-ray image configuration is made of few, simple shapes (mainly, foot-points and loops) and therefore it is straightforward to represent it as the superposition of a small number of basis functions. It is well-known that an advantageous approach to realize compressed sensing is to use a wavelet transform since wavelets can provide, in addition to compression, a multi-scale signal representation. 

Most wavelet implementations are associated to multi-resolution analysis (MRA) \citep{mallat1999wavelet}, mainly because of their computational effectiveness. However such implementations are far from optimal for applications like filtering and deconvolution, owing to the fact that they are not redundant, i.e., the dimension of the image decreases at coarser scales \citep{starck2007undecimated}.  As an alternative to MRA, the Isotropic Undecimated Wavelet Transform (IUWT) is rather often utilized in radio-interferometry \citep{li2011application,gaetal15} and this for two main reasons: first, it provides redundancy and, second, it is better appropriate for restoring astronomical sources (e.g.: stars, galaxies, flares), which are mostly isotropic or quasi-isotropic. 


IUWT relies on a discrete wavelet transform path, which is not fully appropriate when the input data are provided by a rather sparse sampling of the frequency domain, as in visibility-based hard X-ray imaging. Therefore in the present paper we built a 2D isotropic wavelet transform that follows the continuous wavelet transform path. Specifically, our Finite Isotropic Wavelet Transform (FIWT) is inspired by the shearlet transform implementation proposed by \citet{hauser2014fast} and is a redundant transform, which can be effectively applied in deconvolution problems like the {\em{RHESSI}} and {\em{STIX}} ones. This wavelet system is built in the frequency domain by using a 2D isotropic extension of the 1D Meyer mother wavelet \citep{mallat1999wavelet} but other functions can be used with comparable results \citep{mallat1999wavelet,portilla2000parametric}. In order to reduce the numerical instability due to the ill-posedness of the deconvolution problem, in this paper we formulated a sparsity-enhancing regularized version of the FIWT multi-scale sparse decomposition, which we called the Finite Isotropic waVElet Compressed Sensing method (5-CS) and we used it to obtain reconstructions of hard X-ray images from synthetic {\em{STIX}} and experimental {\em{RHESSI}} data. We also point out that our approach adopts the analysis prior formulation instead of the synthesis prior formulation followed in the case of radio-interferometry visibilities \citep{li2011application}.

Finally, it is worth noting that an alternative way to realize compressed sensing in imaging is to use a dictionary made of few shapes, replicated many times for different scales and positions, and then to realize sparsity with respect to this dictionary. However, 5-CS realizes compressed sensing utilizing a continuous wavelet transformation and therefore it does not need any catalogue of basis images to work. This has two advantages: first, the construction of a catalogue requires to know in advance all possible source shapes and, second and more importantly, catalogue-based compressed sensing is computationally more demanding.


The paper is organized as follows. In Section 2 we formally set up the imaging problem for {\em{RHESSI}} and {\em{STIX}}. Section 3 defines the FIWT and its implementation. Section 4 describes the image reconstruction method based on FIWT and compressed sensing and Section 5 discusses the results obtained starting from {\em{STIX}} and {\em{RHESSI}} visibilities. Our conclusions are offered in Section 6.



\section{The hard X-ray reconstruction problem}
\label{sec:xray_rec_problem}

This section provides a quick overview of the model of data formation for {\em{RHESSI}} and {\em{STIX}}. 

\begin{figure}[!th]
  \centering
  \subfloat[RHESSI]{\includegraphics[width=0.415\textwidth]{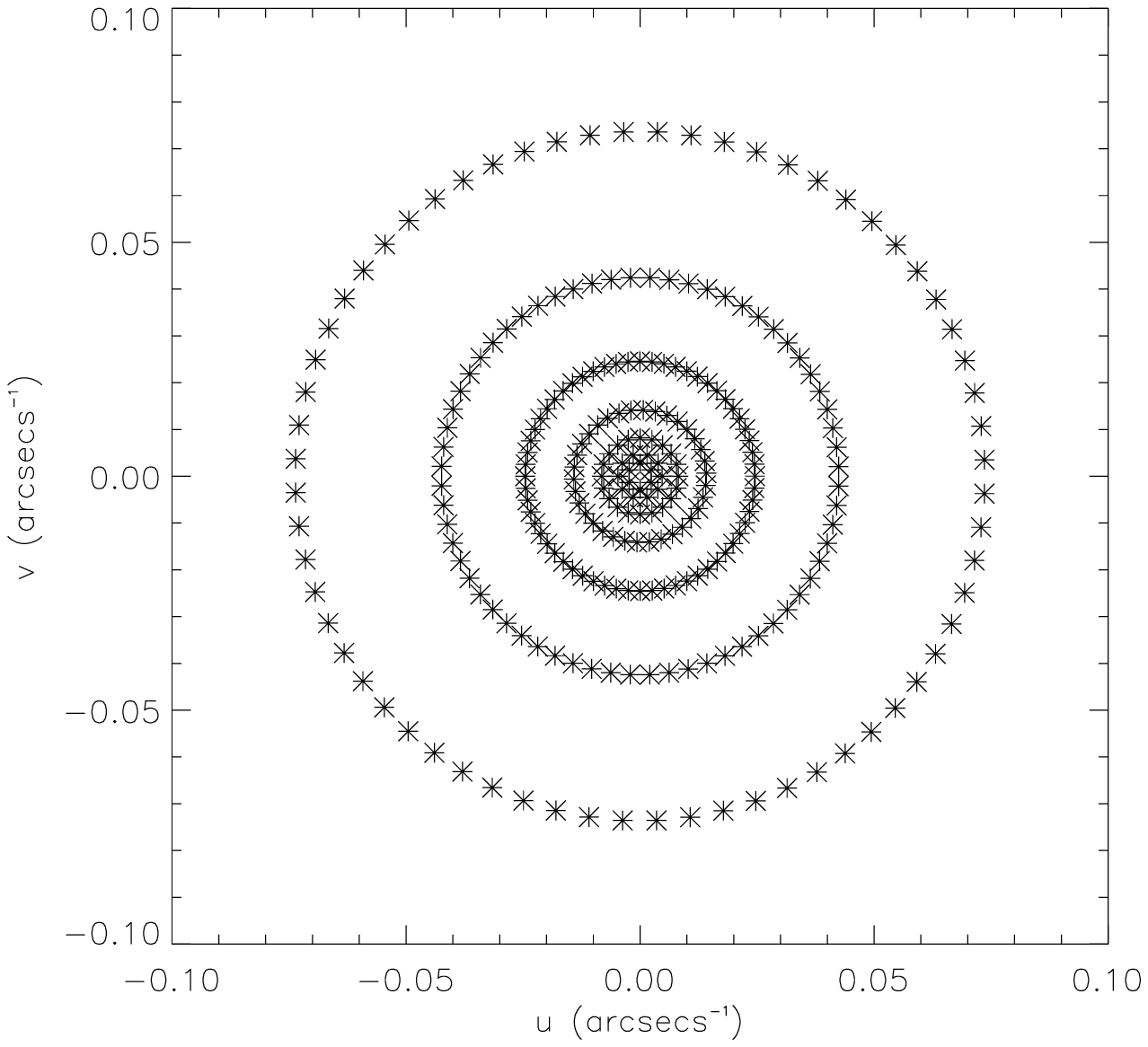}}
  \subfloat[STIX]{\includegraphics[width=0.415\textwidth]{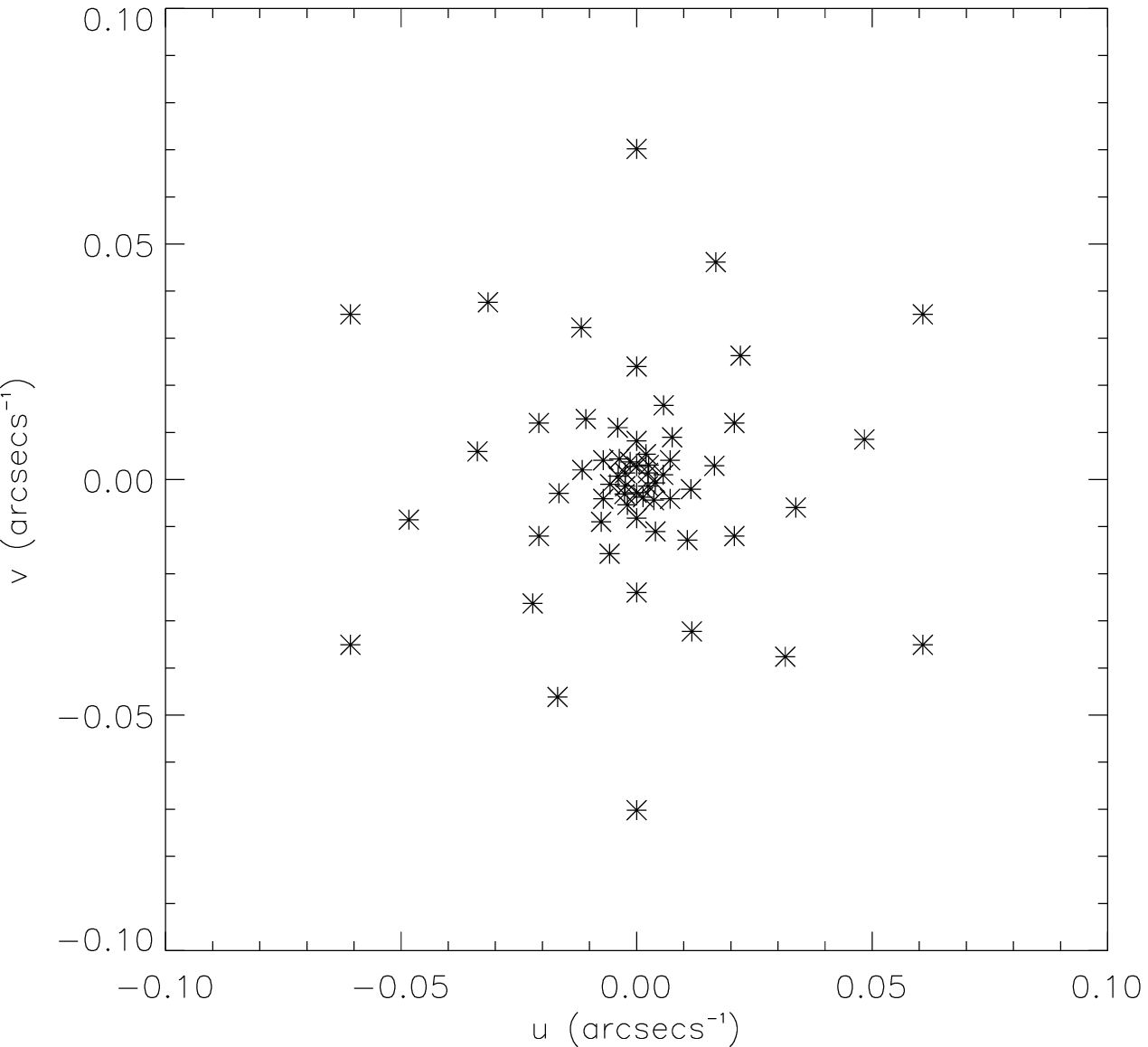}}
  \caption{Sampling of the $(u, v)$ plane performed by {\em{RHESSI}} (left panel) and {\em{STIX}} (right panel). For better visualization, panel (a) shows the sampling starting from detector 3.}
  \label{fig:visibility_sampling}
\end{figure}

In the energy domain, {\em{RHESSI}} data range from some keV to some MeV with energy resolution of around 1 keV. The imaging module is composed by nine pairs of Rotating Modulation Collimators (RMCs), each one formed by a pair of equally spaced fine grids, placed in front of the detecting device. Each pair is composed by identical grids characterized by a given pitch, different from the ones characterizing all the other pairs of grids. The rotational motion of the spacecraft around its own axis causes a periodic modulation of the incident flux. As a result \citep{hurford2003rhessi} the instrument 
samples the $(u,v)$ plane according to nine circles, as shown in Figure \ref{fig:visibility_sampling} (a). It is worth mentioning that the number of samples in each circle is not fixed but determined in an optimal way during the computation procedure of the visibilities.

{\em{STIX}} is formed by 30 detectors recording X-ray photons in the range 4 -- 150 keV. On each detector, the incident flux is modulated by means of a sub-collimator formed by two distant grids with slightly different pitches and slightly different orientations. The effect of this grid configuration is to create the superposition of two spatial modulations, named Moir{\'e} pattern. The recording process on the detector associated with each Moir{\'e} pattern provides a specific {\em{STIX}} visibility. Therefore {\em{STIX}} recording hardware allows sampling the frequency domain in 30 different $(u,v)$ points placed on spirals as shown in Figure \ref{fig:visibility_sampling} (b). A rigorous description of the data formation process in {\em{STIX}} can be found in \citet{giordano2015process}.

As a consequence of these kinds of hardware design, the mathematical model for data formation in {\em{RHESSI}} and {\em{STIX}} can be formulated in a matrix form as
\begin{equation}
    \HH \cdot \FF\ff=\vv,
    \label{eq:stix_linsystem}
\end{equation}
\noindent where the original $N \times N$ photon flux image to reconstruct is lexico-graphically re-ordered to define the vector $\ff\in \RR^M$, with $M=N^2$. Further, $\vv\in \CC^M$ is a sparse vector whose non-zero components correspond to the measured visibilities, $\FF\in \CC^{M\times M}$ is the discretized Fourier transform, $\HH$ is a sparse binary matrix that realizes the sampling in the $(u,v)$ plane, and $\cdot$ denotes the entry-wise product. If we apply the discretized inverse Fourier transform $\FF^{-1}$ on both sides of \eqref{eq:stix_linsystem} we obtain
\begin{equation}
    \FF^{-1}\HH \ast \ff=\FF^{-1}\vv,
\end{equation}
\noindent where $\ast$ is the convolution operator. Therefore, the reconstruction of the flux image $\ff$ from the given visibilities $\vv$ can be essentially viewed as a deconvolution problem, where $\FF^{-1}\HH$ is the point spread function and $\FF^{-1}\vv$ is the \emph{dirty map} from which the PSF blurring effect must be subtracted.

\section{The Finite Isotropic Wavelet Transform}
\label{sec:fiwt}

Let $\psi_{a,{\bf{t}}}$ be a family of functions defined by the translation and dilation of a \emph{mother wavelet} function $\psi({\bf{x}}) \in L^2(\RR^2)$, i.e.
\begin{equation}
\label{eq:spacial_wavelet}
    \psi_{a,{\bf{t}}}({\bf{x}})= a^{-1/2}\psi\prt{\frac{{\bf{x}}-{\bf{t}}}{a}}
\end{equation}
\noindent where ${\bf{x}}=(x_1,x_2)$ is a point in $\RR^2$, ${\bf{t}}=(t_1,t_2) \in \RR^2$ and $a\in \RR^+$ are the translation and dilation parameters, respectively. The normalization factor $a^{-1/2}$ ensures that $\|\psi_{a,{\bf{t}}}\|=\|\psi\|$, where $\| \cdot \|$ is the norm in $L^2(\RR^2)$. The wavelet transform $\mathcal{W}(f)$ of a function $f \in L^2(\RR^2)$ is defined by \citep{mallat1999wavelet}
\begin{align}\label{eq:a1}
    \mathcal{W}(f)(a,{\bf{t}}) &= \langle f, \psi_{a,{\bf{t}}} \rangle \nonumber\\
        &= \langle \hat f, \hat\psi_{a,{\bf{t}}} \rangle \nonumber\\
        &= \int_{\RR^2} \hat f(u,v) \overline{\hat\psi_{a,{\bf{t}}}(u,v)}\ du dv 
\end{align}
\noindent where $\langle f,\psi_{a,{\bf{t}}}\rangle$ is the scalar product in $L^2(\RR^2)$ and $\hat f$ and $\hat\psi$ are the Fourier transform of $f$ and $\psi$, respectively. If the \emph{admissibility condition} is satisfied, i.e.
\begin{equation}
    C = \int_{\RR^2}\frac{|\hat\psi(u,v)|^2}{u^2 + v^2}\ du dv < \infty,
\end{equation}
\noindent then it is possible to define the inverse wavelet transform as
\begin{equation}\label{aa1}
    f({\bf{x}}) = \frac{1}{C}\int_{\RR^2} \int_0^{\infty} \mathcal{W}(f)(a,{\bf{t}}) \psi_{a,{\bf{t}}}({\bf{x}}) \ d{\bf{t}}\frac{da}{a^3}~.
\end{equation}

We constructed a new specific wavelet transform, named the Finite Isotropic Wavelet Transform (FIWT), using the 2D isotropic extension, in the Fourier domain, of a 1D symmetric function. Specifically, we started from the 1D Meyer mother function $\psi_M(x)$ \citep{mallat1999wavelet} and constructed the 2D isotropic mother wavelet function as
\begin{equation}\label{minnie1}
{\hat{\psi}}(u,v) = {\hat{\psi}}_M(\sqrt{u^2 + v^2})~.
\end{equation}
Consequently, from \eqref{eq:spacial_wavelet} we obtained
\begin{equation}
    \hat\psi_{a,{\bf{t}}}(u,v) = a^{1/2}\hat\psi_M\prt{a\sqrt{u^2+v^2}}e^{-2\pi i(u,v)\cdot {\bf{t}}}.
    \label{eq:squre_psi}
\end{equation}
\noindent Further, in order to span the whole $(u,v)$ plane, we included in the wavelet framework the \emph{scaling function} $\phi$ and constructed it again in the Fourier domain as
\begin{equation}\label{minnie2}
{\hat{\phi}}(u,v) = {\hat{\phi}}_M(\sqrt{u^2 + v^2})~,
\end{equation}
where $\phi_M(x)$ is the 1D Meyer scaling function. The FIWT is finally defined from (\ref{eq:a1}), i.e.
\begin{equation}\label{fiwt}
    \mathcal{W}(f)(a,{\bf{t}}) = a^{1/2} \mathcal{F}^{-1}\prt{\hat f(u,v) \hat\psi_M\prt{a\sqrt{u^2+v^2}}}({\bf{t}})~,
\end{equation}
where $\mathcal{F}^{-1}$ is the inverse Fourier transform.

In the imaging framework the flux distribution $f({\bf{x}})$ is pixelized at positions
\begin{equation}\label{aaaa1}
(x_1)_{p} = c_1 - \frac{FOV}{2} + p \delta ~~~p=0,\ldots,N-1~~,
\end{equation}
\begin{equation}\label{aaaa1}
(x_2)_{q} = c_2 - \frac{FOV}{2} + q \delta ~~~q=0,\ldots,N-1~~,
\end{equation}
where $(c_1,c_2)$ is the image center, $FOV$ is the image field-of-view and $\delta$ is the pixel dimension in arcsec. With the same lexico-graphical re-ordering used in section 2 we obtain the vector ${\bf{f}}=(f_1,\ldots,f_M)$ from $f((x_1)_p,(x_2)_q)~~p,q=0,\ldots,N-1$, where $M=N^2$ is the number of pixels of the image to restore and the components correspond to the flux intensity in each pixel. Accordingly to what done is Section 2, the vector ${\bf{f}}$ is again the unknown of the image reconstruction problem.


In the wavelet transformation we will consider $j_0$ scales obtained according to the discretization
\begin{equation}\label{eq:b1}
a_j = 2^{-j}~~~,~~~~j=0,\ldots,j_0-1~.
\end{equation}
On the other hand, the discretization of the traslation parameter ${\bf{t}}$ is made according to
\begin{equation}\label{eq:ttt1}
(t_1)_{n} = n \delta~~,~~n=0,\ldots,N-1~,
\end{equation}
\begin{equation}\label{eq:ttt2}
(t_2)_{l} = l \delta~~,~~l=0,\ldots,N-1~.
\end{equation}

\noindent Then, for each scale $a_j$ and for each translation $((t_1)_n,(t_2)_l)$, we construct the $M-$dimensional vector {\boldmath${\bf{\psi}}$}$_{j,n,l}$ which corresponds to the pixelization and re-ordering of the wavelet (\ref{eq:spacial_wavelet}) for $a=a_j$, $t_1=(t_1)_n$, and $t_2=(t_2)_l$. Analogously, for each translation we construct the $M-$dimensional vector {\boldmath${\bf{\phi}}$}$_{n,l}$, which is the result of the pixelization and re-ordering of the scaling function $\phi$. This leads to the construction of the set of $M \cdot (j_0 + 1)$ $M$-dimensional vectors $\{{\bf{u}}_i~~,~~i=1,\dots,M \cdot (j_0+1) \} =
$ $\{${\boldmath{${\bf{\psi}}$}}$_{j,{n,l}}~~j=0,\ldots,j_0-1~;~~n,l=0,\dots, N-1\} \cup$ $\{${\boldmath$\phi$}$_{n,l}~~~n,l=0,\ldots,N-1\}$. This set provides a Parseval frame, i.e., given ${\bf{f}} \in \R^M$, then
\begin{equation}\label{eq:bb1}
\|{\bf{f}}\|^2 = \sum_{i=1}^{M \cdot (j_0+1)} |({\bf{f}},{\bf{u}}_i)|^2~,
\end{equation}
and
\begin{equation}\label{eq:bbb1}
{\bf{f}} = \sum_{i=1}^{M \cdot (j_0+1)} ({\bf{f}},{\bf{u}}_i) {\bf{u}}_i~,
\end{equation}
where $(\cdot,\cdot)$ denotes the canonical inner product in $\R^M$. Finally, ${\bf{W}}$ is the $M \cdot (j_0+1) \times M$ matrix whose rows are made of the vectors ${\bf{u}}_i$ for $i=1,\ldots,M \cdot (j_0+1)$. 
The fact that $\{{\bf{u}}_i\}_{i=1}^{M \cdot (j_0+1)}$ is a Parseval frame implies that $\WW^T\WW=I$ and that the forward and inverse discretized FIWT can be written as
\begin{equation}\label{pippo1}
{\bf{w}}= \WW {\bf{f}}~~~,~~~{\bf{f}} = \WW^T {\bf{w}}~,
\end{equation}
respectively, where ${\bf{w}}$ is the column vector of dimension $M \cdot (j_0+1)$ whose components are
\begin{equation}\label{pippo2}
w_i = ({\bf{f}},{\bf{u}}_i)~~~i=1,\ldots,M \cdot (j_0 + 1)~.
\end{equation}
We conclude by observing that the computational complexity of the FIWT is around $\mathcal{O}(N^2\log N)$.

\begin{figure*}[!t]
 \captionsetup[subfigure]{labelformat=empty}
 \centering
 \subfloat[$\hat\phi(u,v)$]{
 \includegraphics[width=0.25\textwidth]{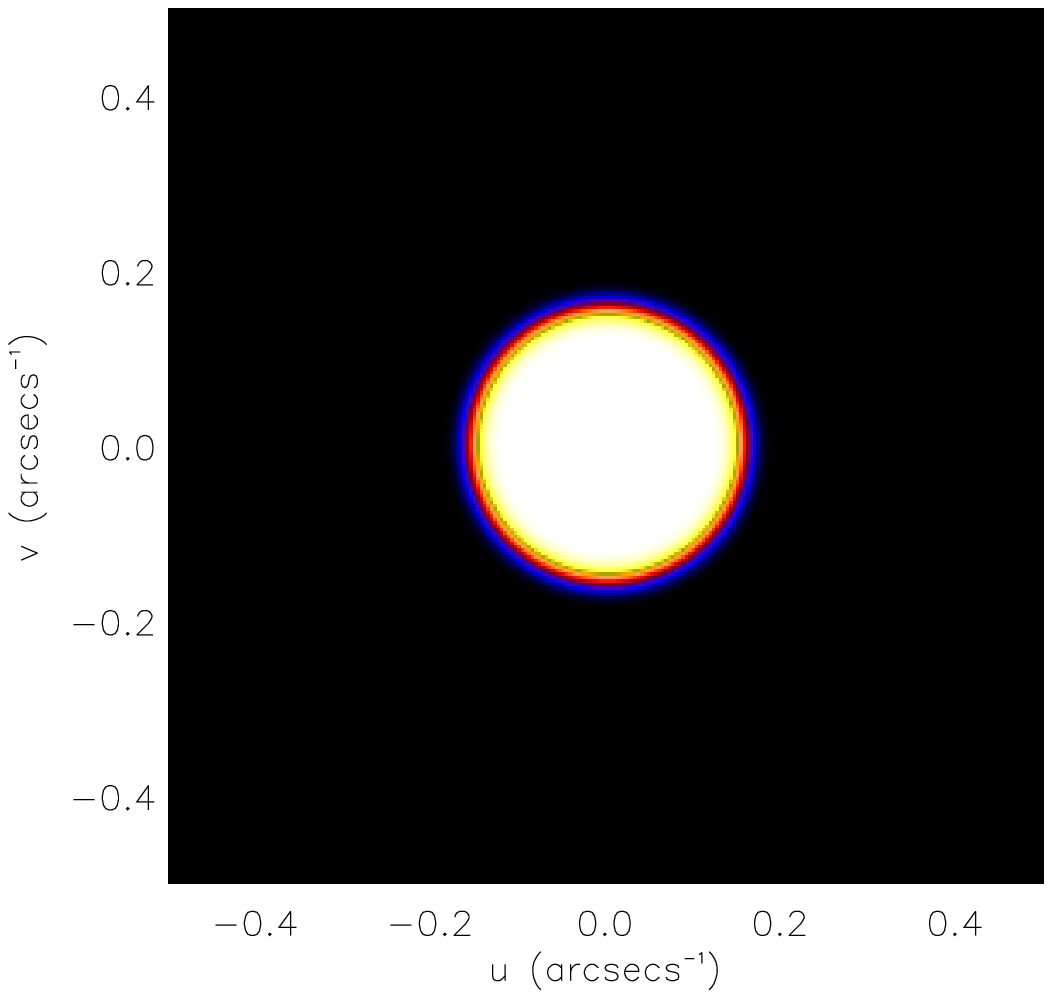}}
 \subfloat[$\hat\psi_{a=1}(u,v)$]{
 \includegraphics[width=0.25\textwidth]{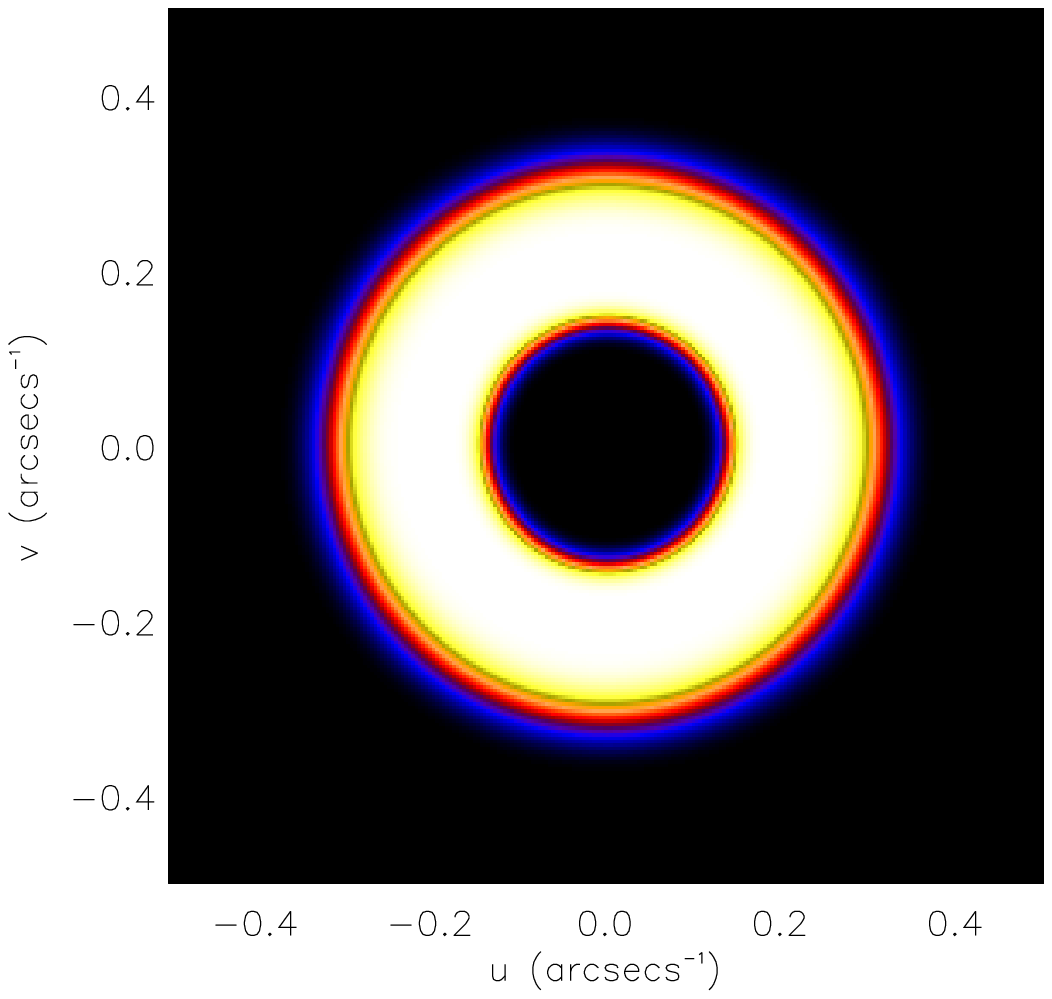}}
 \subfloat[$\hat\psi_{a=2^{-1}}(u,v)$]{
 \includegraphics[width=0.25\textwidth]{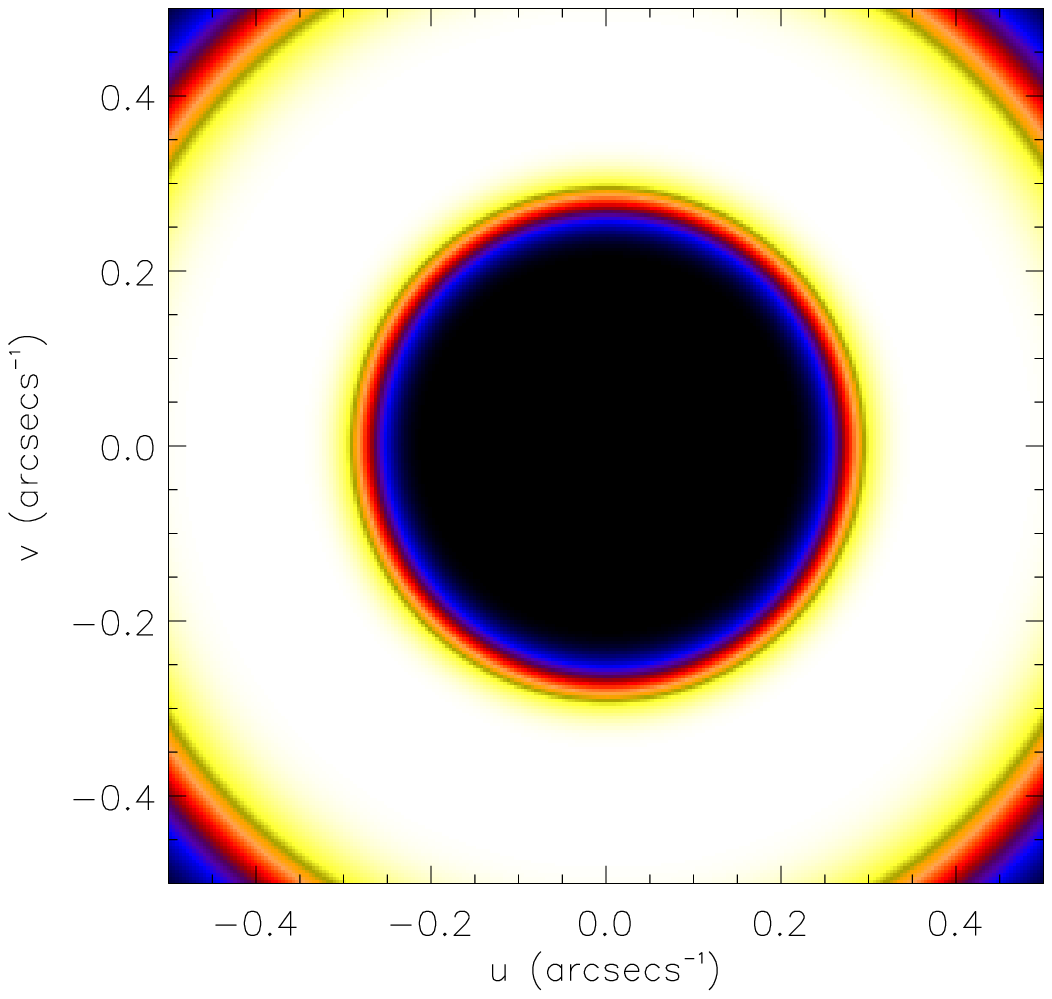}}
 \subfloat[$\hat\psi_{a=2^{-2}}(u,v)$]{
 \includegraphics[width=0.25\textwidth]{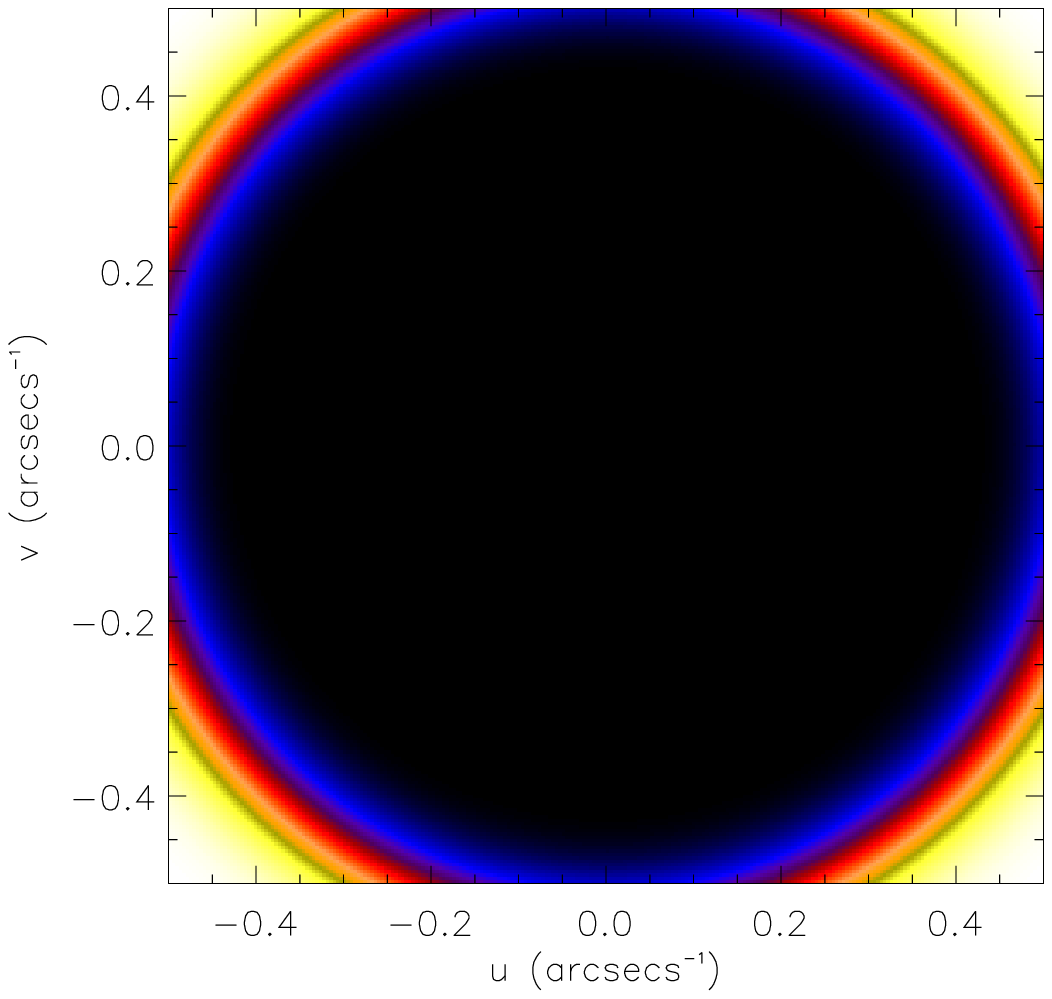}}
 \caption{2D isotropic extensions of the Meyer scaling function and mother wavelet with no translation. The functions are represented in the Fourier domain for 3 scales.}
 \label{fig:meyer_support}
\end{figure*}

\section{The image reconstruction method}

The reconstruction of $\ff$ from $\vv$ is an \emph{ill-posed problem} \citep{bertero1998introduction} and therefore some prior knowledge about the image is required to regularize the reconstruction problem \citep{engl1996regularization}. A valid approach in hard X-ray imaging is to regularize the inverse problem with the $\ell_1$ norm in some transformation domain. In fact, the method we propose in this paper, which we named Finite Isotropic waVElet transform Compressed Sensing (5-CS), addresses the optimization problem
\begin{equation}
  \label{eq:analysis_prior_form}
  \min_\ff \llv{||\HH \cdot \FF\ff-\vv||^2_2 + \lambda||\WW\ff||_1}~.
\end{equation}
The \emph{data term} of the objective function to minimize, $||\HH \cdot \FF\ff-\vv||^2_2$, quantifies the prediction error with respect to the measurements. The \emph{regularization term}, $||\WW\ff||_1$, is designed to penalize an estimate that would not exhibit the sparsity property with respect to FIWT. The regularization parameter, $\lambda > 0$, provides a tradeoff between fidelity to the measurements and sparsity.

%
Problem \eqref{eq:analysis_prior_form} can be numerically solved by means of any algorithm for non-linear optimization. Here we used the Fast Iterative Shrinkage-Thresholding Algorithm (FISTA) \citep{beck2009fast}, which is an $\ell_1$ solver widely used in many fields mainly because of its reliability and rapid convergence. Specifically, here FISTA is implemented by imposing the conservation of the flux at each iteration and by utilizing a standard rule for optimally stopping the iterations.

The only open parameter for method (\ref{eq:analysis_prior_form}) is the regularization parameter $\lambda$, which  plays a crucial role in the reconstruction process. Finding an appropriate value for $\lambda$ is often not a trivial problem, depending on both the criteria adopted for assessing the quality of the reconstructed image and the amount of information known about the original image and its noise. There exist several strategies in the literature to properly estimate the regularization parameter, where the discrepancy principle \citep{morozov2012methods}, the Miller method \citep{miller1970least}, the generalized cross-validation (GCV) method \citep{golub1979generalized}, and the L-curve method \citep{miller1970least,hansen1992analysis} are the most used ones. In this work, we chose the Miller method \citep{miller1970least} because of its simplicity and because it is not computationally demanding. According to this selection criterion, if the following bounds
\begin{equation}
  \label{eq:miller_condition}
    ||\HH \cdot \FF {\bf{f}}-\vv||^2_2 \leq \varepsilon, \quad ||\WW {\bf{f}} ||_1 \leq E
\end{equation}
are known, or can be estimated from the dirty map, then the regularization parameter can be chosen as $\lambda = \varepsilon / E$. In order to satisfy conditions (\ref{eq:miller_condition}), the bounds $\{\varepsilon, E\}$ are estimated performing the first iteration ${\bf{f}}_1$ of the FISTA algorithm with $\lambda=0$. The regularization parameter is then set equal to
\begin{equation}
  \lambda = \frac{||\HH \cdot \FF {\bf{f}}_{1}-\vv||^2_2}{||\WW {\bf{f}}_{1}||_1}.
\end{equation}

\begin{figure*}[!t]
 \centering
 \subfloat[$\lambda_0 = 0$]{
 \includegraphics[width=0.25\textwidth]{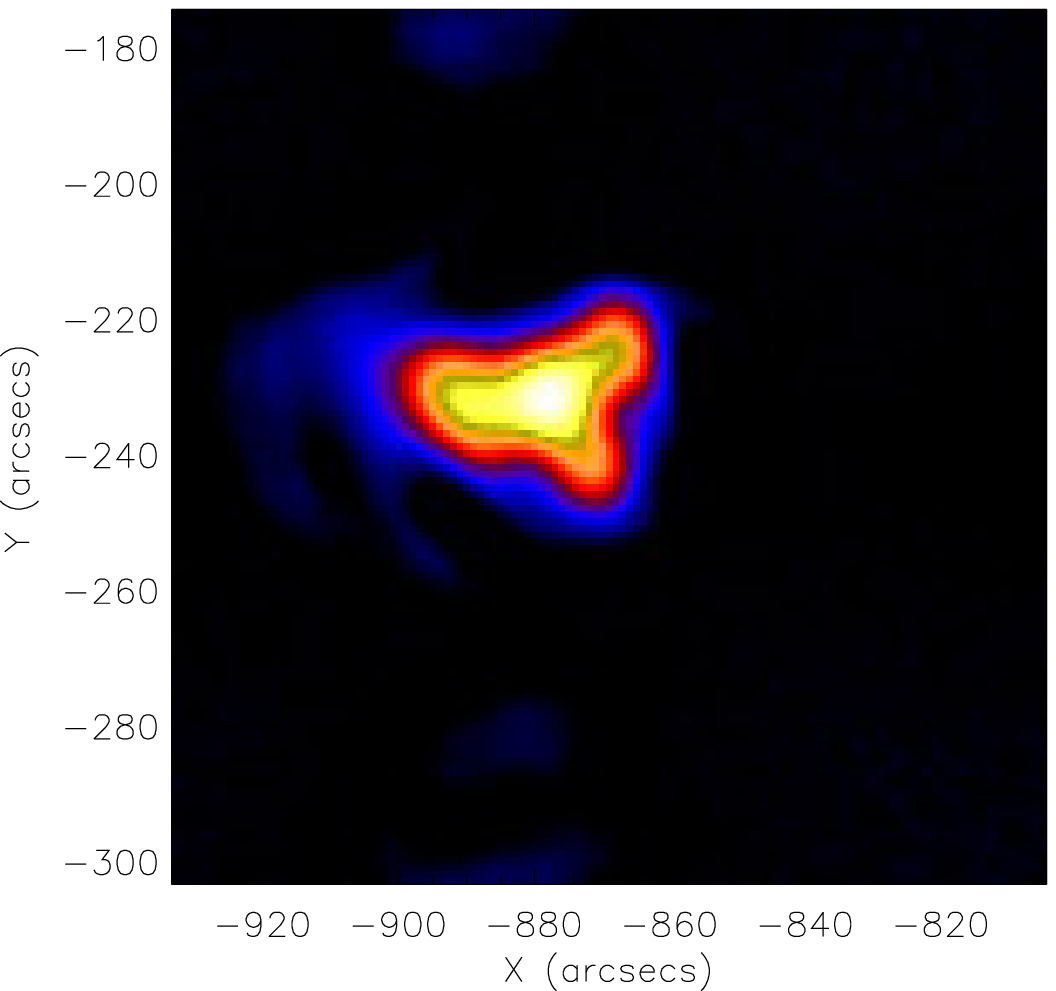}}
 \subfloat[$\lambda_{L} = 3.1 \times 10^{-3}$]{
 \includegraphics[width=0.25\textwidth]{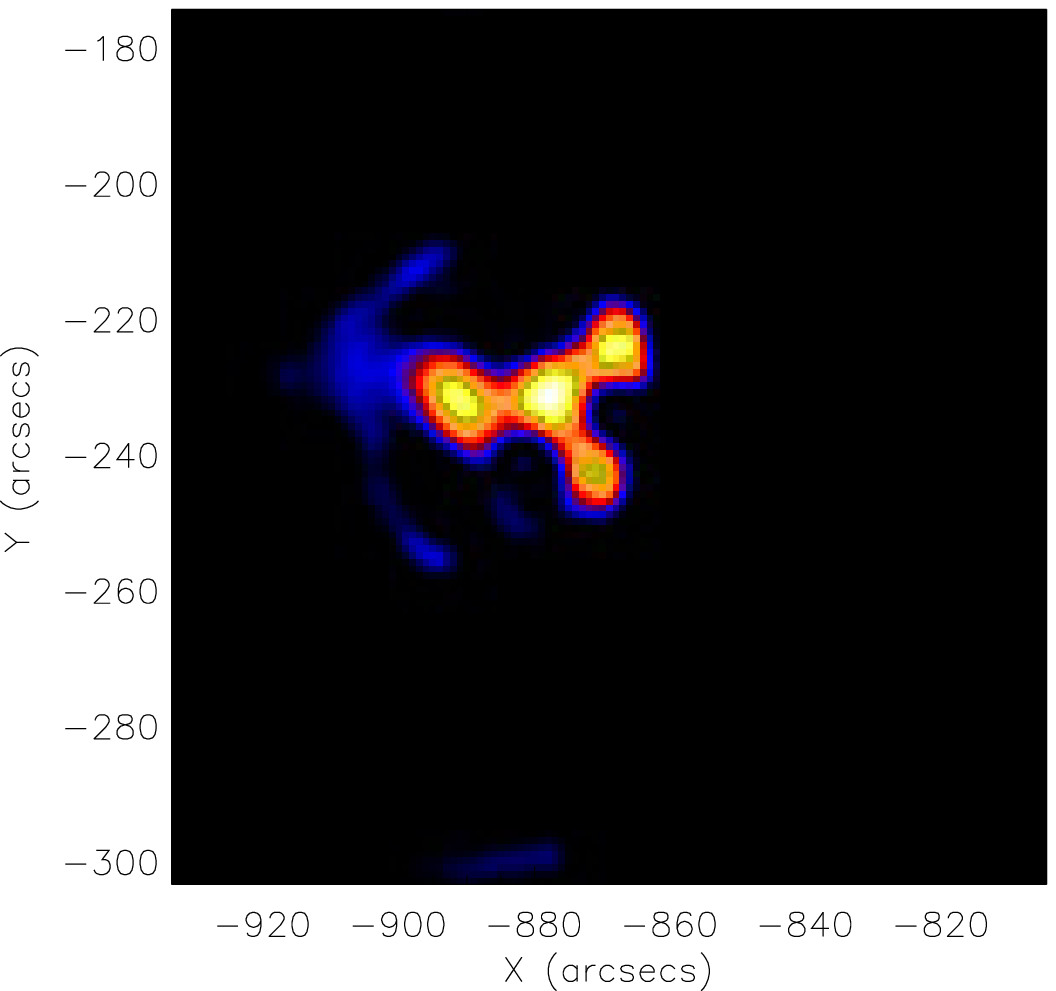}}
 \subfloat[$\lambda_{M} = 5.7 \times 10^{-3}$]{
 \includegraphics[width=0.25\textwidth]{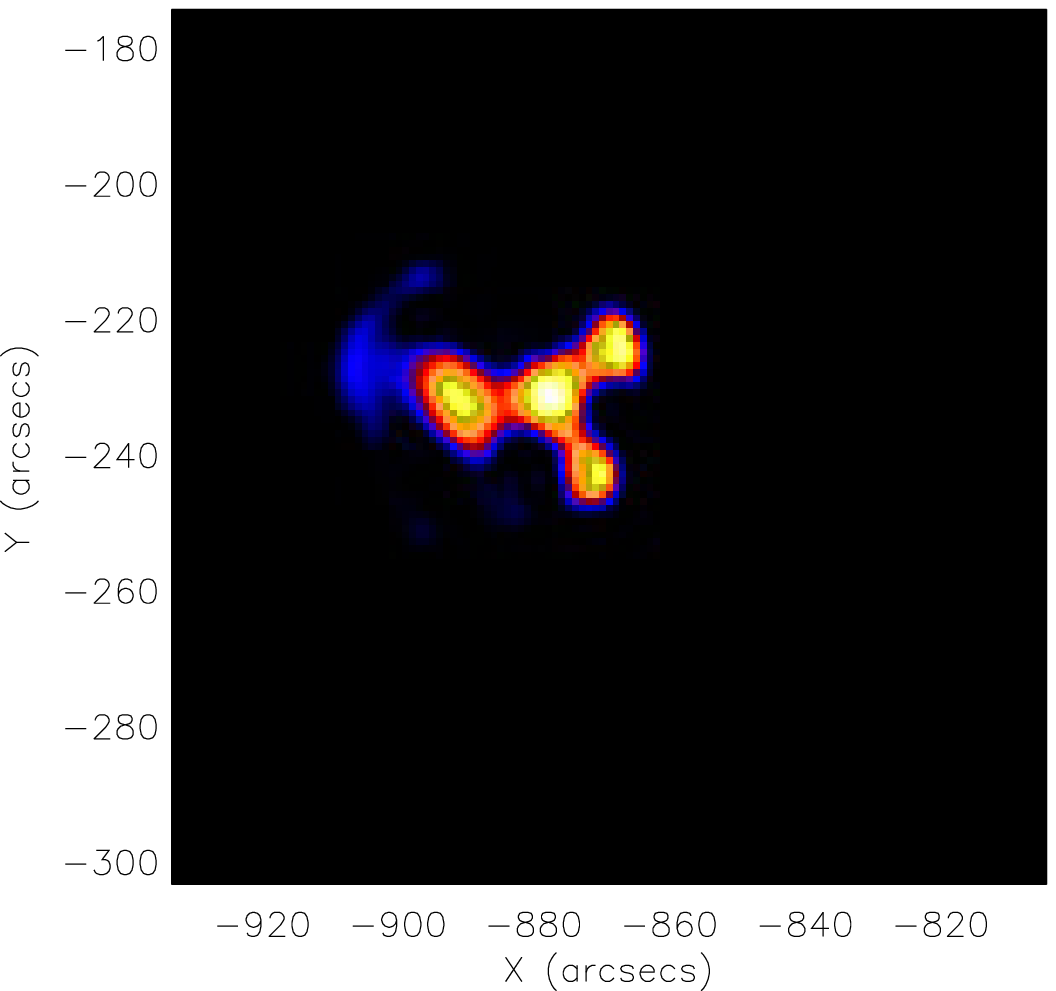}}
 \subfloat[$\lambda_1 = 0.1$]{
 \includegraphics[width=0.25\textwidth]{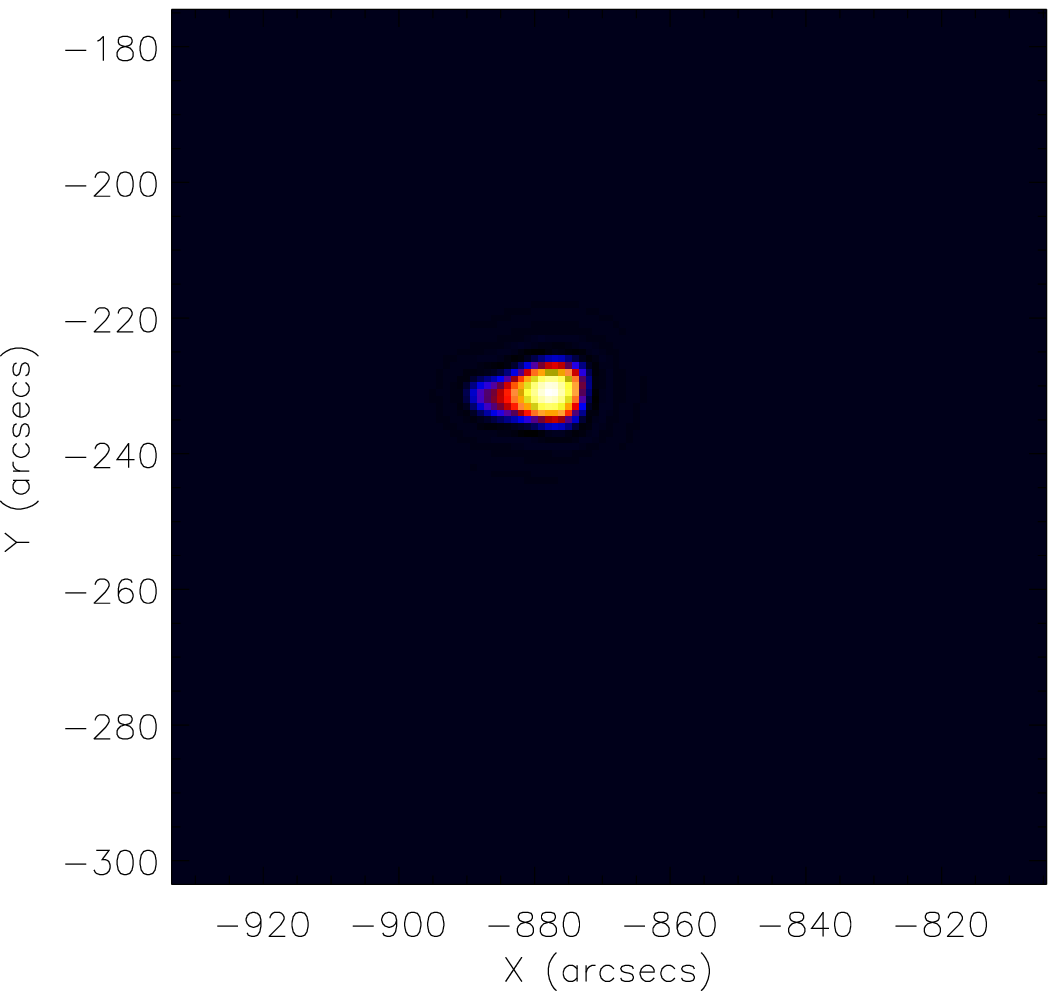}}
 \caption{July 23 2002 solar flare image reconstruction by 5-CS with different choices for $\lambda$: (a) $\lambda=0$ (non-regularized solution), (b) $\lambda=3.1 \times 10^{-3}$ as provided by the L-curve method, (c) $\lambda=5.7 \times 10^{-3}$ as provided by the Miller method, (d) $\lambda=0.1$ (over-regularized solution). The time interval is between 00:29:10-00:30:19 UT while the energy channel is $36-41$ keV. Detectors from $2$ through $9$ have been employed.}
 \label{fig:lambda_examples}
\end{figure*}

%
We compared the performance of the Miller method with the one of the L-curve criterion in the case of the reconstruction of an image of the July 23 2002 event (00:29:10 - 00:30:19 UT; $36-41$ keV). More specifically, Figure \ref{fig:lambda_examples} compares the reconstructions provided by 5-CS for four values of the regularization parameter: $\lambda=\lambda_0=0$, the value $\lambda = \lambda_L$ provided by the L-curve method, the value $\lambda = \lambda_M$ provided by the Miller method, and the value $\lambda=\lambda_1 = 0.1$ realizing over-regularization. The values of $\lambda_L$ and $\lambda_M$ are very close and therefore the corresponding reconstructions are very similar, although the Miller method requires less computational effort.

\section{Experimental results}
\label{sec:experiments}
\newcommand{\fevent}{path}

In this Section we performed an experimental assessment of the proposed reconstruction algorithm. First, we evaluated our method with {\em{STIX}} synthetic visibilities where the simulations are performed by means of the {\em{STIX}} Data Processing Software\footnote{https://stix.cs.technik.fhnw.ch/}. Next, 5-CS was validated against experimental visibilities produced by real flare measurements captured by {\em{RHESSI}}.
 
\subsection{The case of {\em{STIX}} synthetic data}

\graphicspath{ {./experiments/stix_evaluation/} }
\begin{figure*}[!th]
 \centering
 \renewcommand{\fevent}{sim_2point_1}
 \subfloat[Simulated Double Footpoint Flare (SDFF1)]{
 \includegraphics[width=0.25\textwidth]{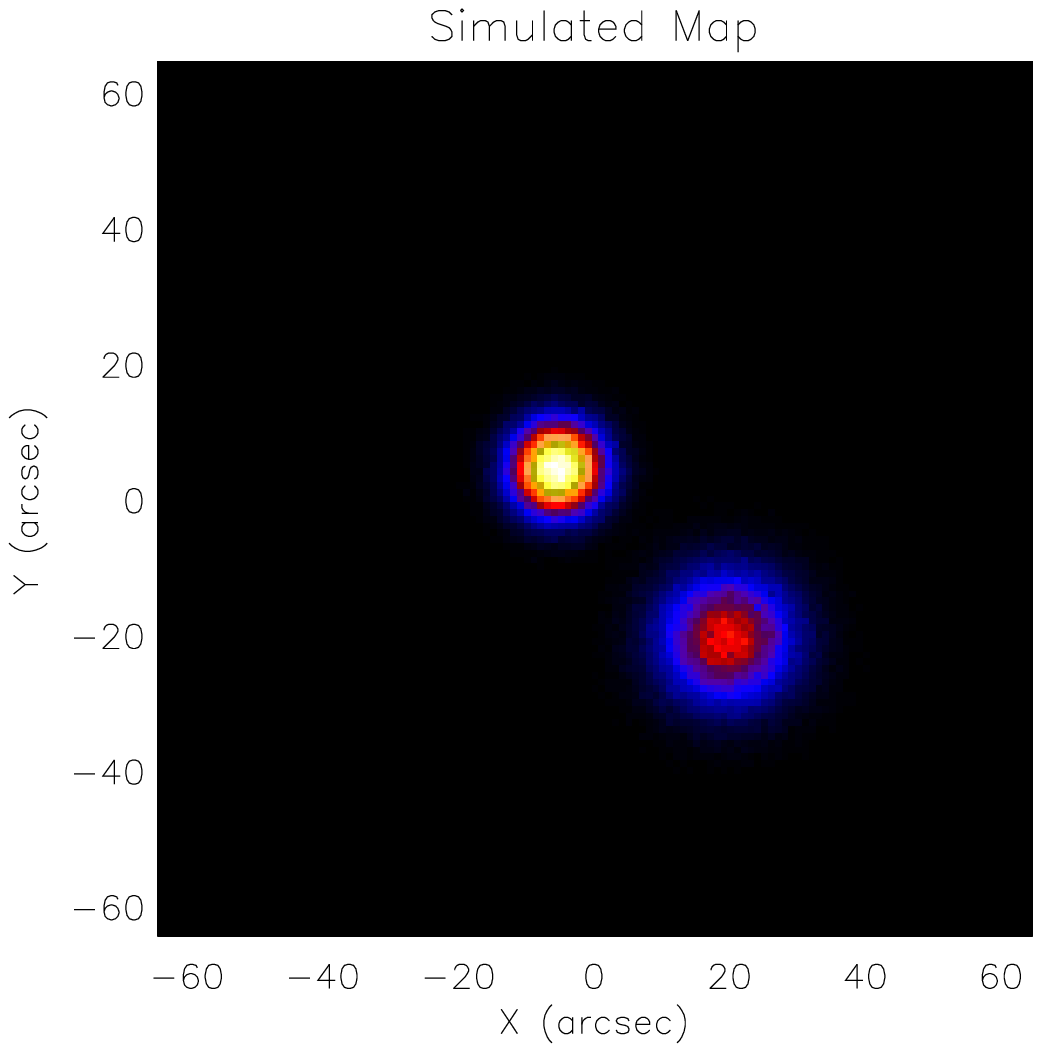}
 \includegraphics[width=0.25\textwidth]{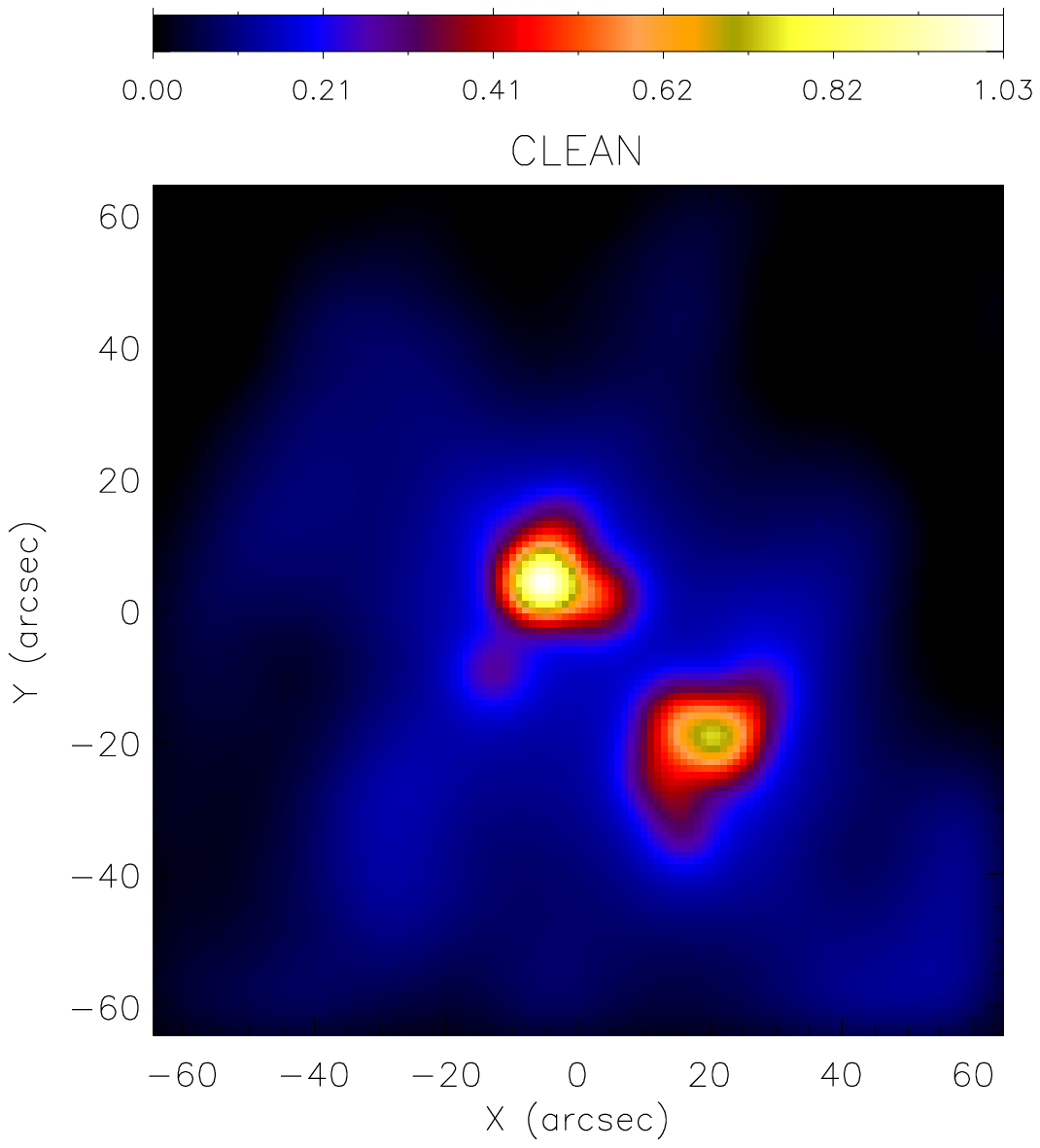}
 \includegraphics[width=0.25\textwidth]{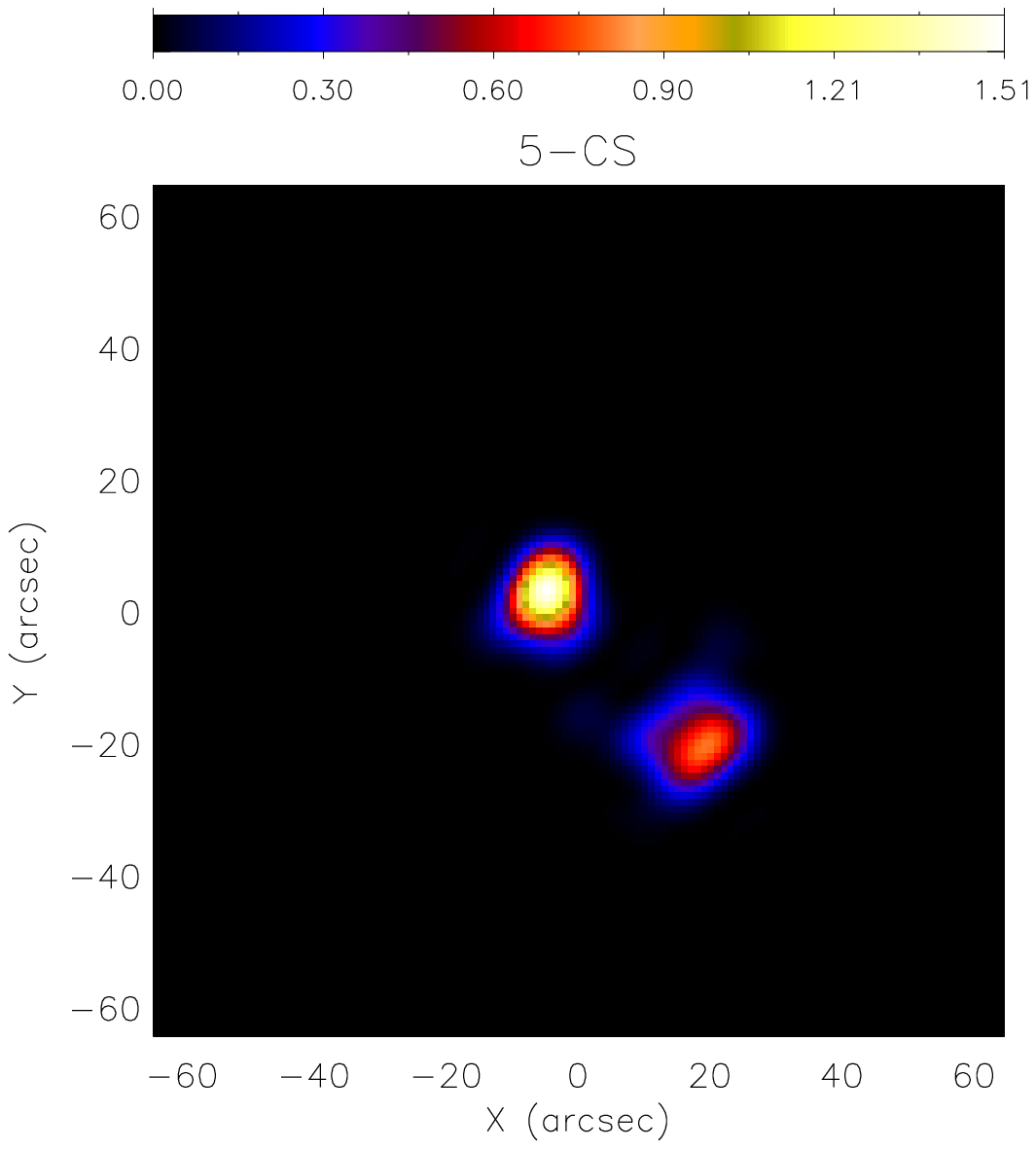}}\\[0.1cm]
 \renewcommand{\fevent}{sim_2point_2}
 \subfloat[Simulated Double Footpoint Flare (SDFF2)]{
 \includegraphics[width=0.25\textwidth]{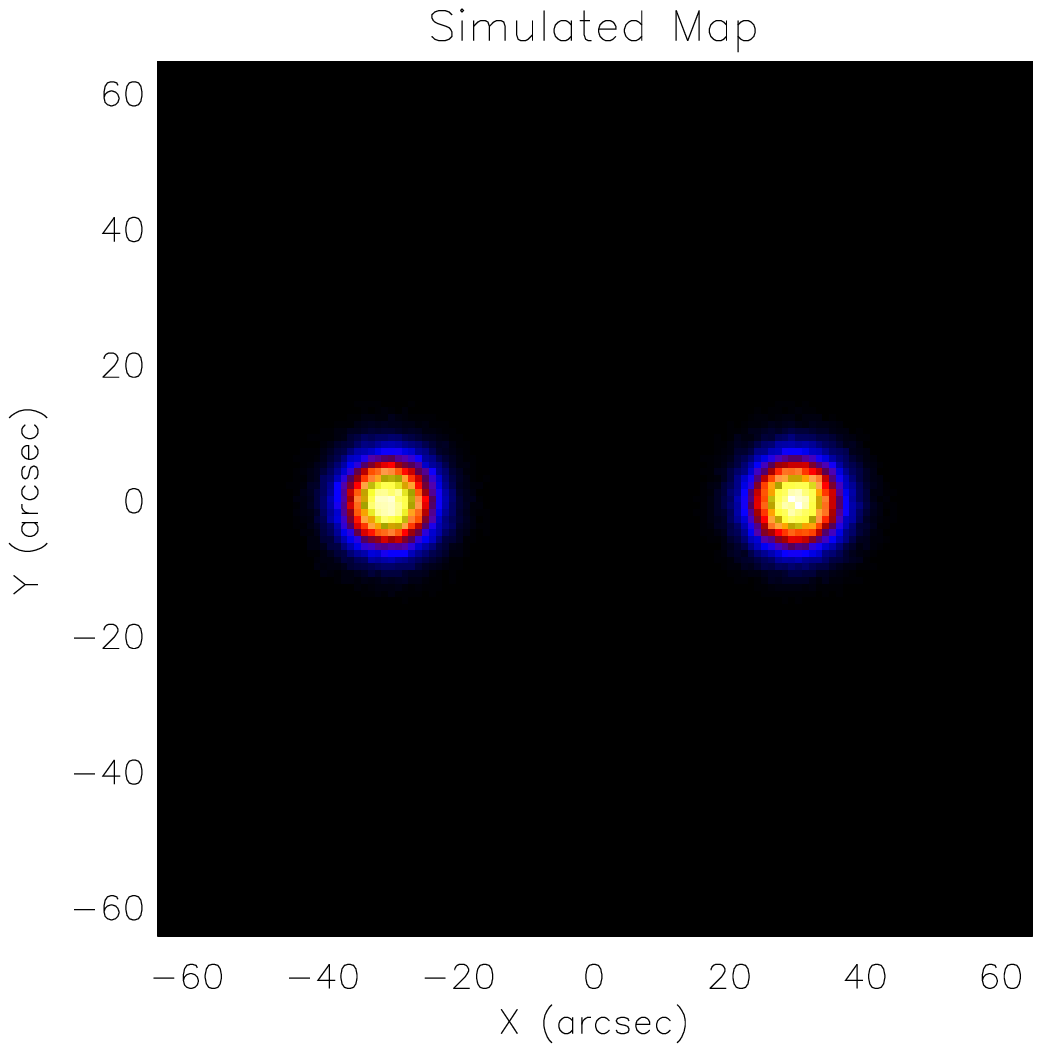}
 \includegraphics[width=0.25\textwidth]{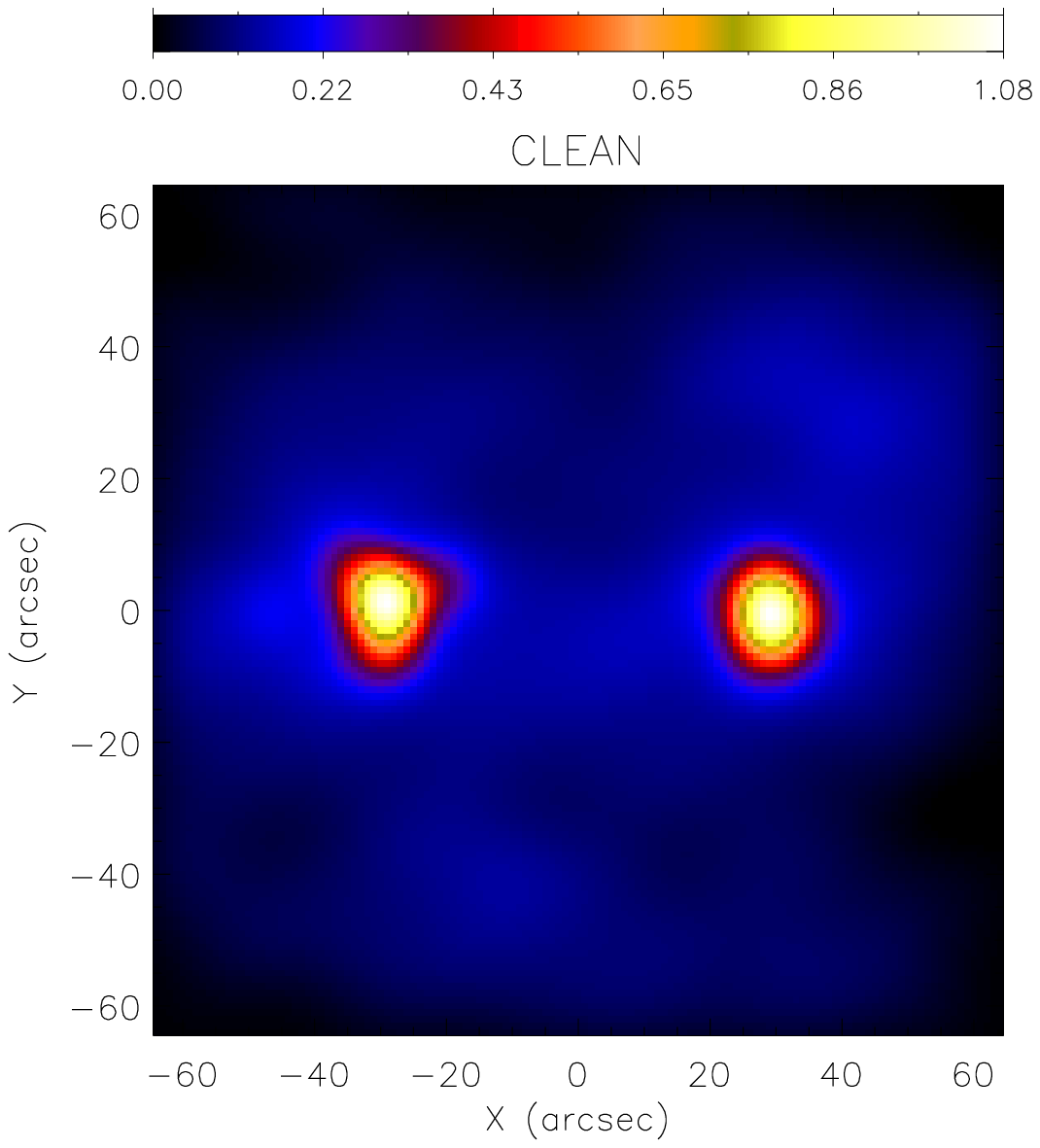}
 \includegraphics[width=0.25\textwidth]{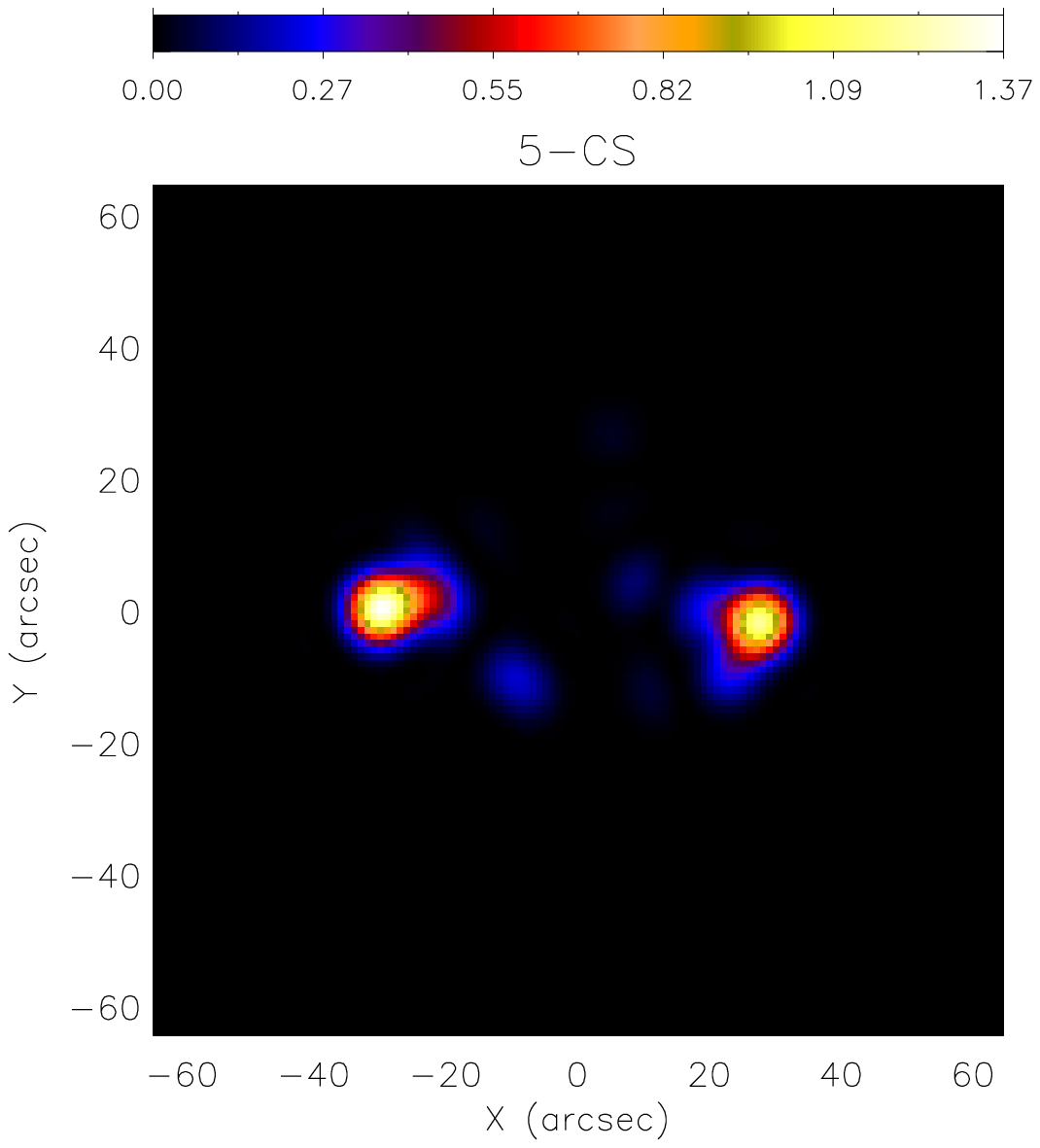}}\\[0.1cm]
 \renewcommand{\fevent}{sim_loop_1}
 \subfloat[Simulated Loop Flare (SLF1)]{
 \includegraphics[width=0.25\textwidth]{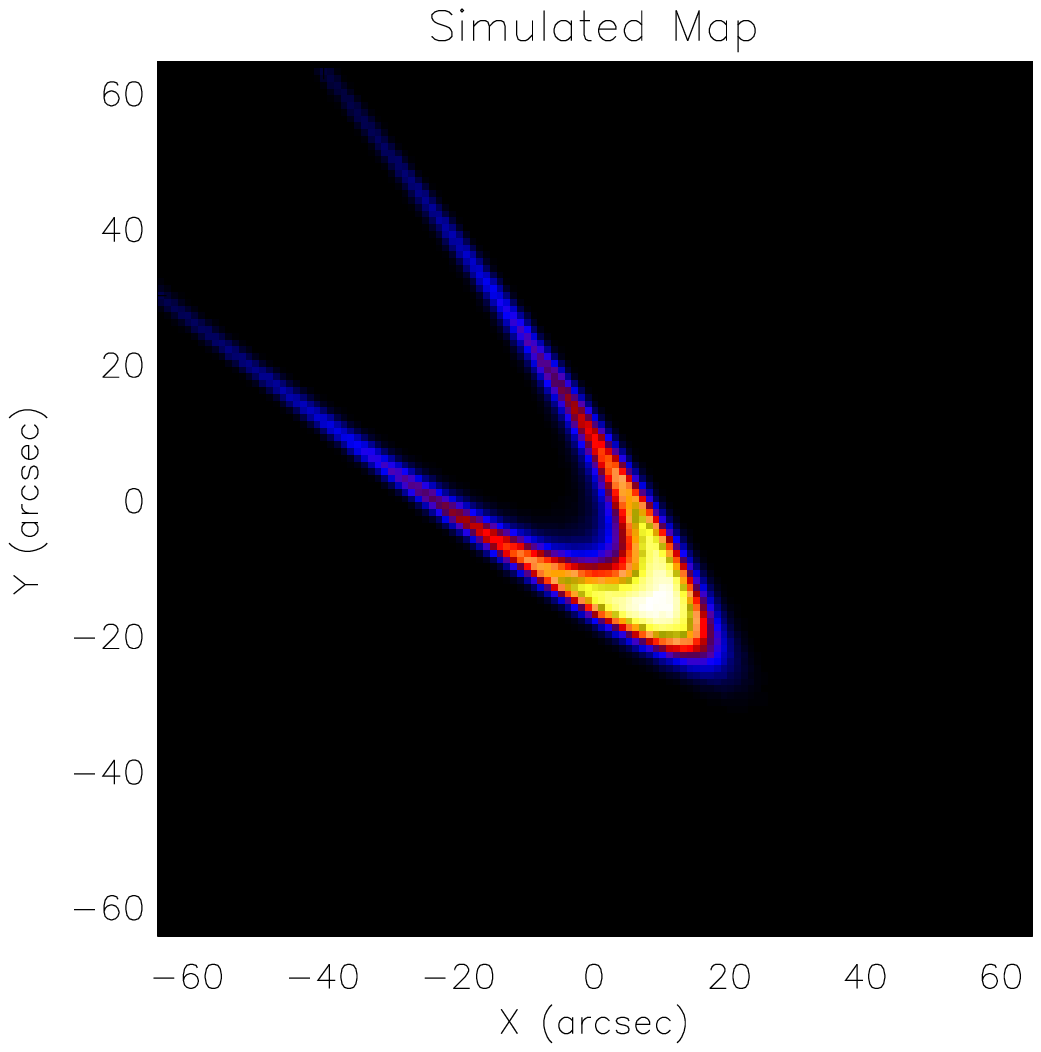}
 \includegraphics[width=0.25\textwidth]{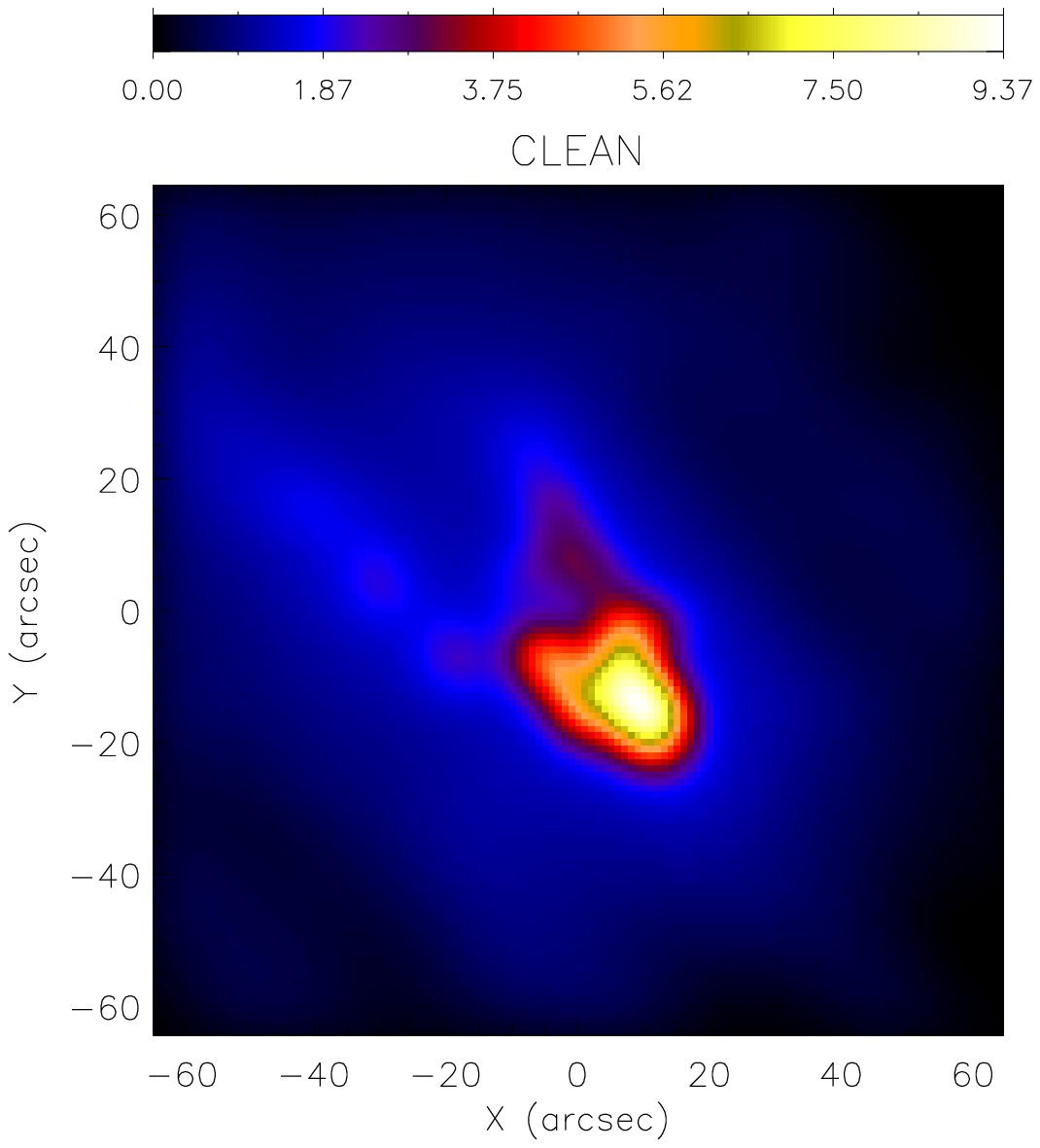}
 \includegraphics[width=0.25\textwidth]{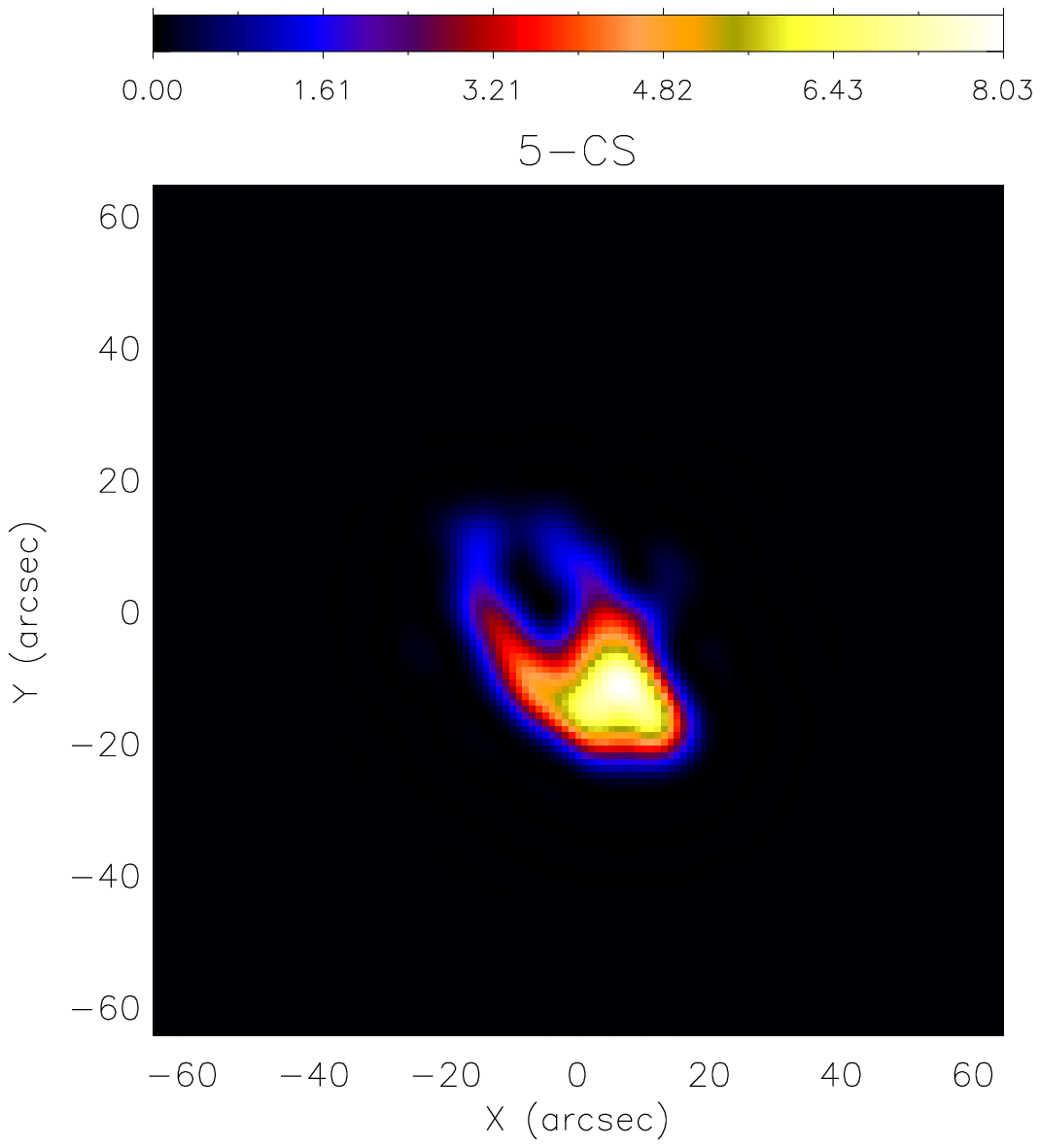}}\\[0.1cm]
 \renewcommand{\fevent}{sim_loop_2}
 \subfloat[Simulated Loop Flare (SLF2)]{
 \includegraphics[width=0.25\textwidth]{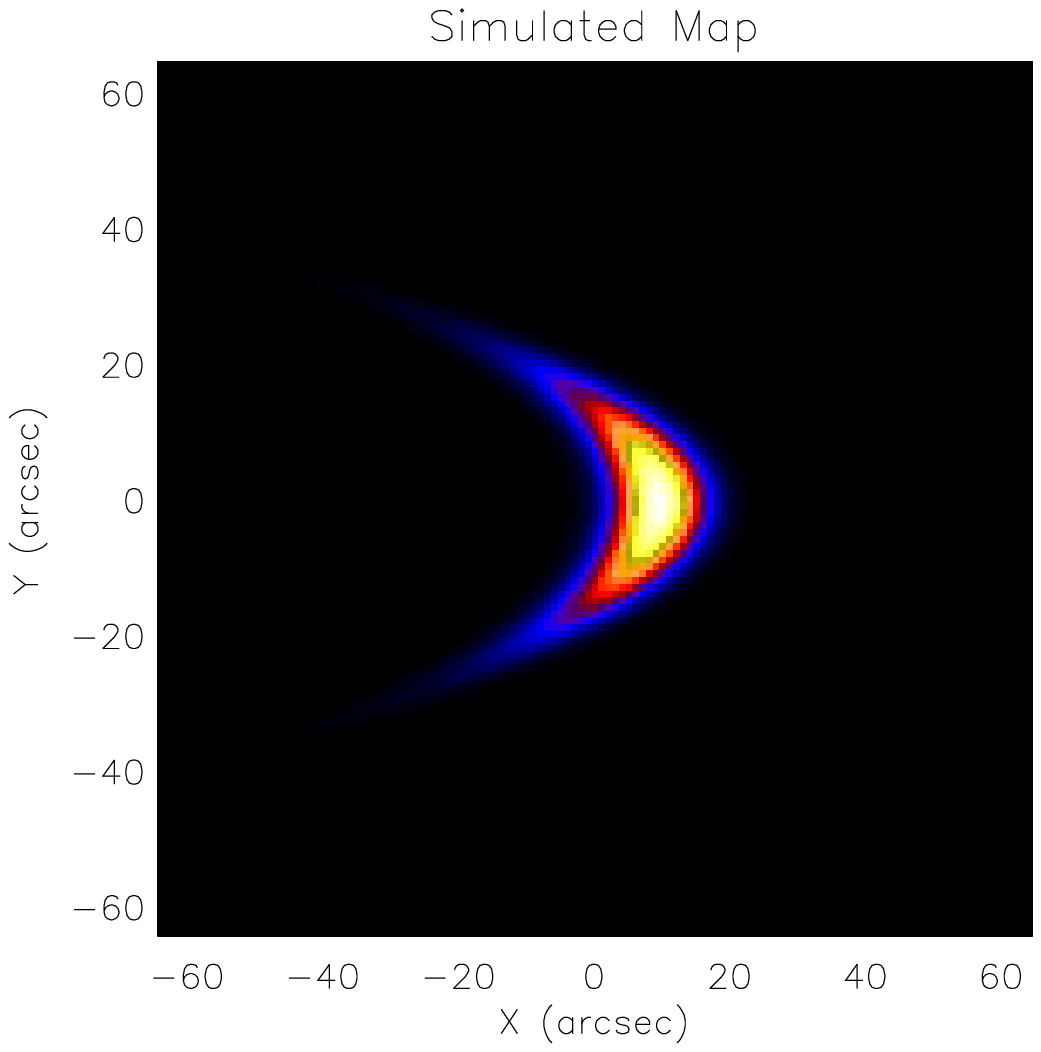}
 \includegraphics[width=0.25\textwidth]{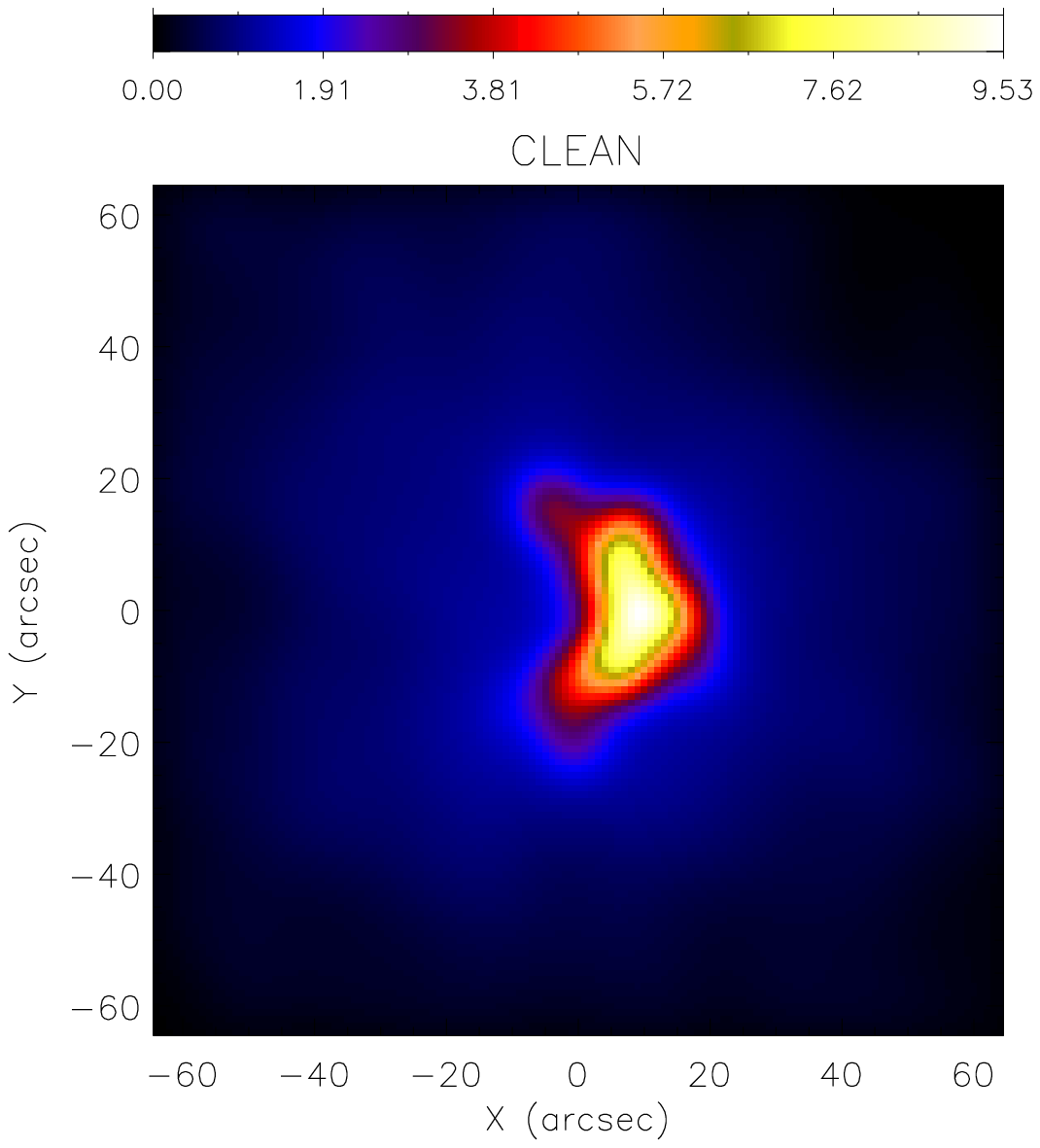}
 \includegraphics[width=0.25\textwidth]{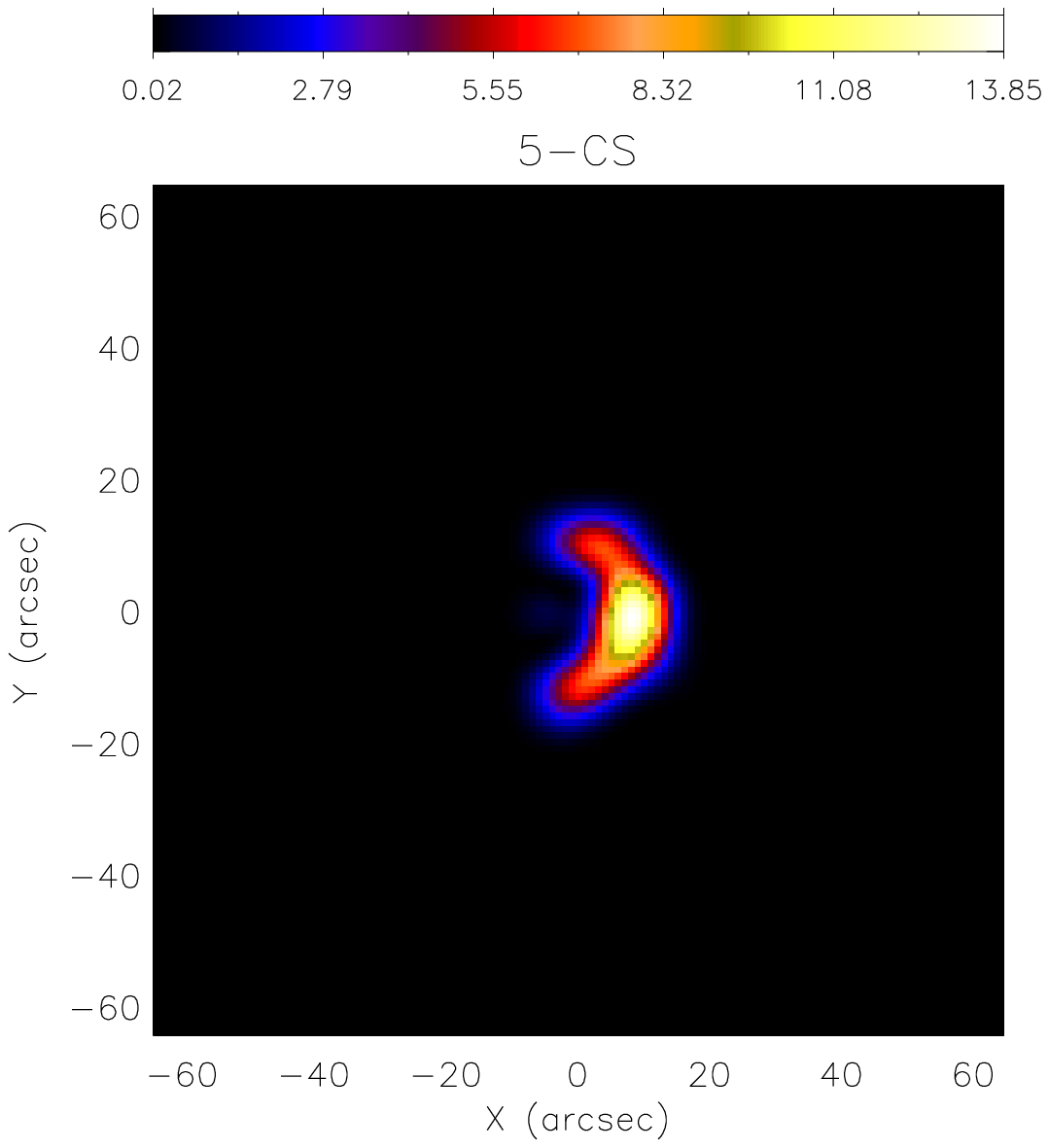}}
 \caption{Reconstructed images for four different STIX simulated flare events. Left column: the original source shape. Middle column: reconstructions provided by CLEAN using visibilities. Right column: reconstructions provided by 5-CS.}
 \label{fig:stix_eval_mixt}
\end{figure*}

We simulated four flaring events with different configurations. In the first two images (Figure 4, (a) and (b)) the overall photon flux is $10^4$ photons/cm$^2$, but in the first image the two foot-points have different intensity and different size, while in the second one the brightness and dimension of the two sources are the same. The other two images (Figure 4, (c) and (d)) are characterized by a higher overall intensity ($10^5$ photons/cm$^2$) and contain two loops with significantly different curvatures. 

Figure \ref{fig:stix_eval_mixt} compares the images reconstructed by 5-CS with the ones provided by CLEAN \citep{hogbom1974aperture} and Table \ref{tab:stix_eval_stats} contains some physical parameters characterizing the simulated and reconstructed images (in this Table the positions and the full width at half maximum (FWHM) are measured in arcsec). These results imply that 5-CS and CLEAN recover the physical parameters with comparable accuracy, although 5-CS is able to significantly reduce the impact of spurious artifacts. 

\begin{table}[!t]
\renewcommand{\arraystretch}{1.4}
\caption{Physical parameters inferred from the reconstructed maps in (Figure \ref{fig:stix_eval_mixt}) and compared to the ones in the simulated maps (position and FHWM are measured in arcsec). The reconstructions have been made $10$ times, in correspondence with $10$ realizations of the {\em{STIX}} simulated data. The Table indicates the average values of the parameters and the corresponding standard deviations.}
\label{tab:stix_eval_stats}
\centering
\begin{tabular}{l|l|r|r|r}
	\hline\hline
	& & Simulated Map & CLEAN & 5-CS\\
  \hline
  \multicolumn{5}{c}{\emph{Simulated Double Footpoint Flare (SDFF1)}}\\
  \hline
  & X & 20.0 & 20.4 $\pm$0.9 & 19.2 $\pm$1.5\\
  & Y & -20.0 & -18.9 $\pm$0.9 & -21.8 $\pm$1.1\\
  \multirow{-3}{*}{\begin{sideways}Peak 1\end{sideways}} & FWHM & 15.0 & 9.6 $\pm$3.4 & 9.0 $\pm$4.4\\
  \hline
  & X & -5.0 & -4.0 $\pm$0.7 & -5.0 $\pm$1.2\\
  & Y & 5.0 & 4.8 $\pm$0.6 & 3.5 $\pm$2.9\\
  \multirow{-3}{*}{\begin{sideways}Peak 2\end{sideways}} & FWHM & 10.0 & 12.8 $\pm$0.8 & 11.9 $\pm$1.6\\
  \hline
  & Flux Ratio & 2.00 & 1.41 $\pm$0.10 & 1.78 $\pm$0.42\\
  \hline
  \multicolumn{5}{c}{\emph{Simulated Double Footpoint Flare (SDFF2)}}\\
  \hline
  & X & -30.0 & -28.9 $\pm$0.6 & -29.3 $\pm$0.8\\
  & Y & 0.0 & 1.0 $\pm$0.7 & 0.4 $\pm$0.7\\
  \multirow{-3}{*}{\begin{sideways}Peak 1\end{sideways}} & FWHM & 10.0 & 13.2 $\pm$0.9 & 9.6 $\pm$0.8\\
  \hline
  & X & 30.0 & 29.5 $\pm$0.5 & 28.0 $\pm$0.8\\
  & Y & 0.0 & -0.1 $\pm$0.6 & -1.4 $\pm$0.5\\
  \multirow{-3}{*}{\begin{sideways}Peak 2\end{sideways}} & FWHM & 10.0 & 12.9 $\pm$0.6 & 9.3 $\pm$0.6\\
  \hline
  & Flux Ratio & 1.00 & 0.99 $\pm$0.10 & 1.02 $\pm$0.09\\
	\hline
  \multicolumn{5}{c}{\emph{Simulated Loop Flare (SLF1)}}\\
  \hline
  & X & 10.0 & 9.0 $\pm$0.5 & 9.6 $\pm$0.5\\
  \multirow{-2}{*}{\begin{sideways}Peak\end{sideways}} & Y & -15.0 & -13.6 $\pm$0.5 & -14.6 $\pm$0.7\\
	\hline
  \multicolumn{5}{c}{\emph{Simulated Loop Flare (SLF2)}}\\
  \hline
  & X & 10.0 & 9.8 $\pm$0.4 & 8.7 $\pm$0.7\\
  \multirow{-2}{*}{\begin{sideways}Peak\end{sideways}} & Y & 0.0 & -0.1 $\pm$0.3 & -0.4 $\pm$0.7\\
	\hline
\end{tabular}
\end{table}

\subsection{The case of {\em{RHESSI}} experimental measurements}

\begin{figure*}[!th]
 \centering
 \subfloat[February 20, 2002 -- 11:06:02-11:06:34 UT -- (22-26 keV)]{
 \renewcommand{\fevent}{2002-Feb-20_22-26keV}
 \includegraphics[width=0.25\textwidth]{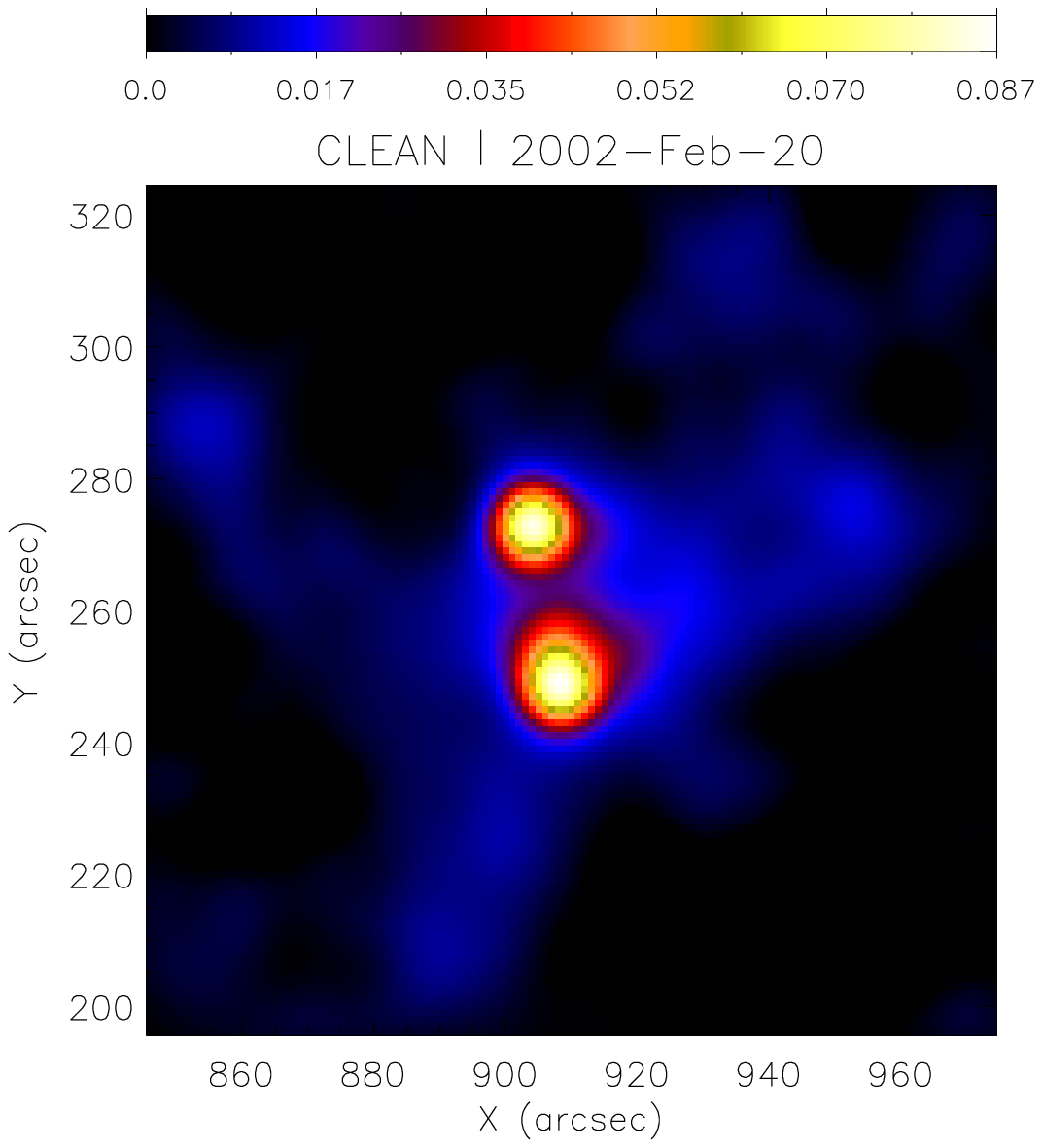}
 \includegraphics[width=0.25\textwidth]{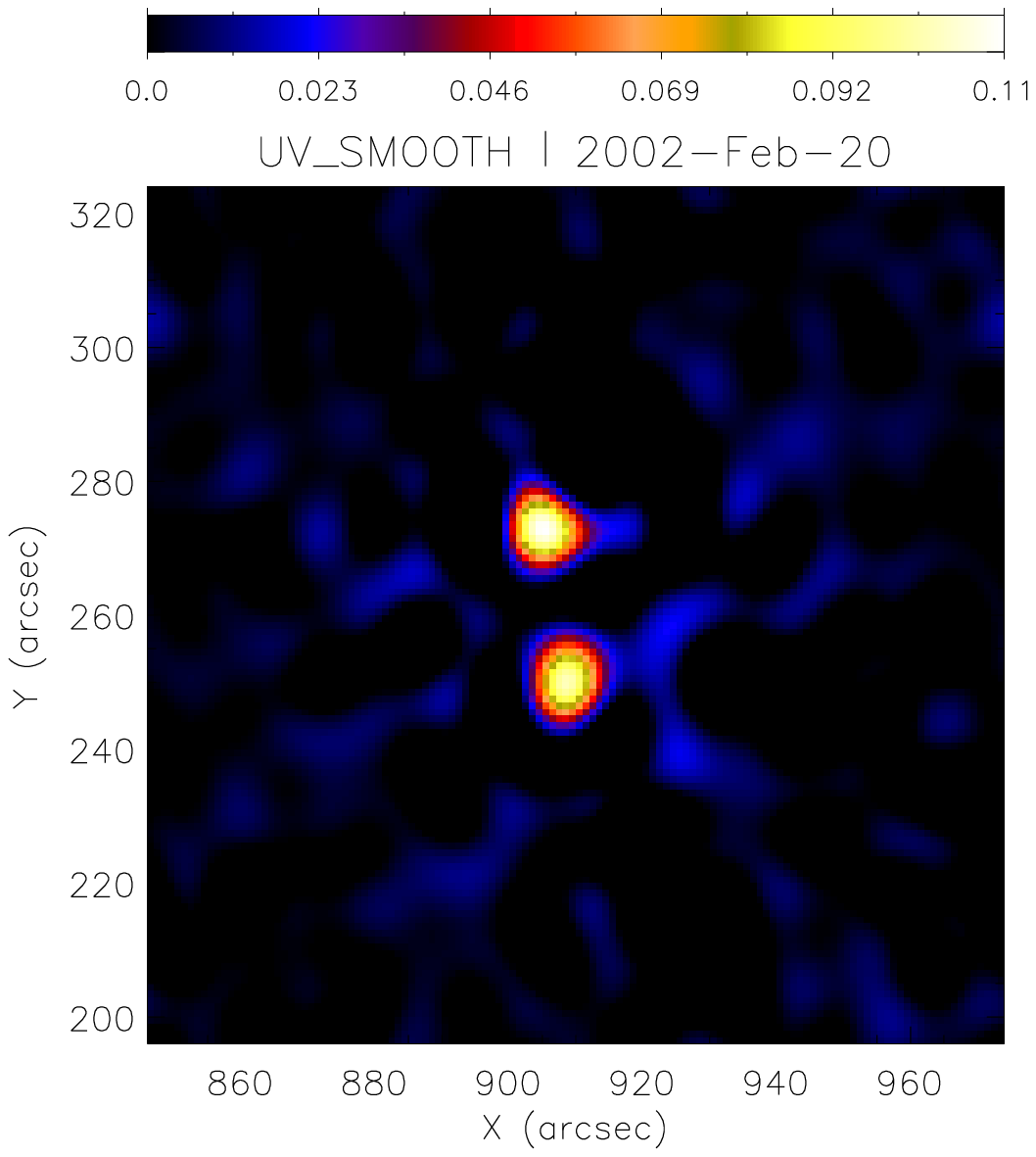}
 \includegraphics[width=0.25\textwidth]{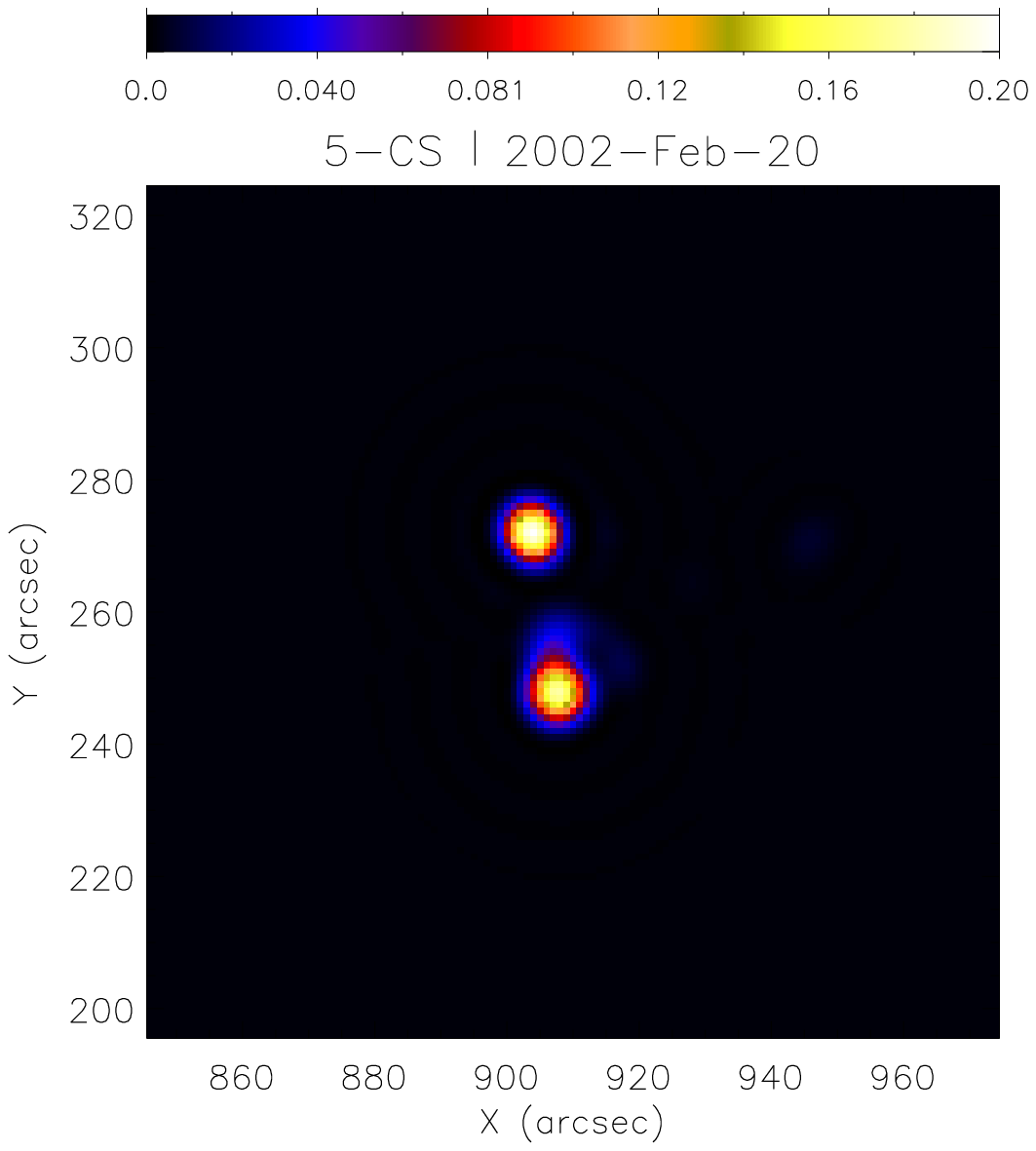}}\\[0.1cm]
 \renewcommand{\fevent}{15-Apr-2002_12-14keV}
 \subfloat[April 15, 2002 -- 00:05:00-00:10:00 UT -- (12-14 keV)]{
 \includegraphics[width=0.25\textwidth]{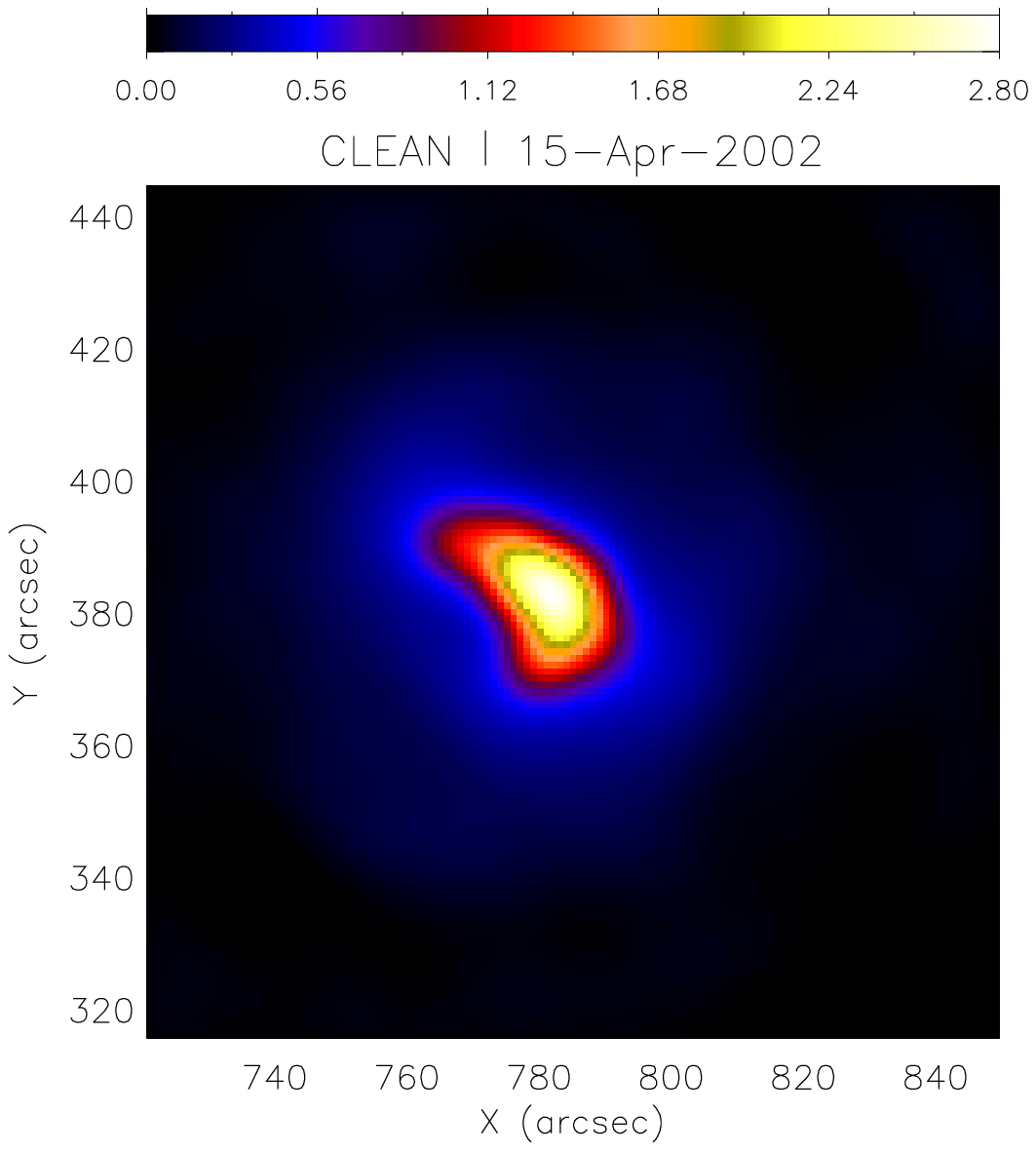}
 \includegraphics[width=0.25\textwidth]{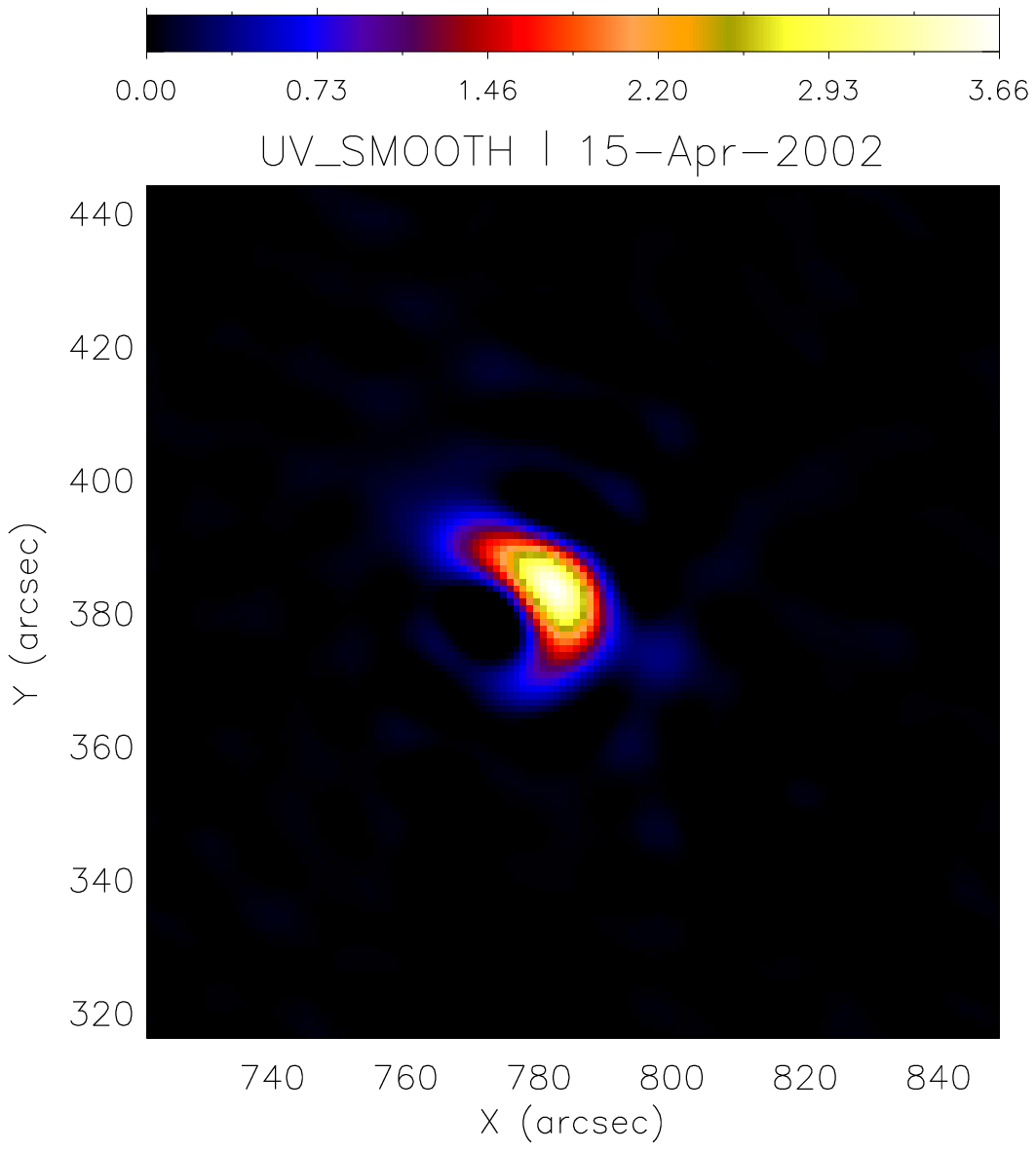}
 \includegraphics[width=0.25\textwidth]{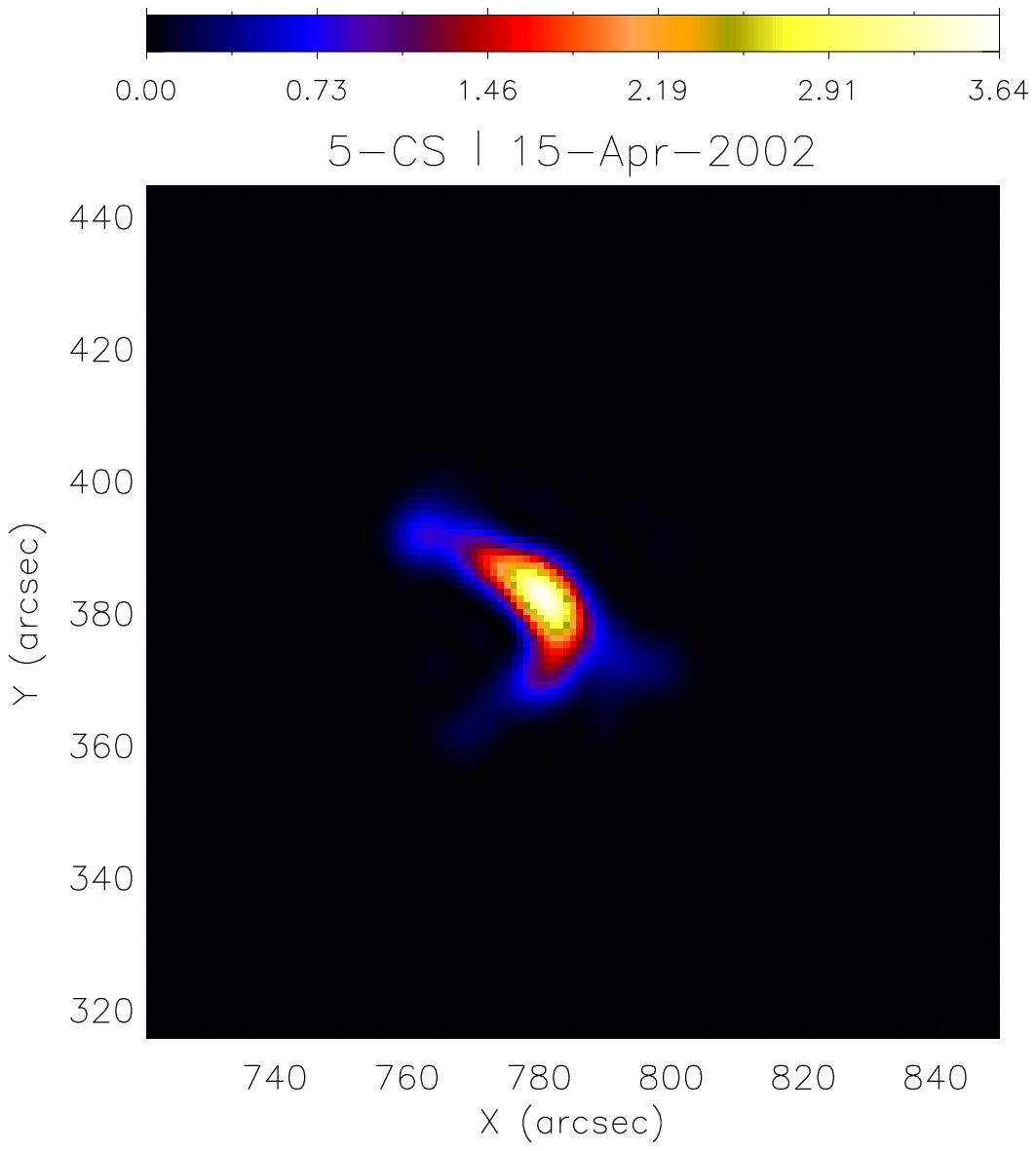}}\\[0.1cm]
 \renewcommand{\fevent}{02-Dec-2003_18-20keV}
 \subfloat[December 02, 2003 -- 22:54:00-22:58:00 UT -- (18-20 keV)]{
 \includegraphics[width=0.25\textwidth]{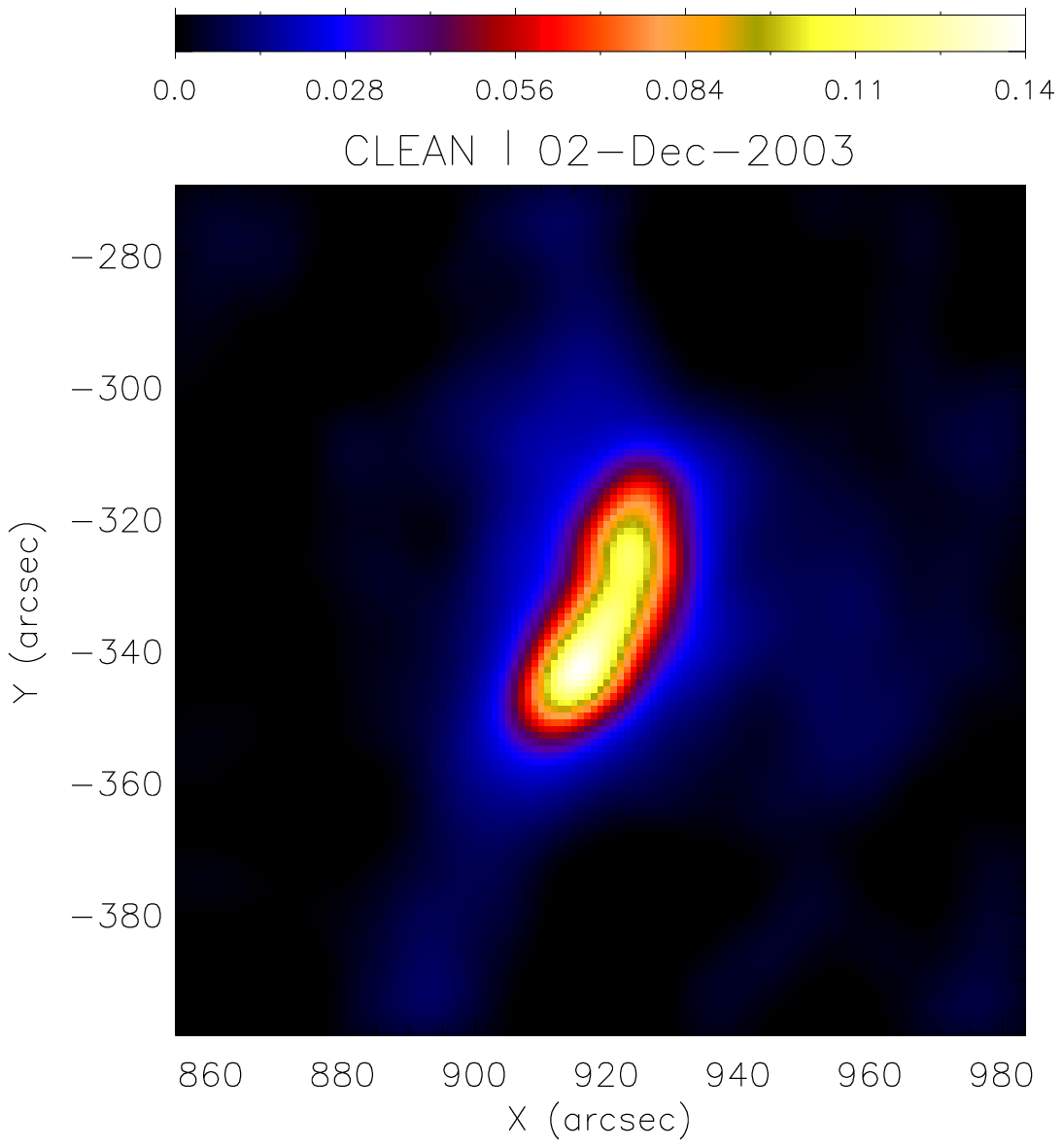}
 \includegraphics[width=0.25\textwidth]{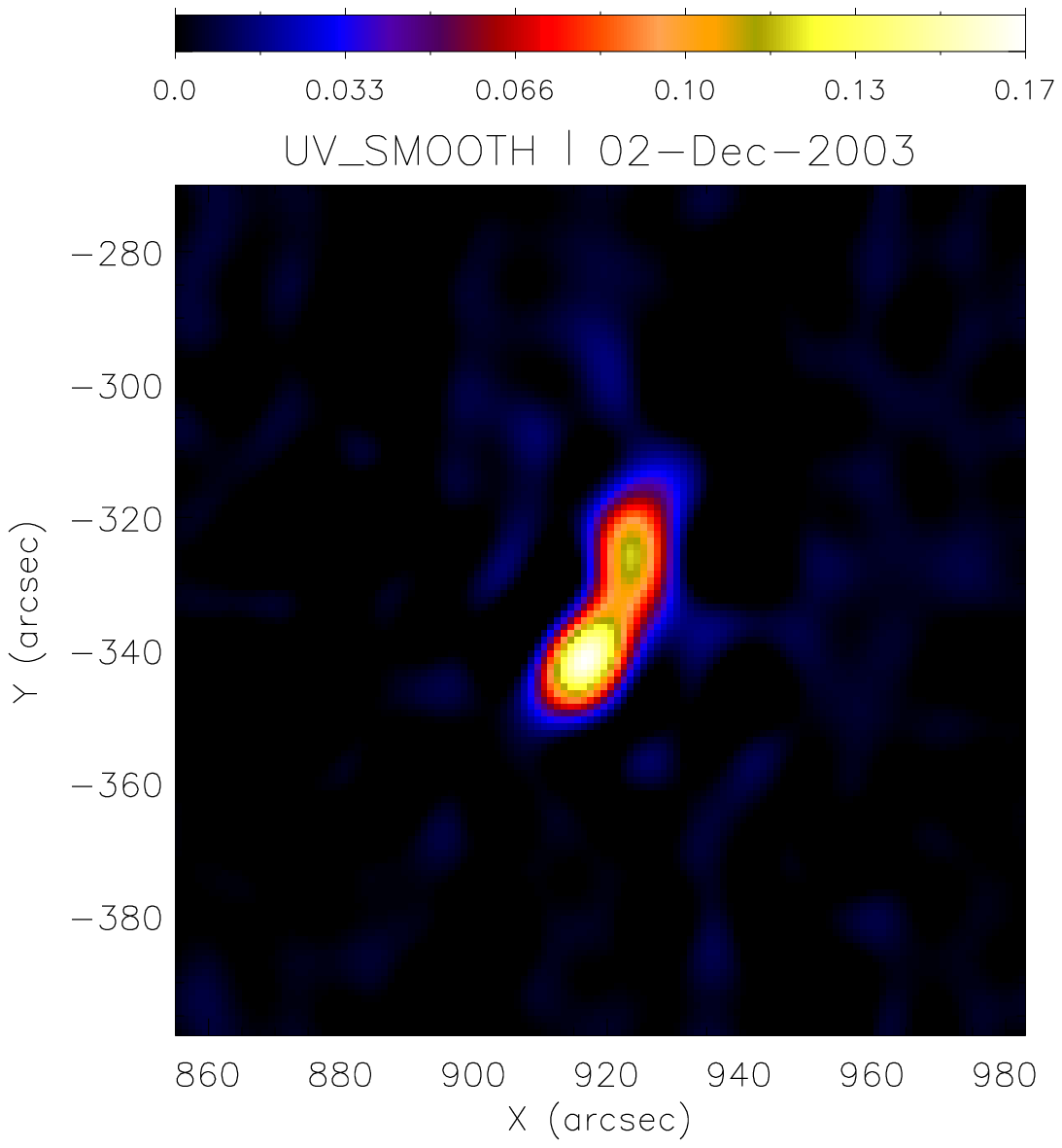}
 \includegraphics[width=0.25\textwidth]{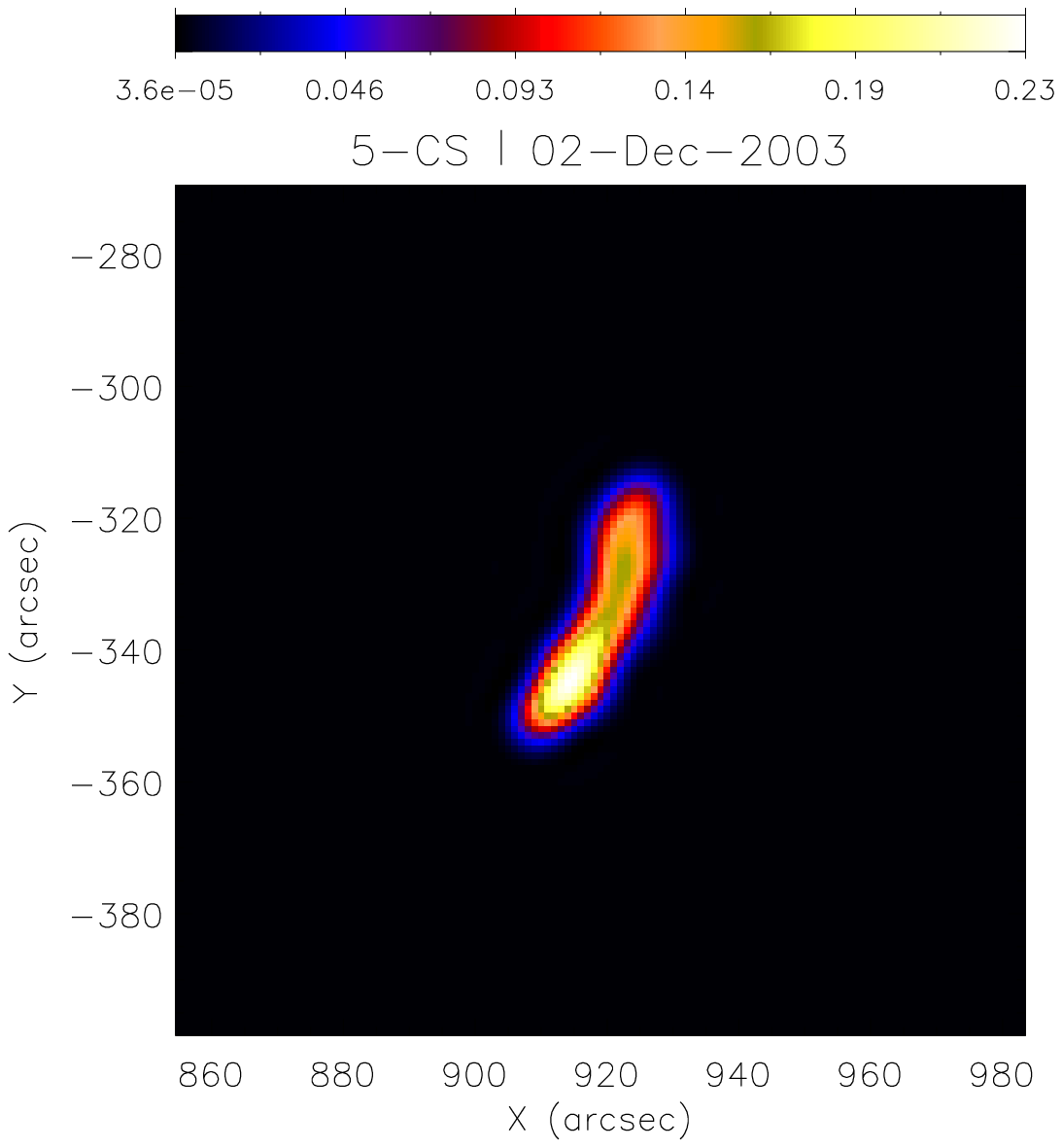}}\\[0.1cm]
 \renewcommand{\fevent}{31-Aug-2004_10-12keV}
 \subfloat[August 31, 2004 -- 05:33:00-05:38:00 UT -- (10-12 keV)]{
 \includegraphics[width=0.25\textwidth]{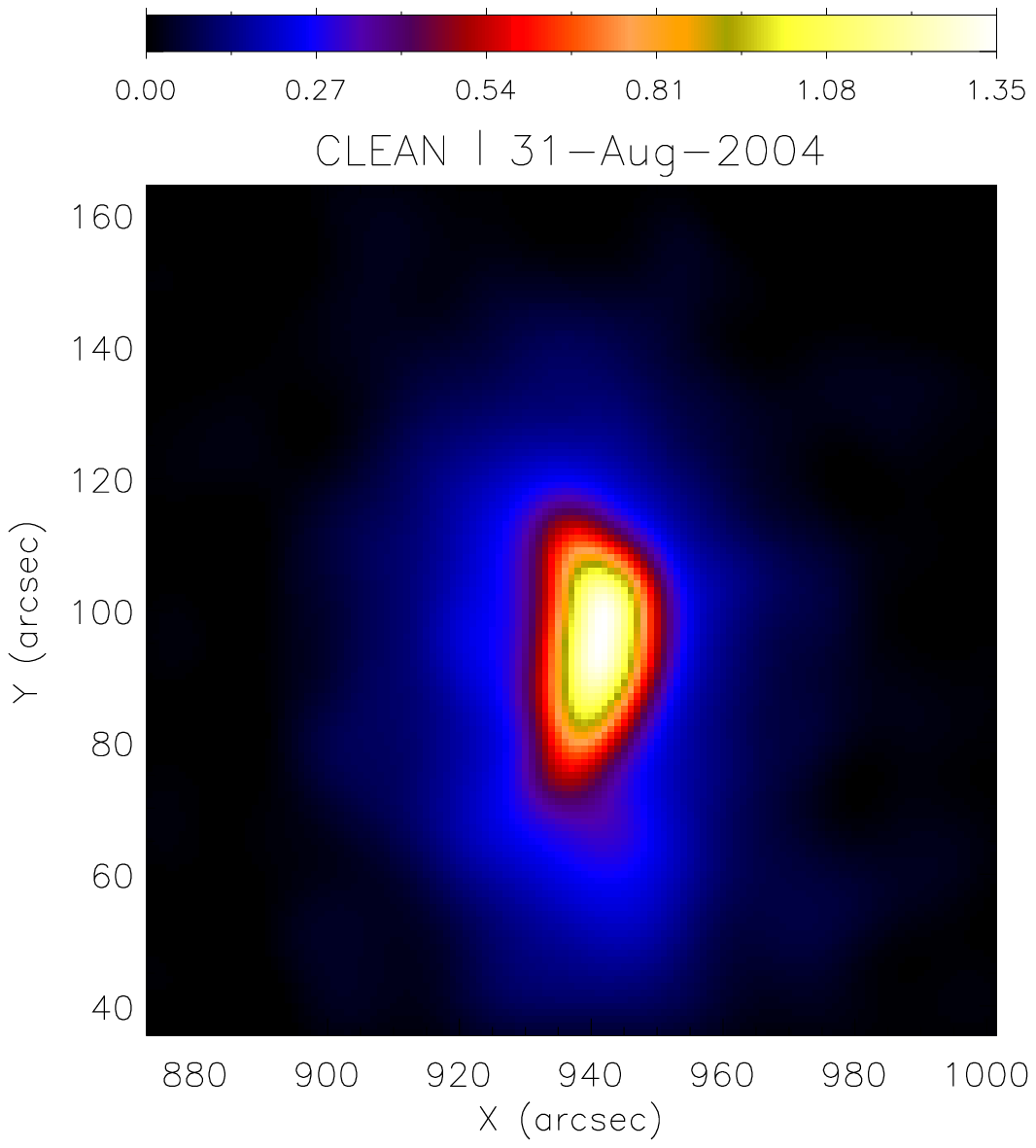}
 \includegraphics[width=0.25\textwidth]{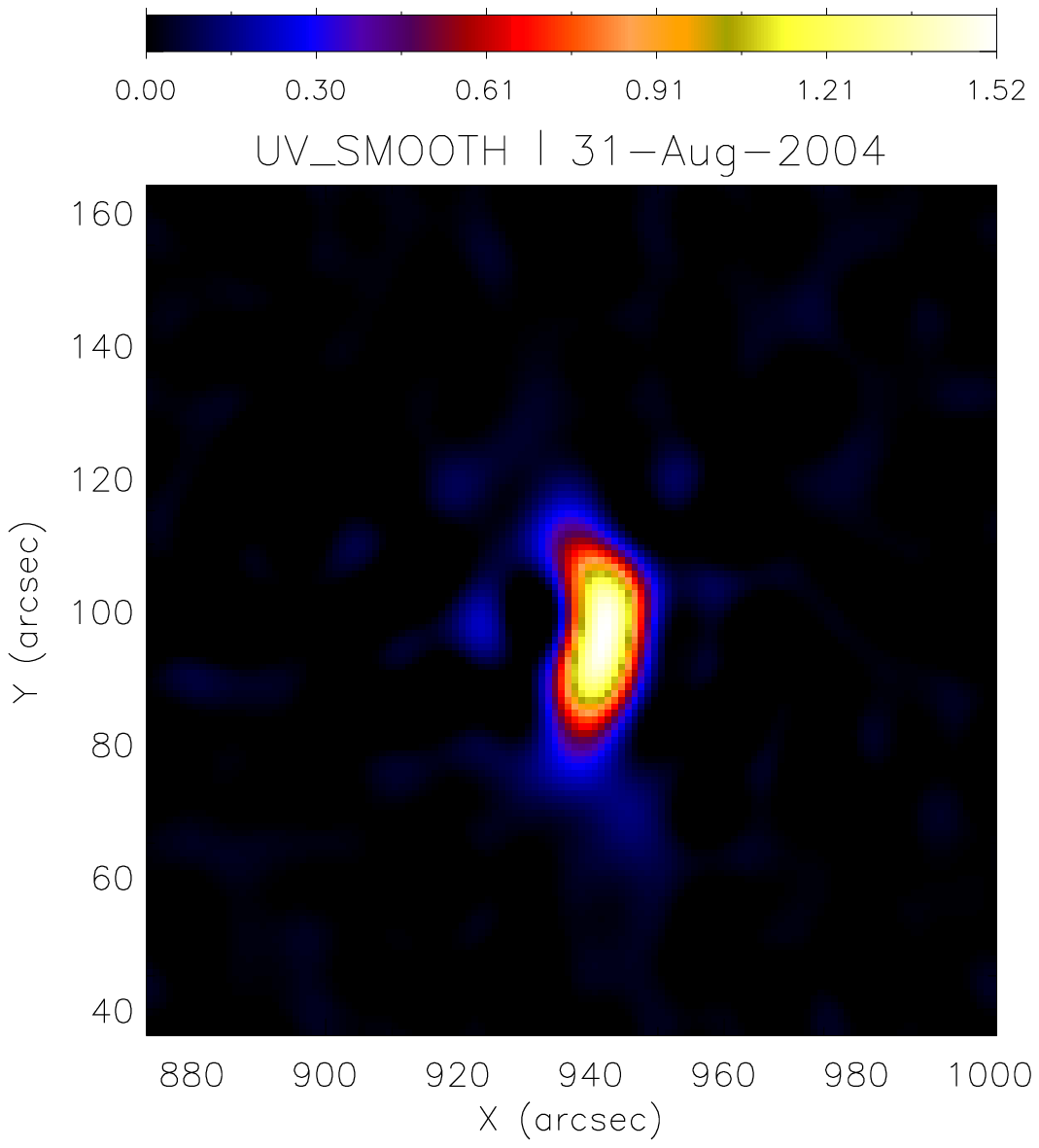}
 \includegraphics[width=0.25\textwidth]{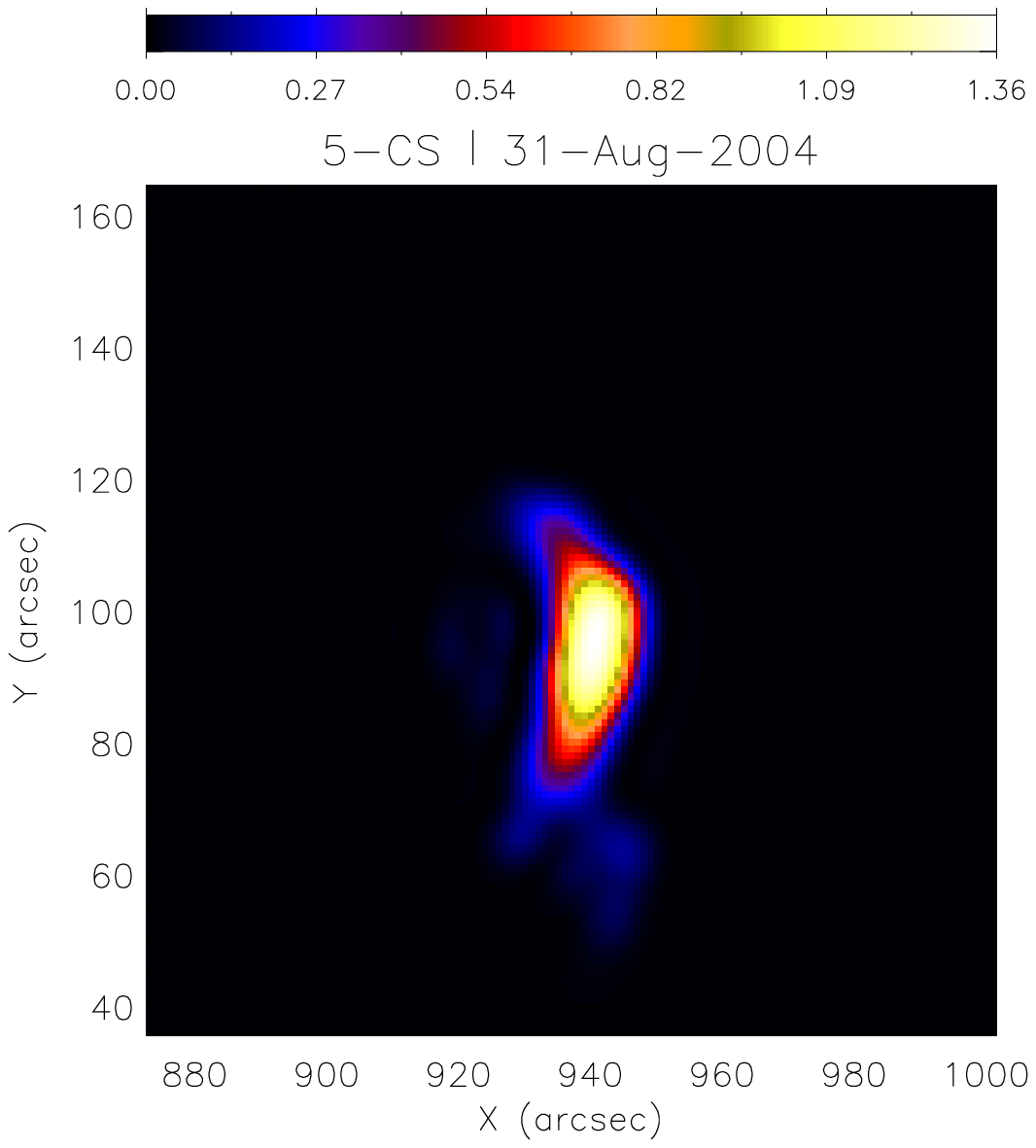}}\\[0.1cm]
 \renewcommand{\fevent}{13-May-2013_6-12keV}
 \subfloat[May 13, 2013 -- 02:04:16-02:04:48 UT -- (6-12 keV)]{
 \includegraphics[width=0.25\textwidth]{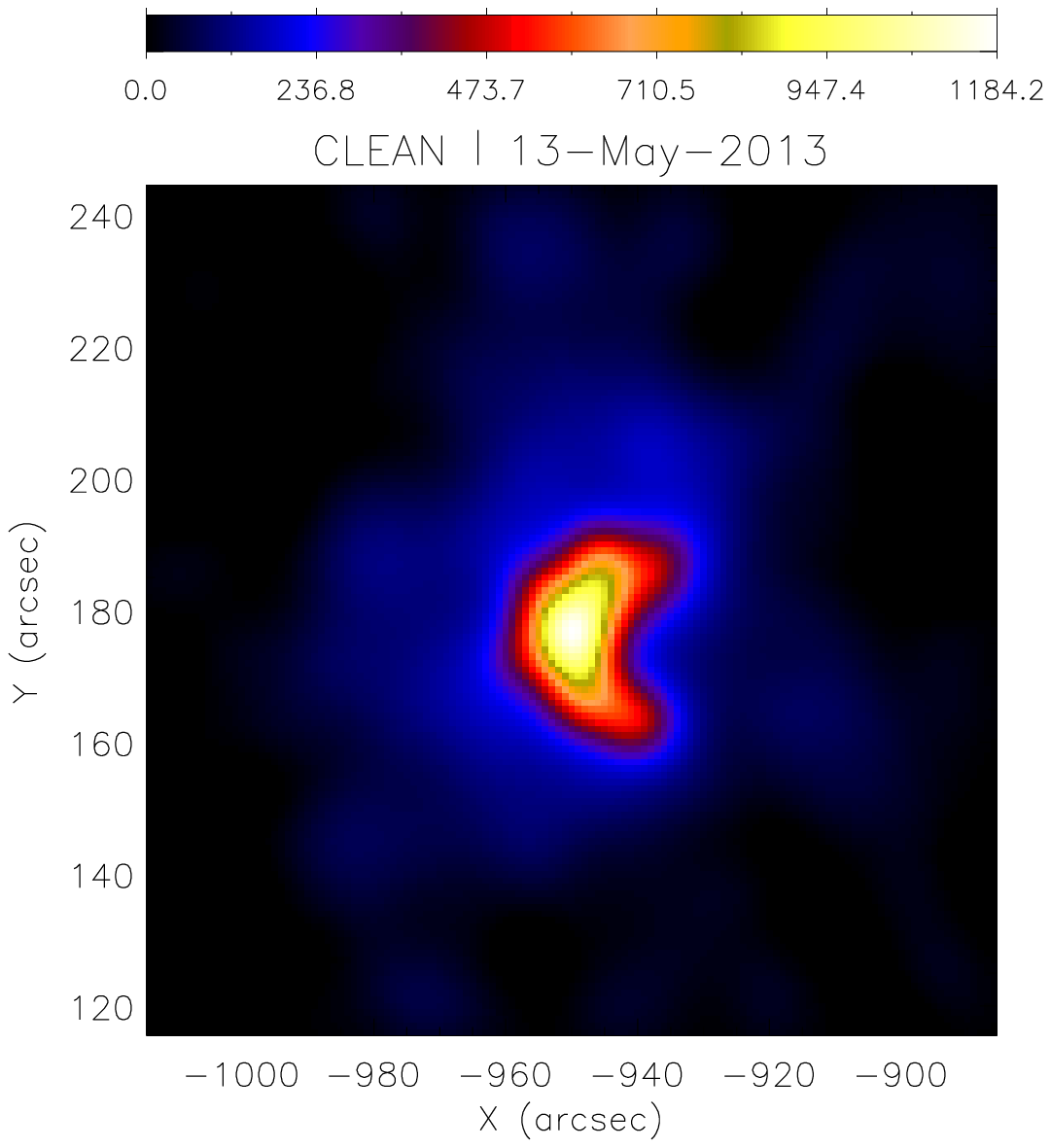}
 \includegraphics[width=0.25\textwidth]{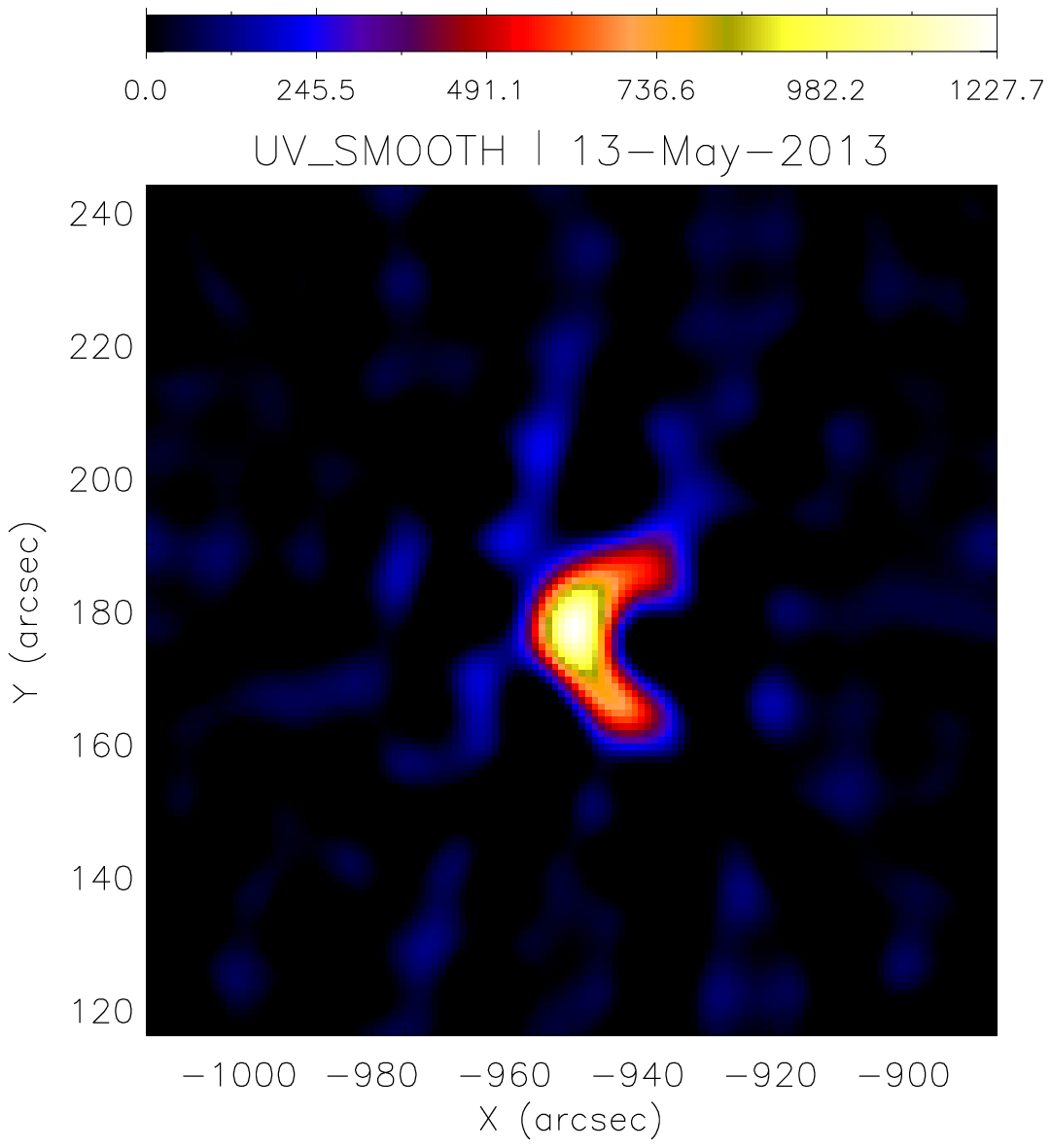}
 \includegraphics[width=0.25\textwidth]{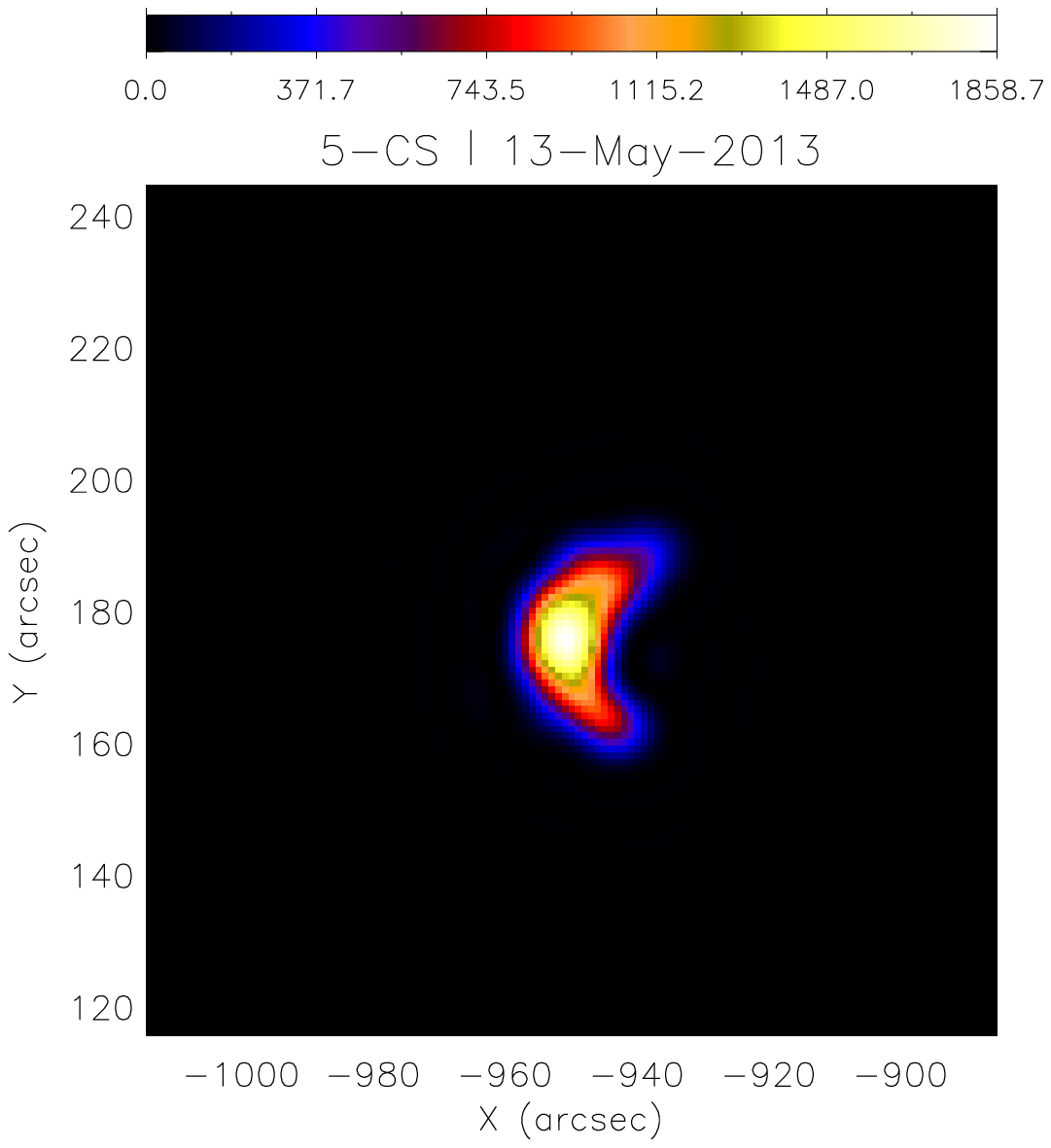}}
 \caption{Reconstructed images from different flaring events using three reconstruction methods: CLEAN (left), uv$\_$smooth (center), and 5-CS (Right). {\em{RHESSI}} detectors from 2 to 9 have been used to generate the visibilities.}
 \label{fig:rhessi_eval_mixt}
\end{figure*}

\begin{figure*}[!th]
 \centering
 \includegraphics[width=0.2\textwidth]{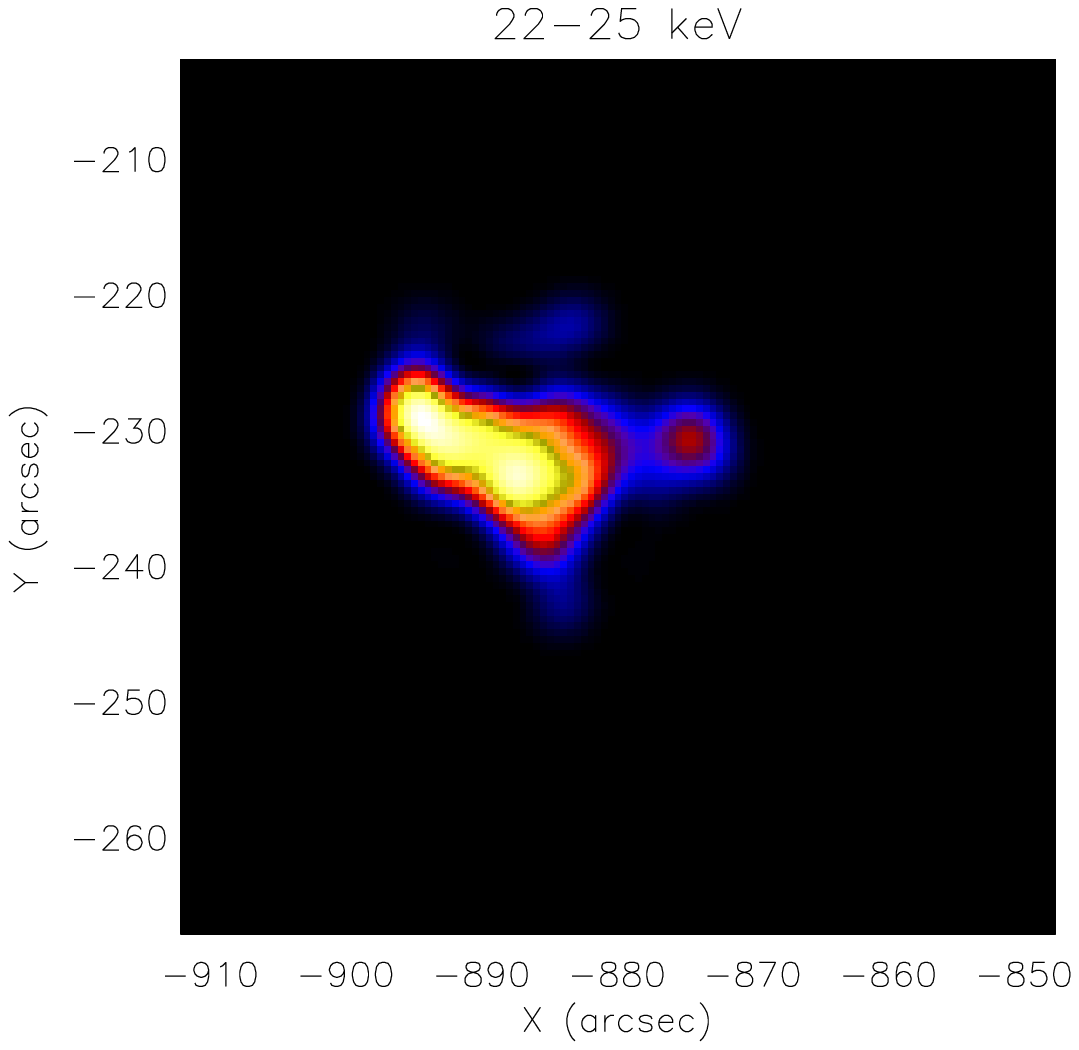}
 \includegraphics[width=0.2\textwidth]{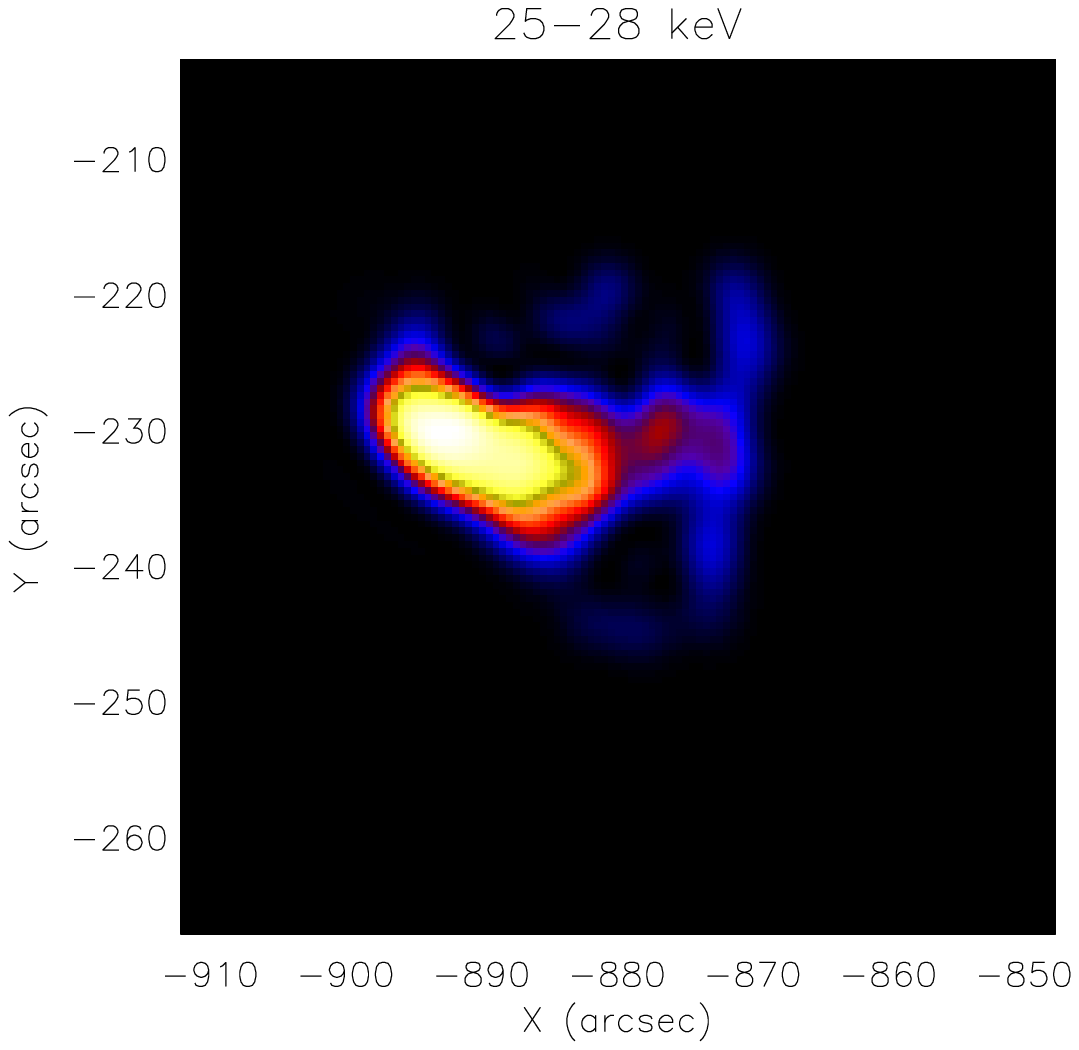}
 \includegraphics[width=0.2\textwidth]{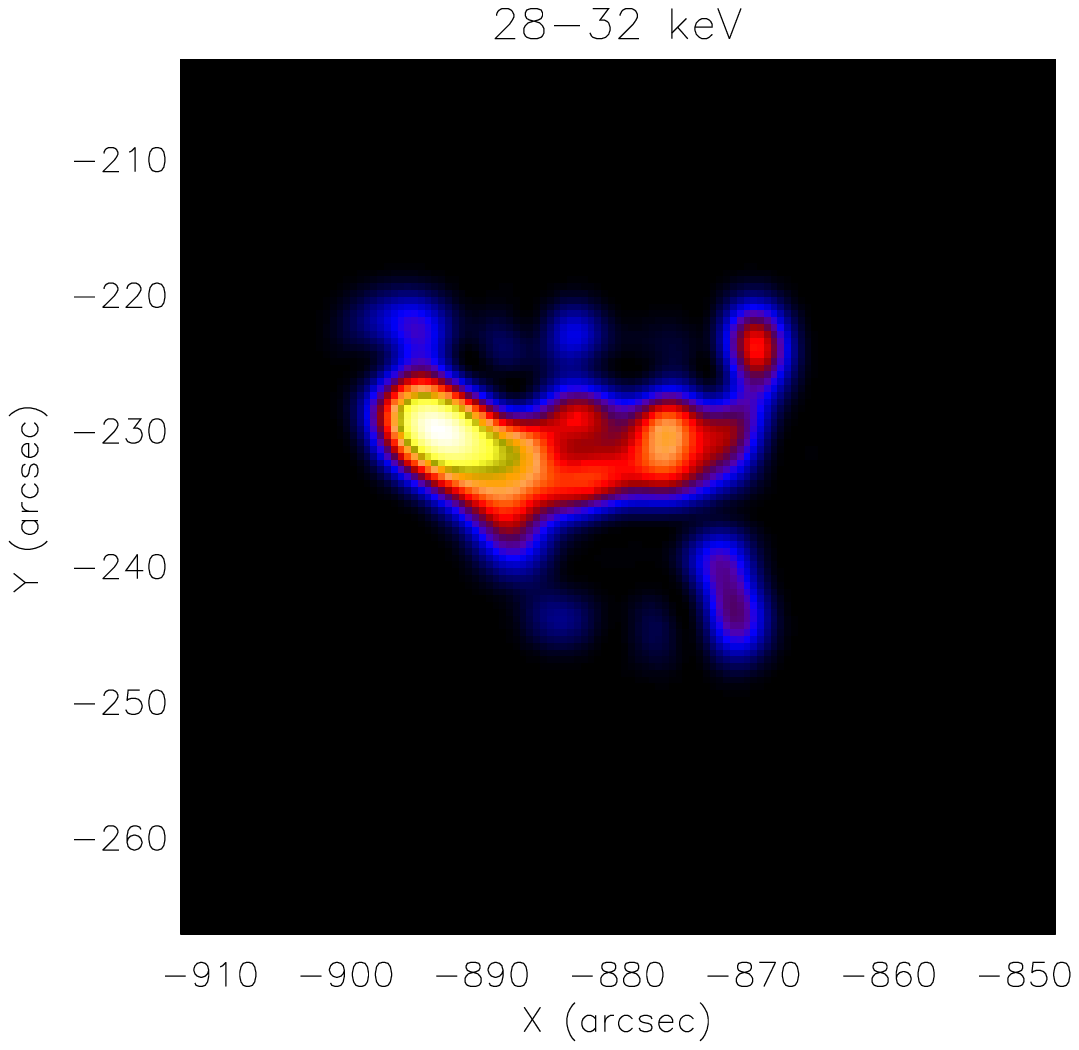}
 \includegraphics[width=0.2\textwidth]{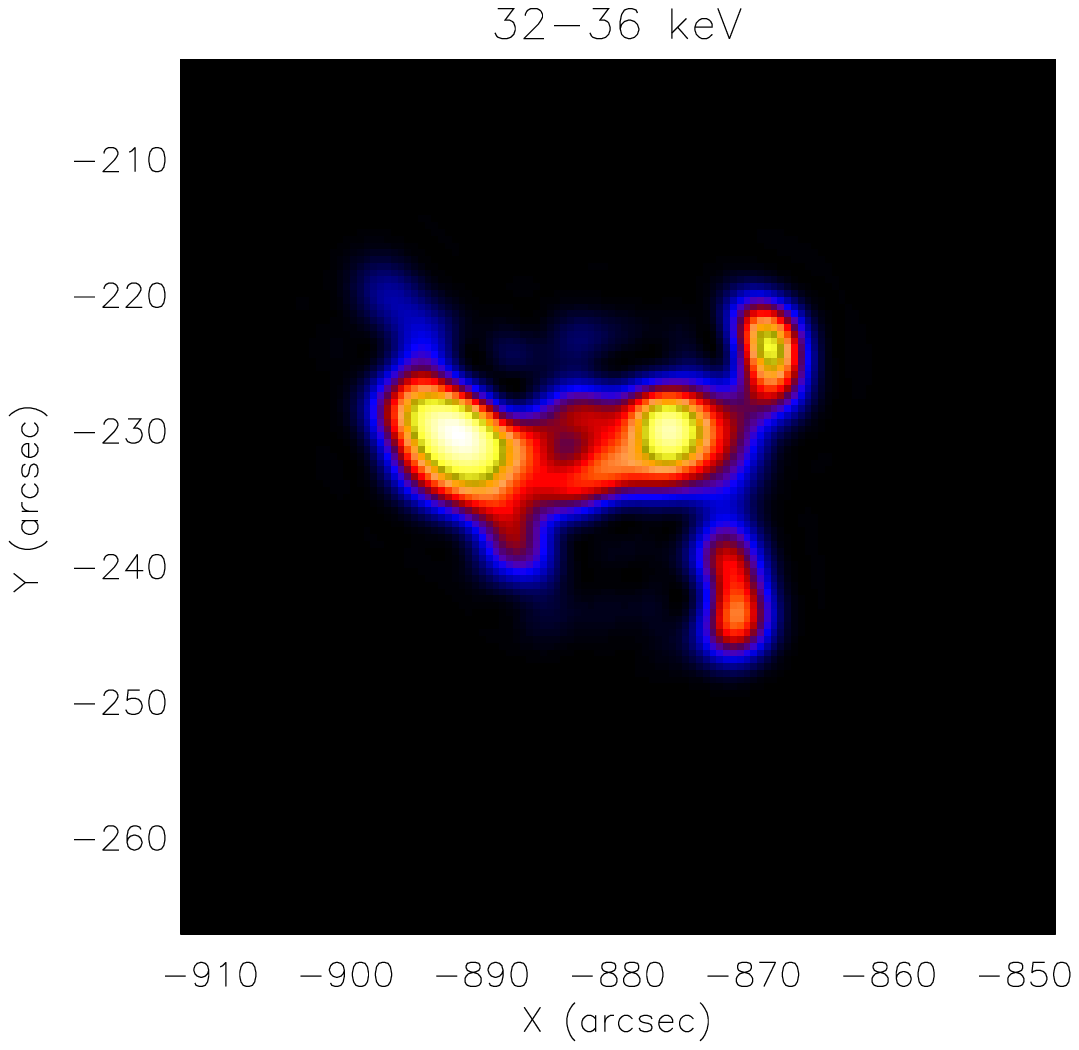}\\[0.1cm]
 \includegraphics[width=0.2\textwidth]{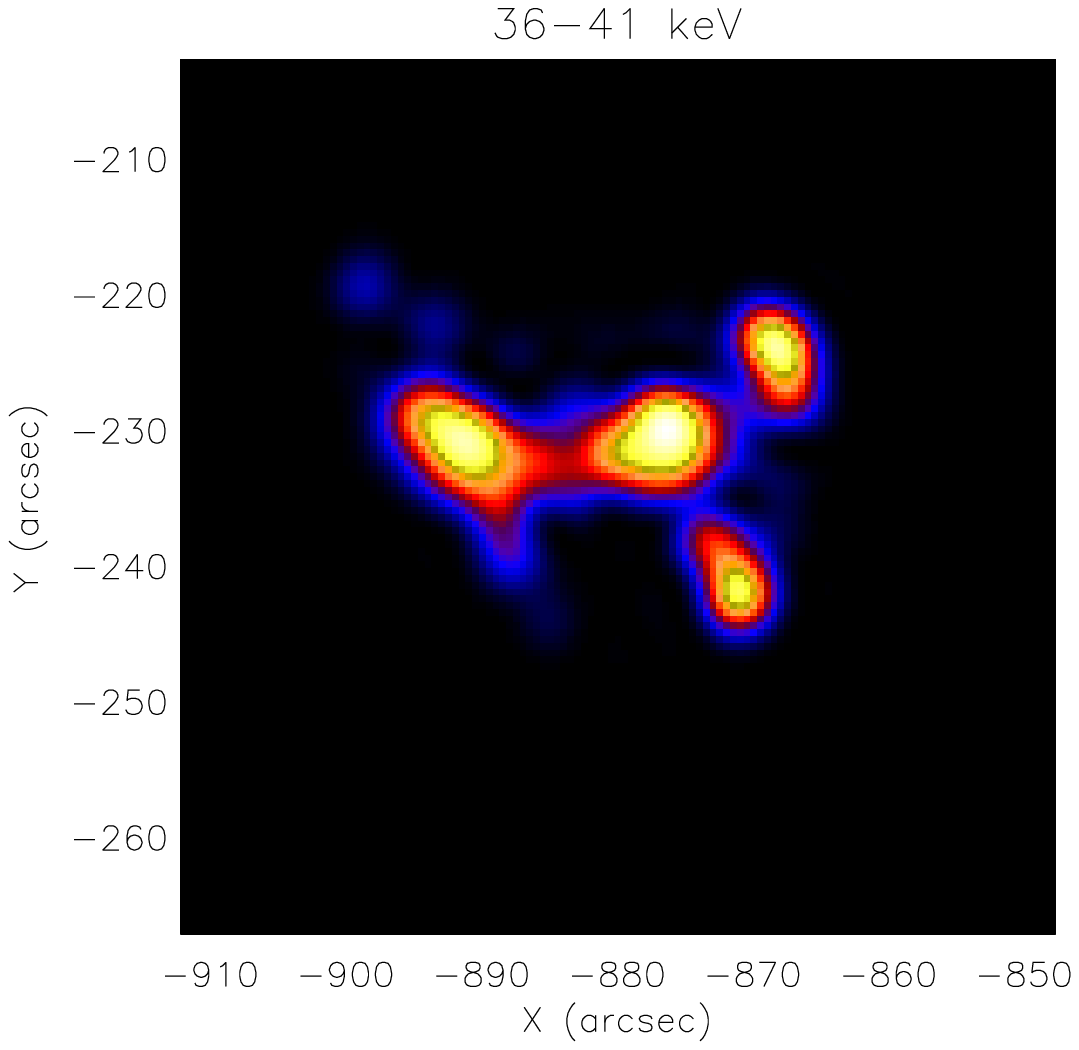}
 \includegraphics[width=0.2\textwidth]{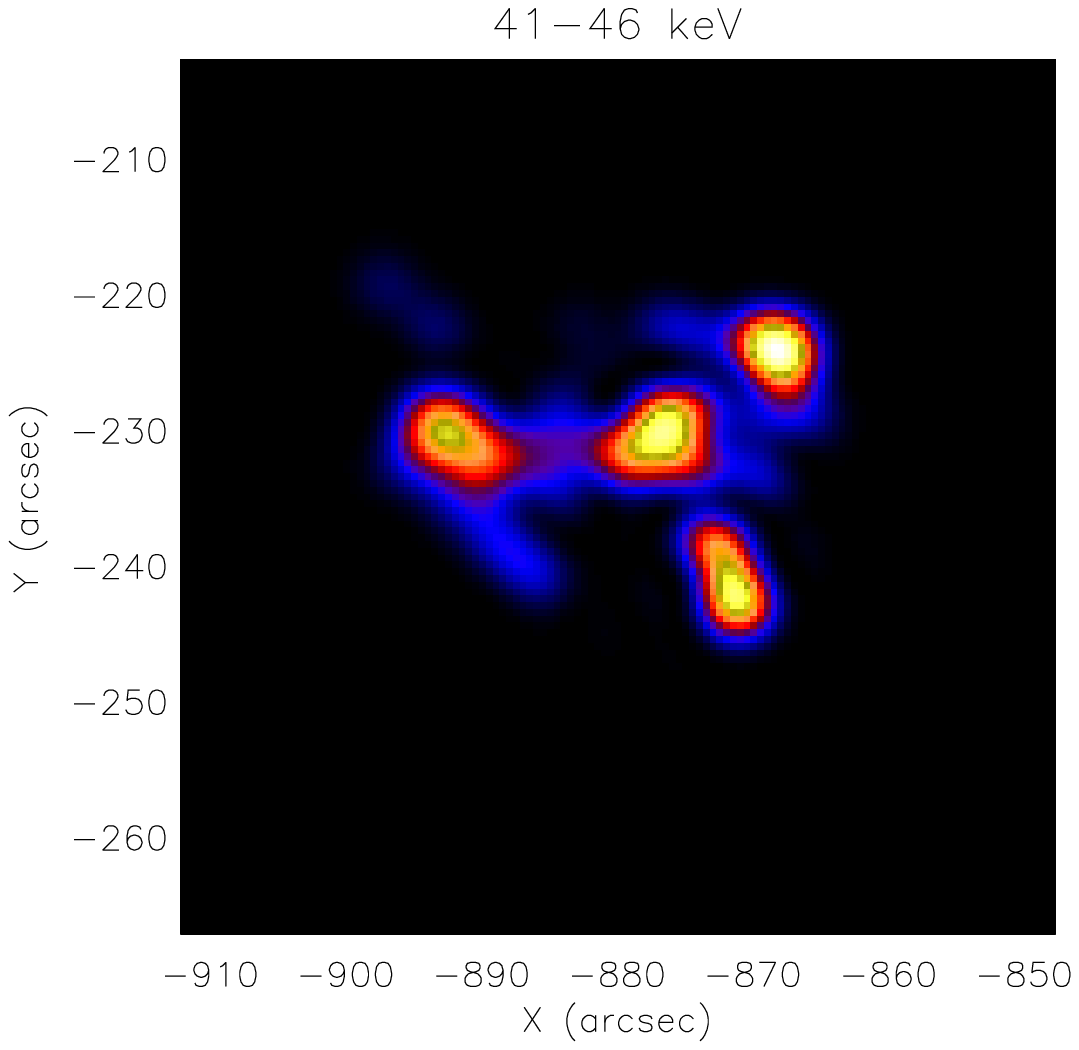}
 \includegraphics[width=0.2\textwidth]{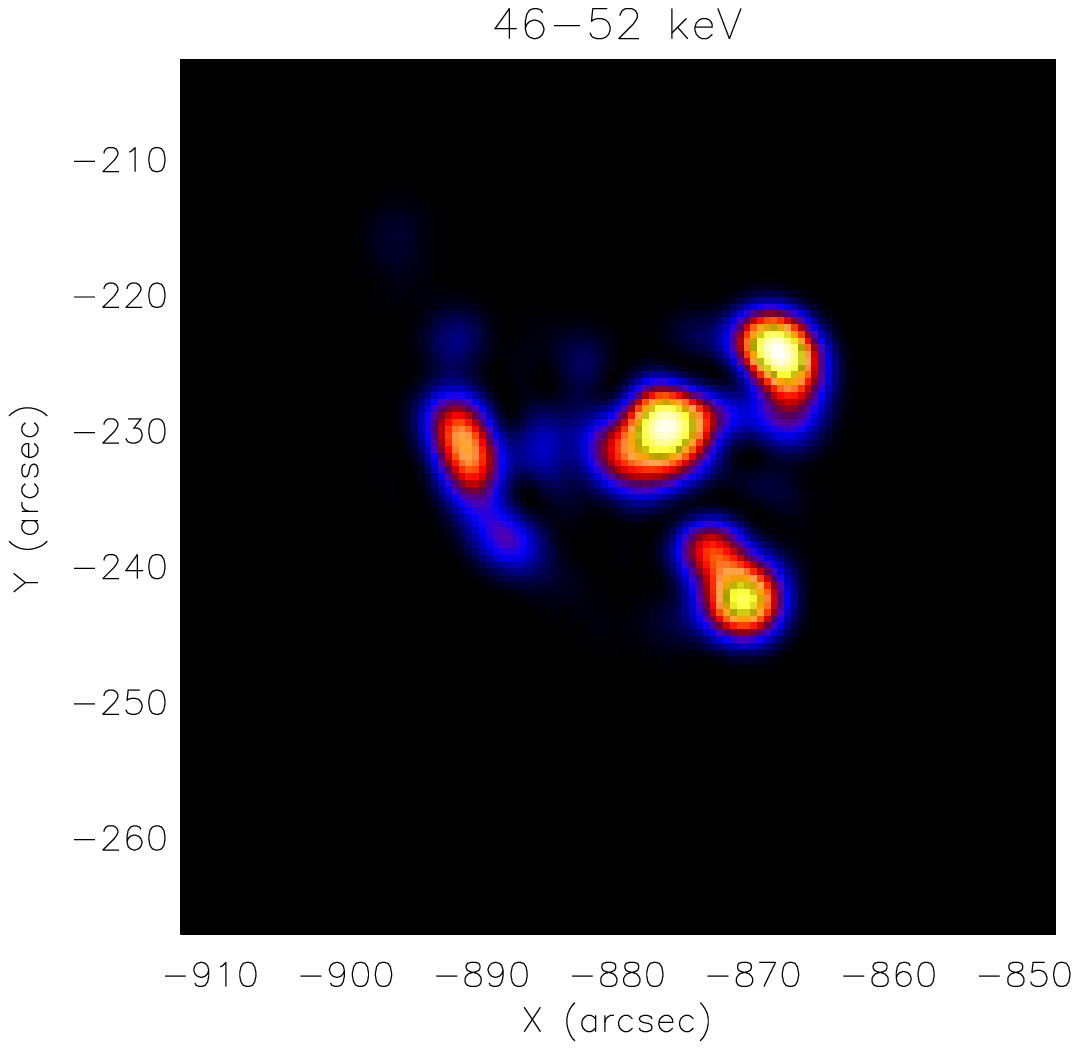}
 \includegraphics[width=0.2\textwidth]{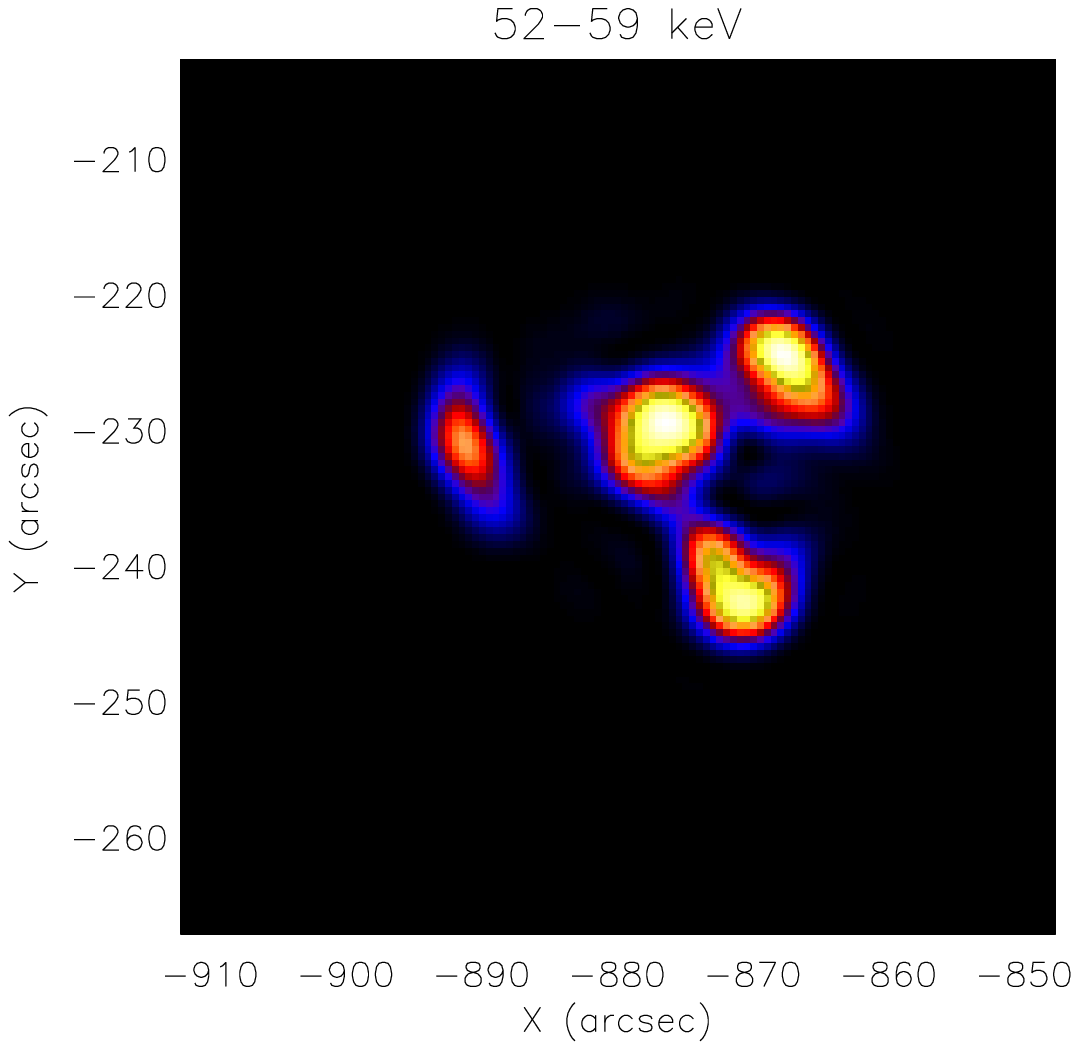}\\[0.1cm]
 \includegraphics[width=0.2\textwidth]{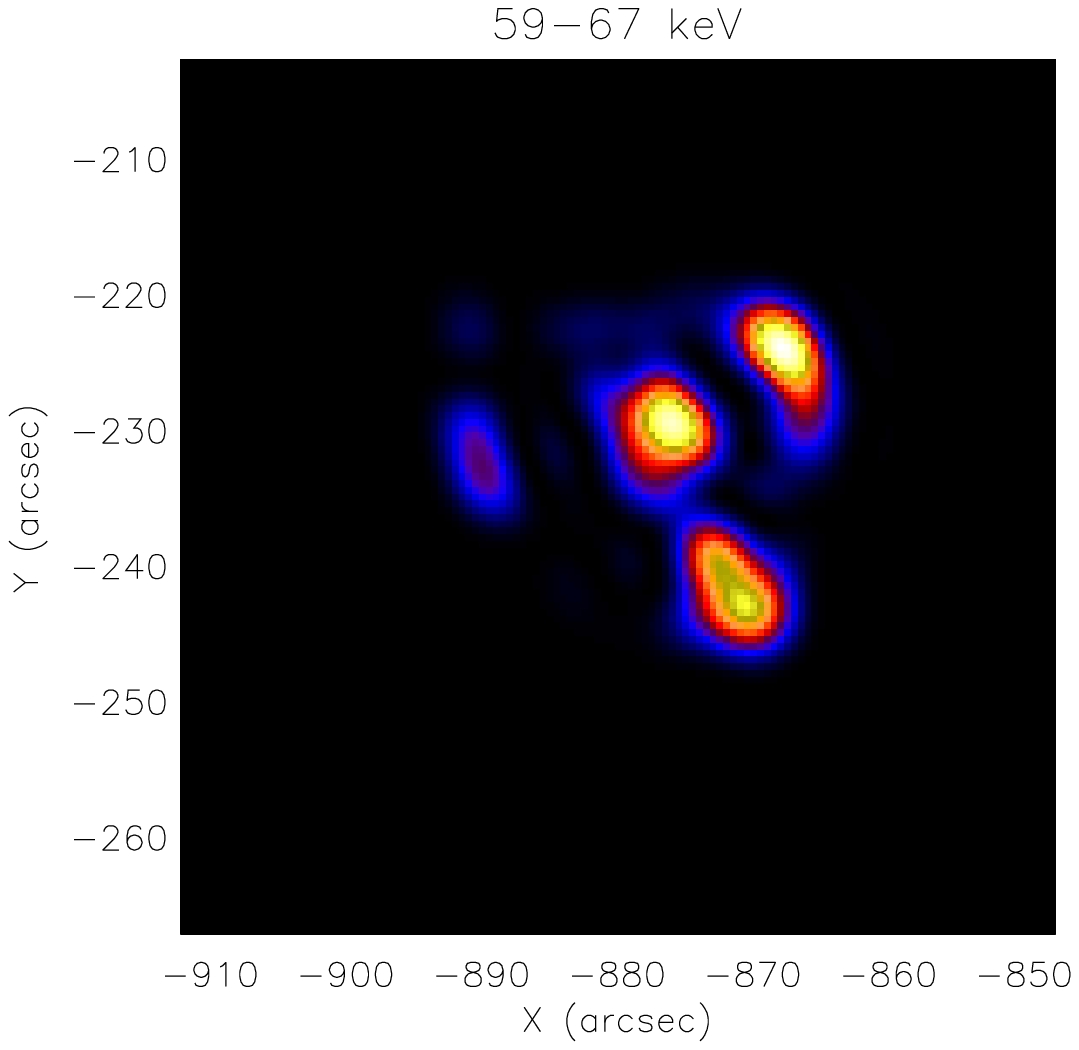}
 \includegraphics[width=0.2\textwidth]{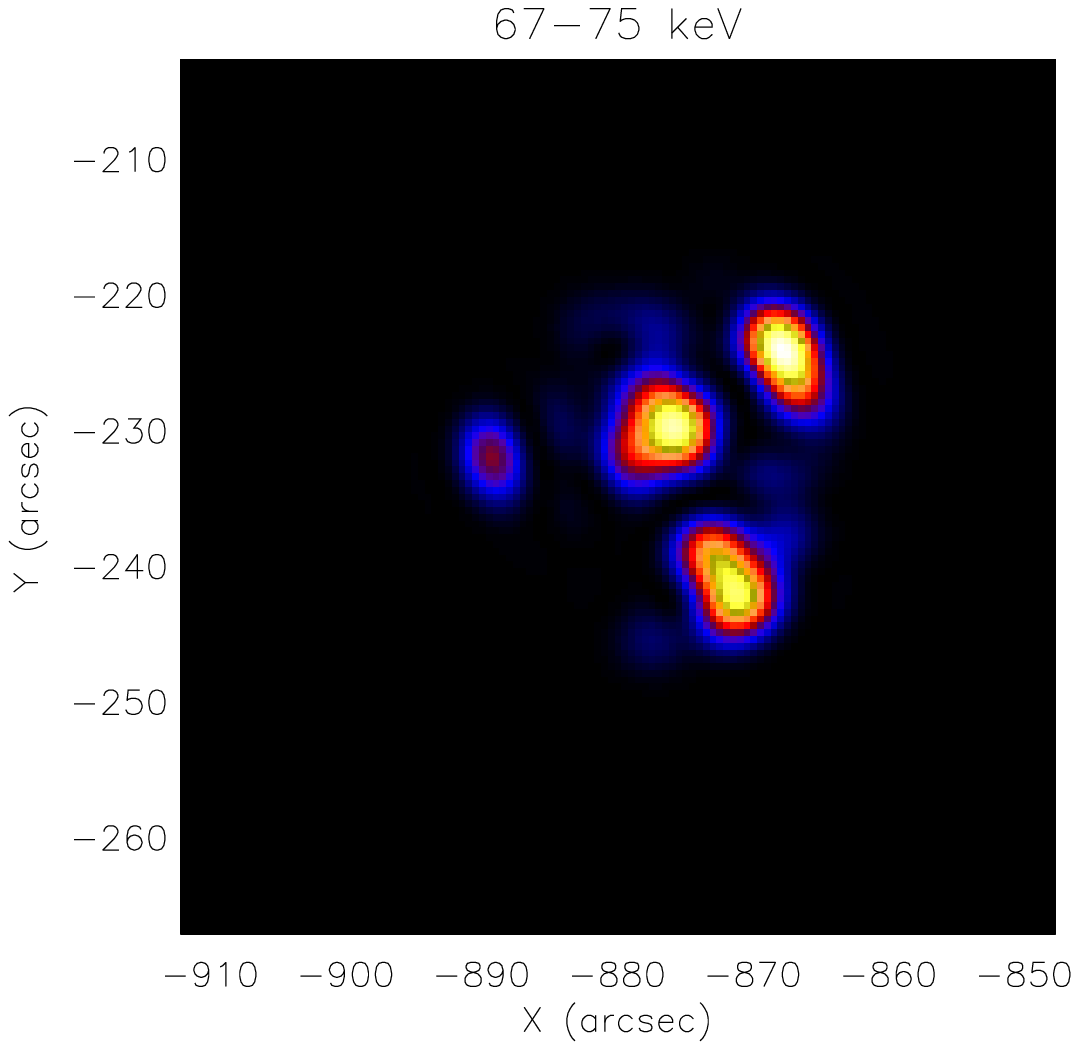}
 \includegraphics[width=0.2\textwidth]{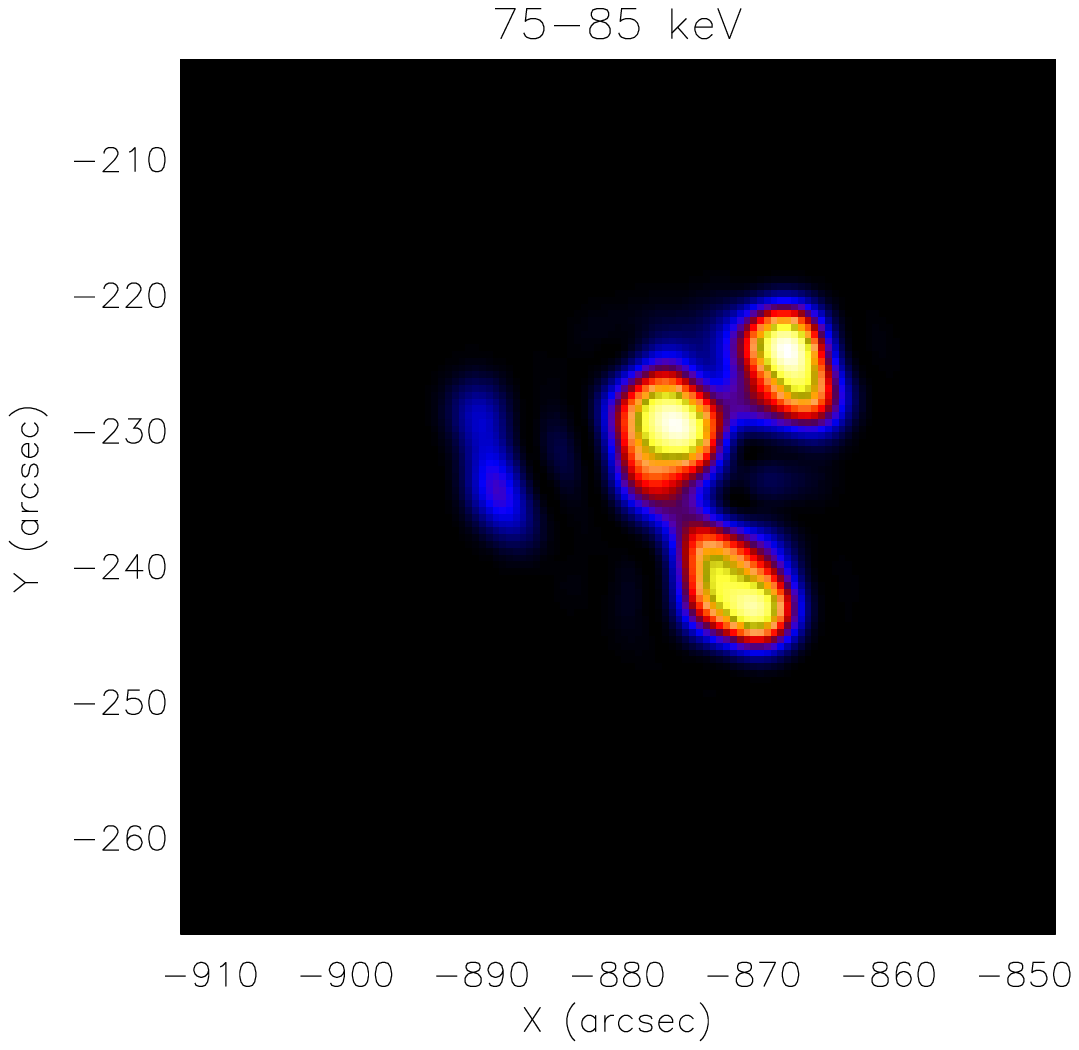}
 \includegraphics[width=0.2\textwidth]{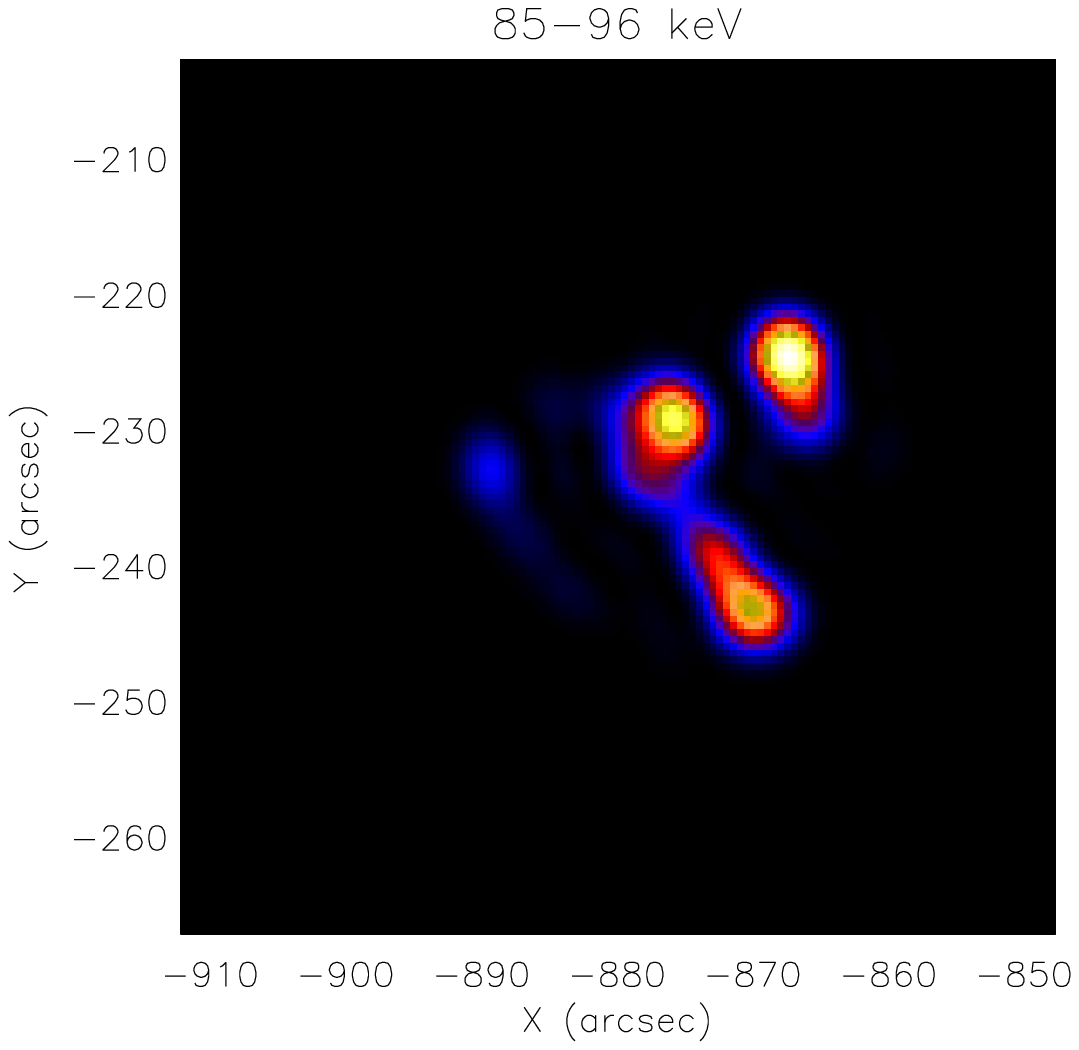}\\[0.1cm]
 \includegraphics[width=0.2\textwidth]{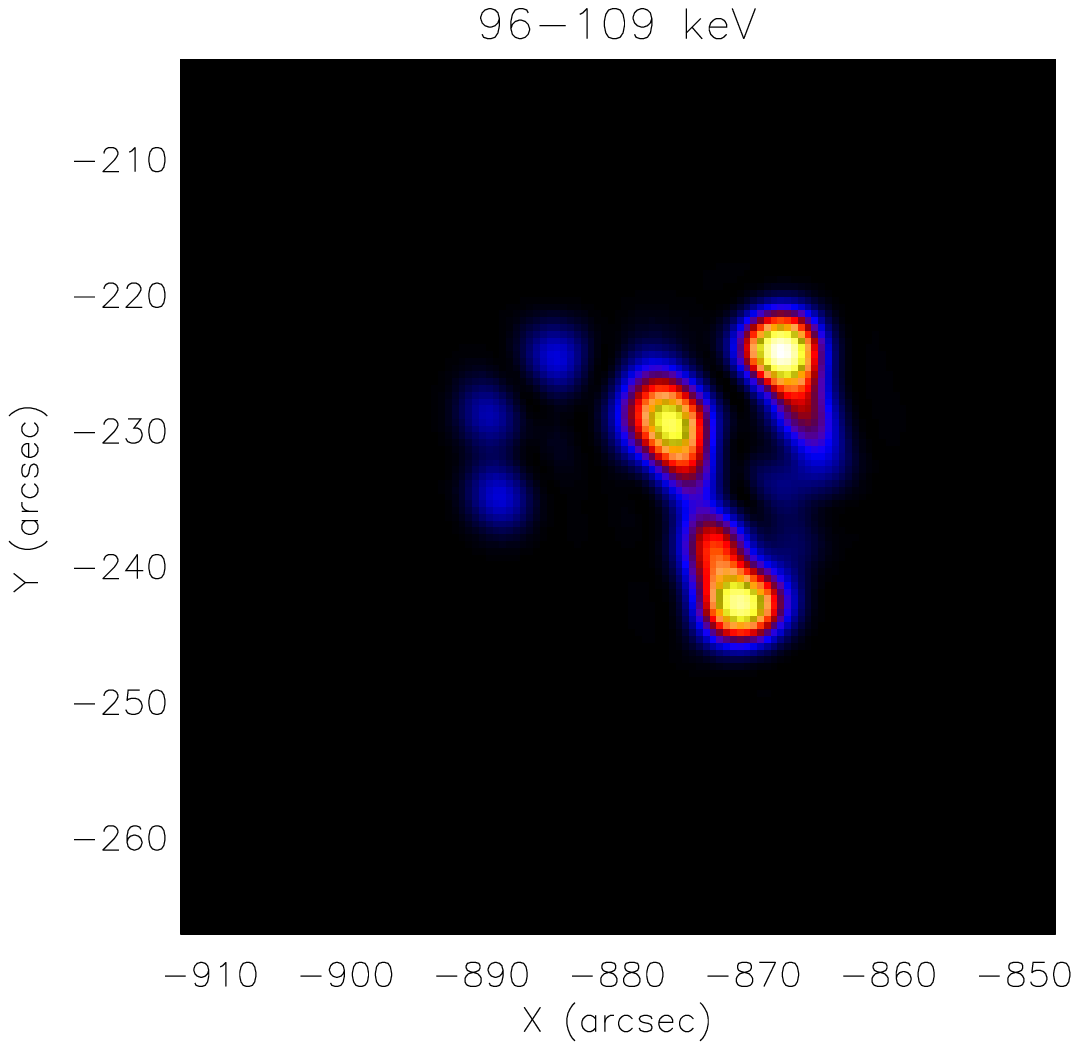}
 \includegraphics[width=0.2\textwidth]{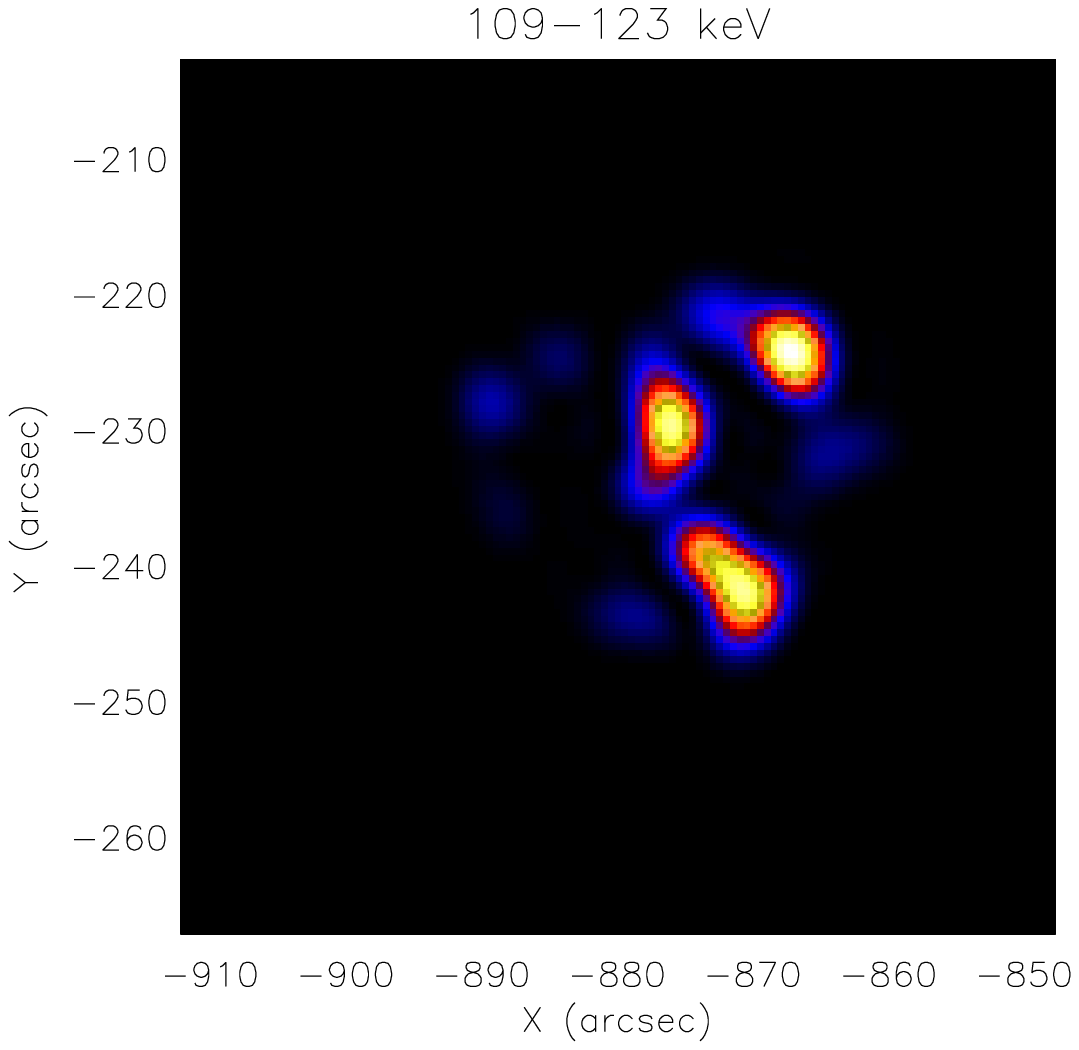}
 \includegraphics[width=0.2\textwidth]{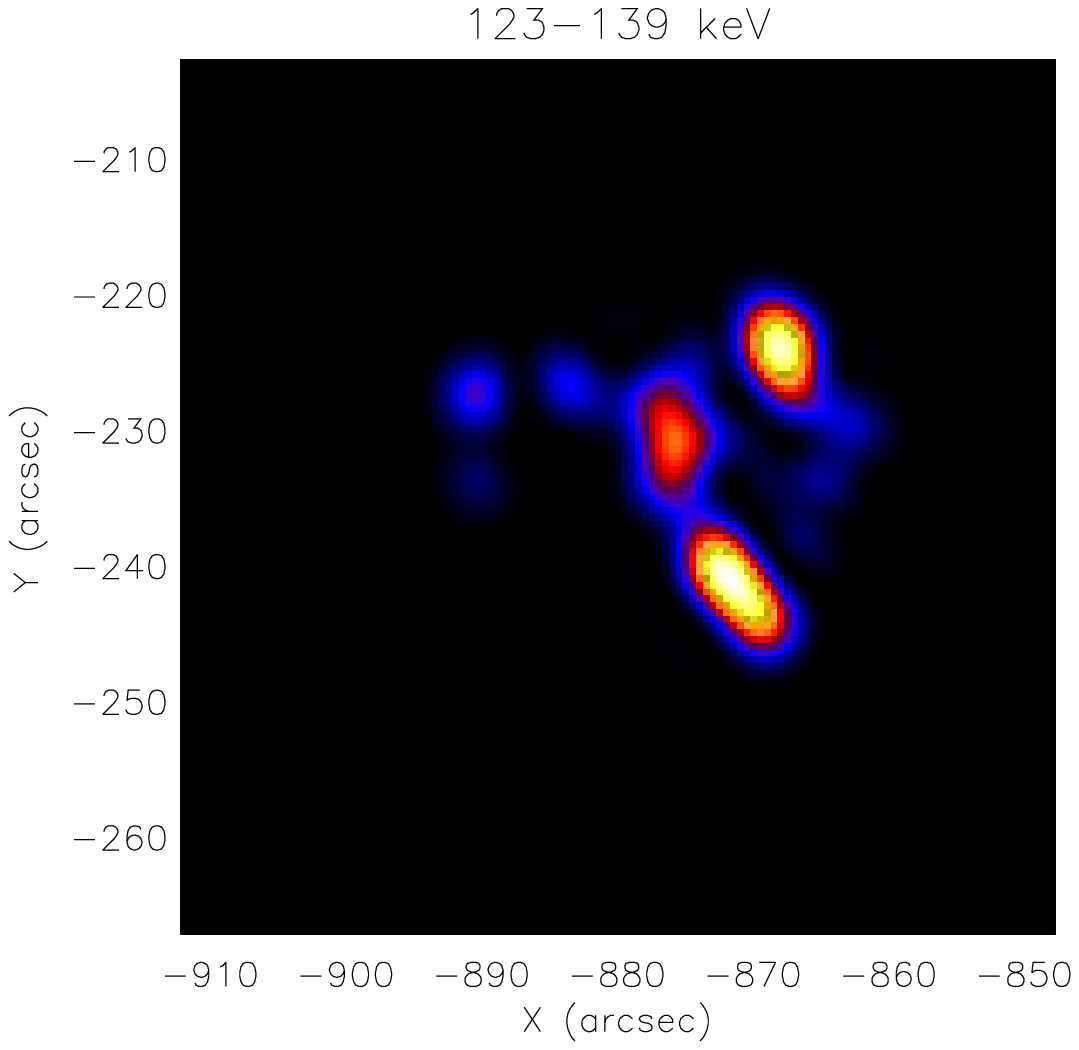}
 \includegraphics[width=0.2\textwidth]{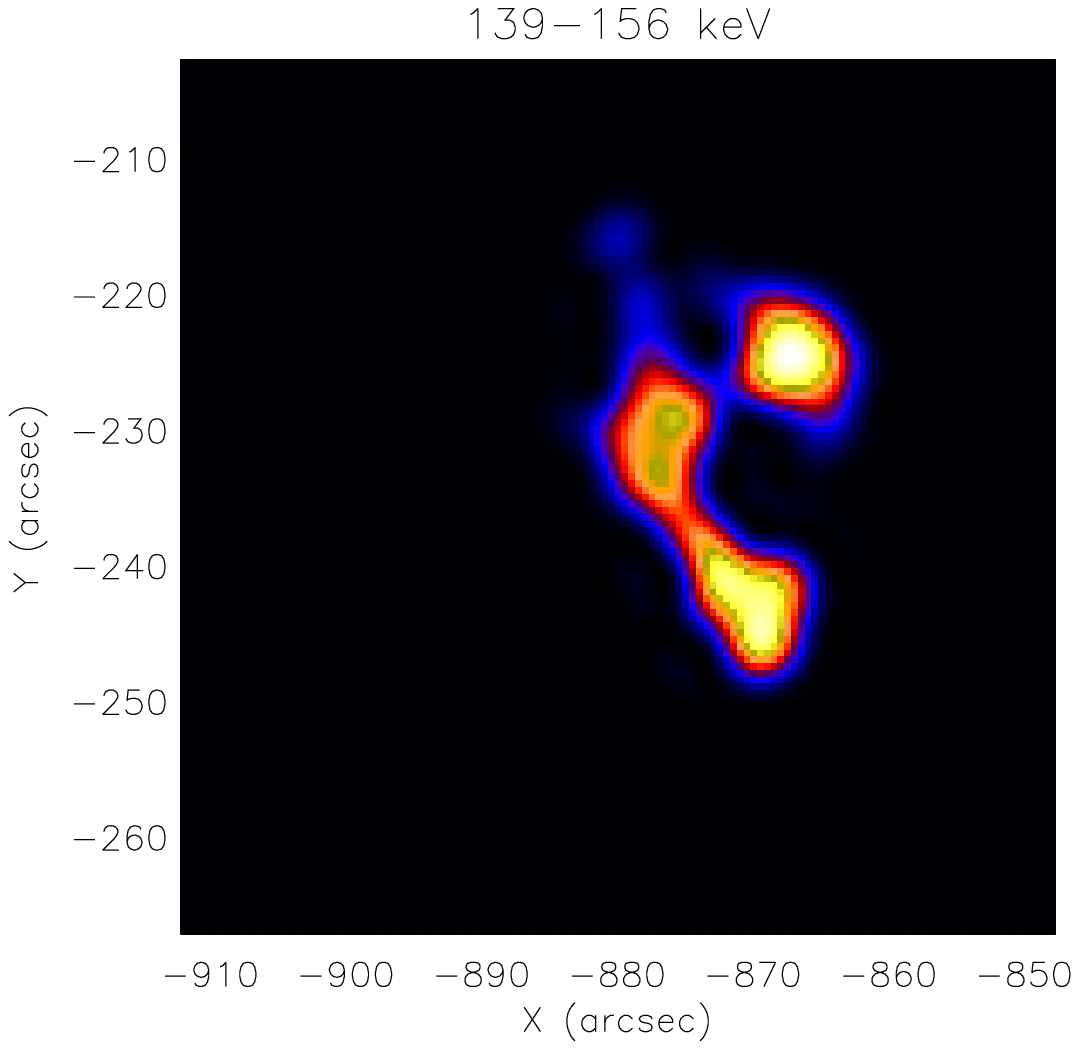}
 \caption{Images reconstructed by 5-CS for the July 23, 2002 flaring event through different energy ranges (00:29:10-00:30:19 UT). Visibilities collected by {\em{RHESSI}} collimators from 2 to 9 have been used for the reconstructions.}
 \label{fig:rhessi_23-Jul-2002_energy_analysis}
\end{figure*}

\begin{figure*}[!th]
 \centering
 \includegraphics[width=0.2\textwidth]{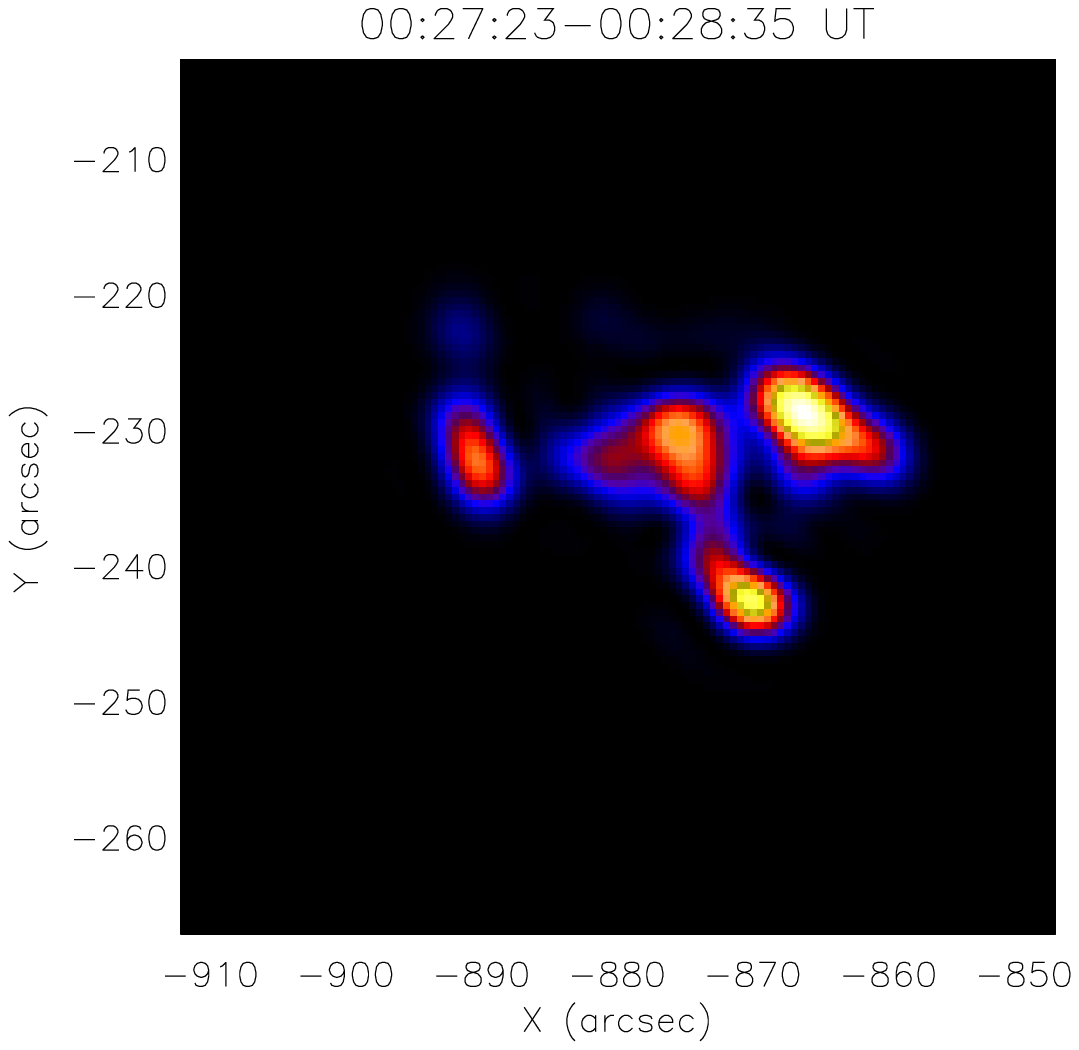}
 \includegraphics[width=0.2\textwidth]{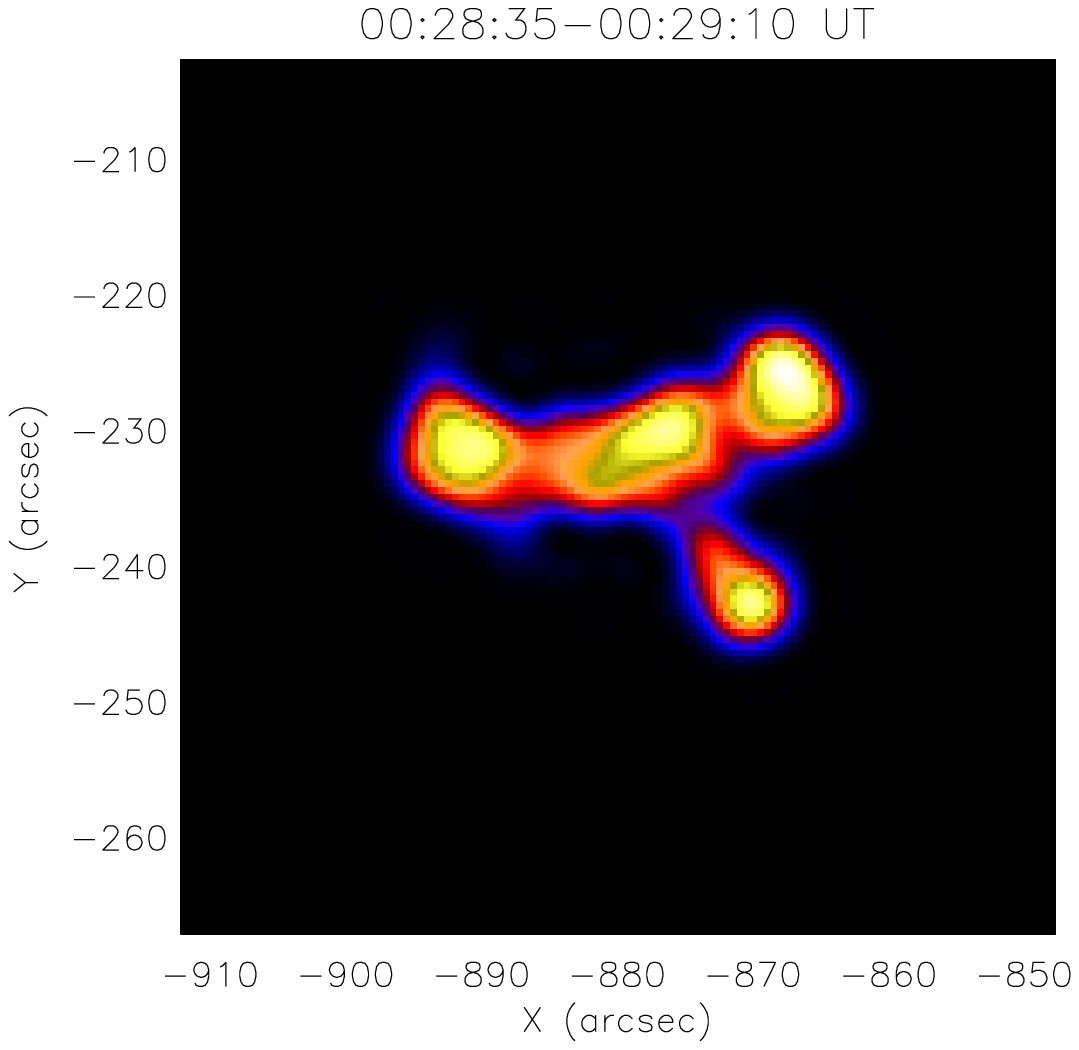}
 \includegraphics[width=0.2\textwidth]{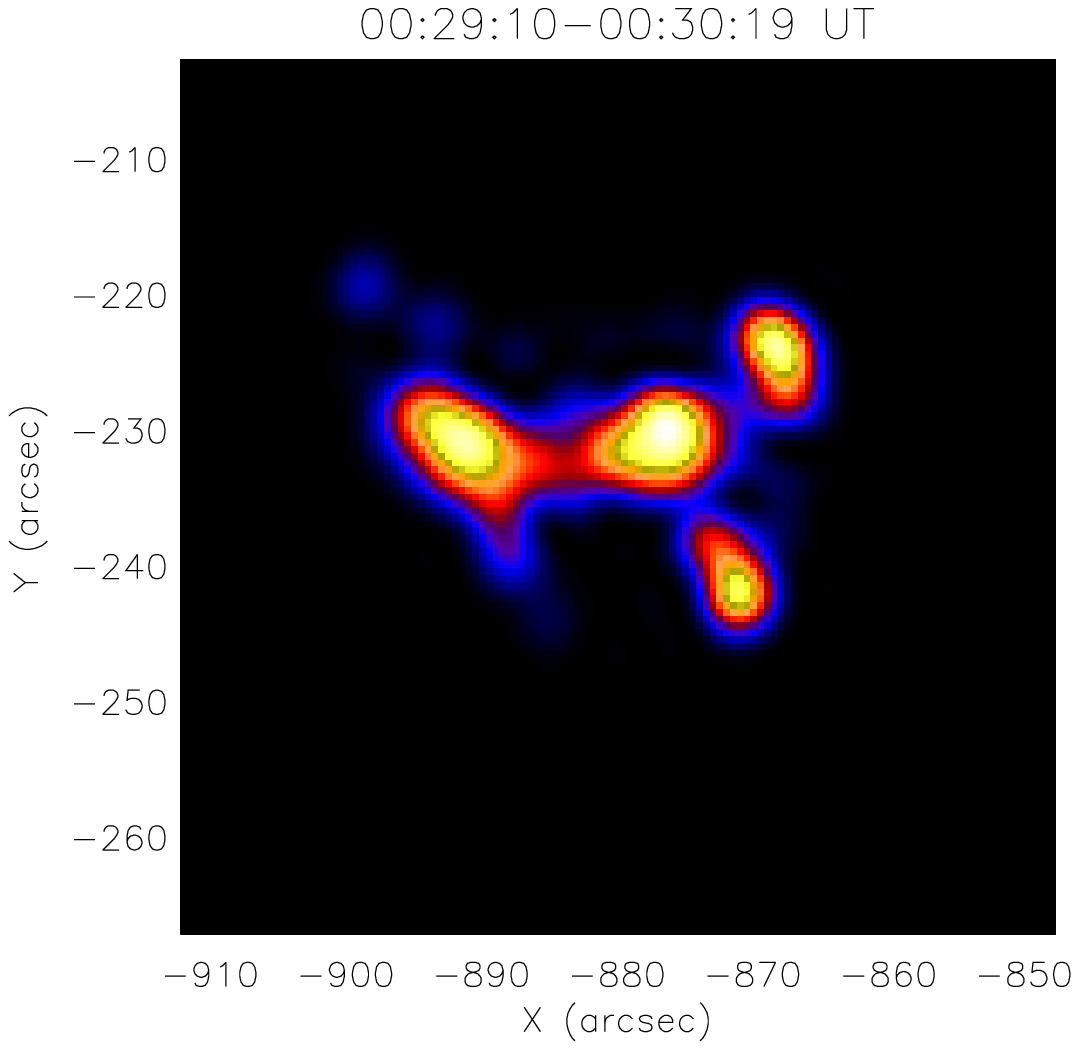}
 \includegraphics[width=0.2\textwidth]{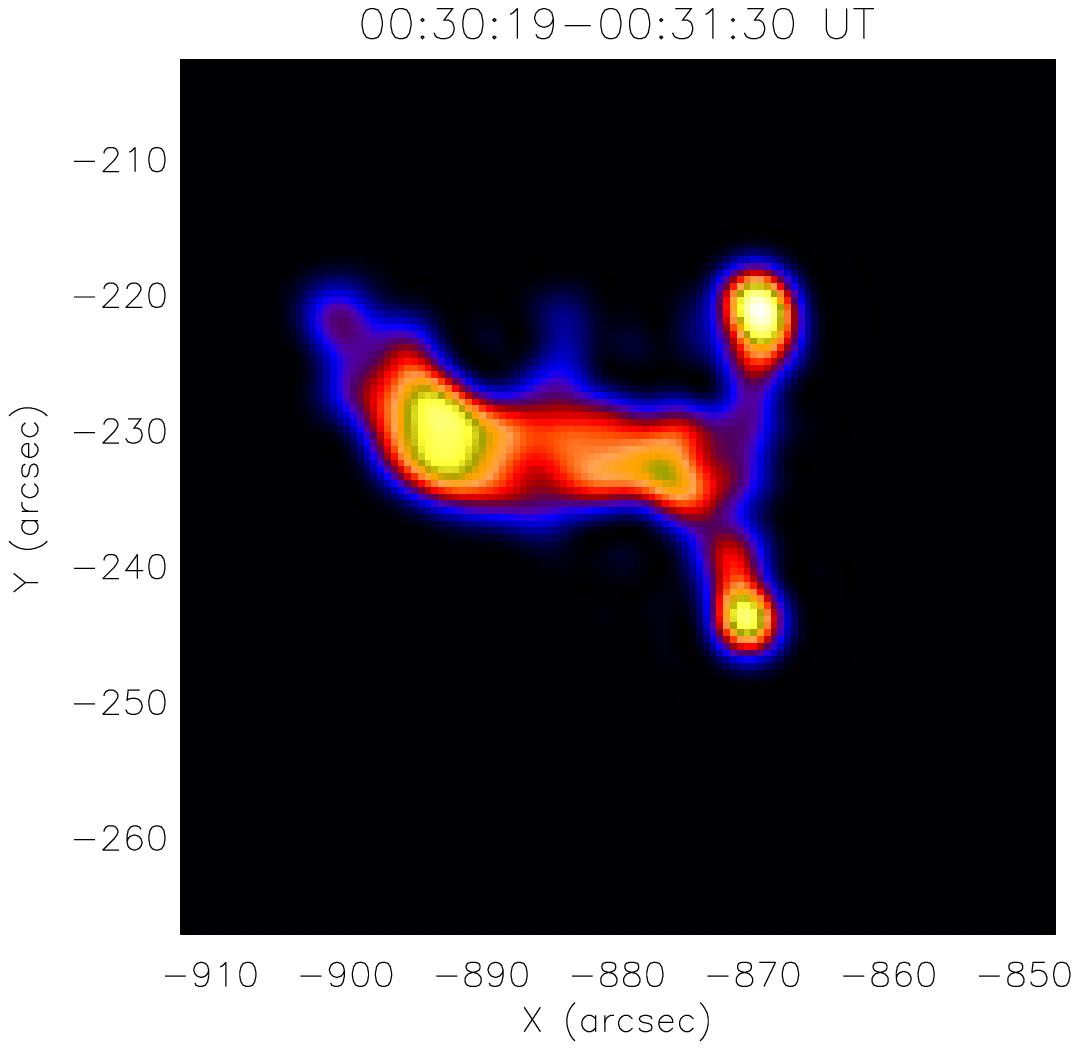}\\[0.1cm]
 \includegraphics[width=0.2\textwidth]{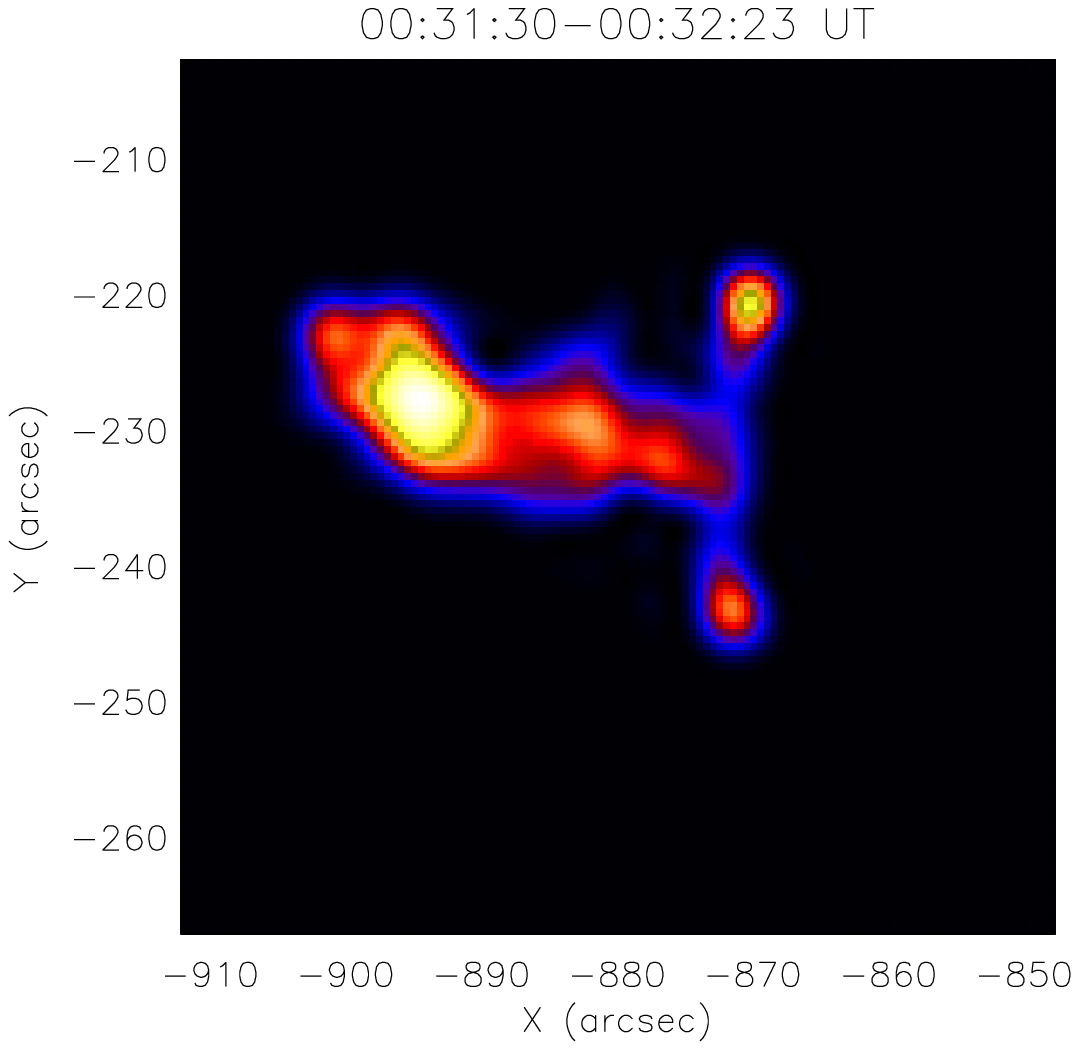}
 \includegraphics[width=0.2\textwidth]{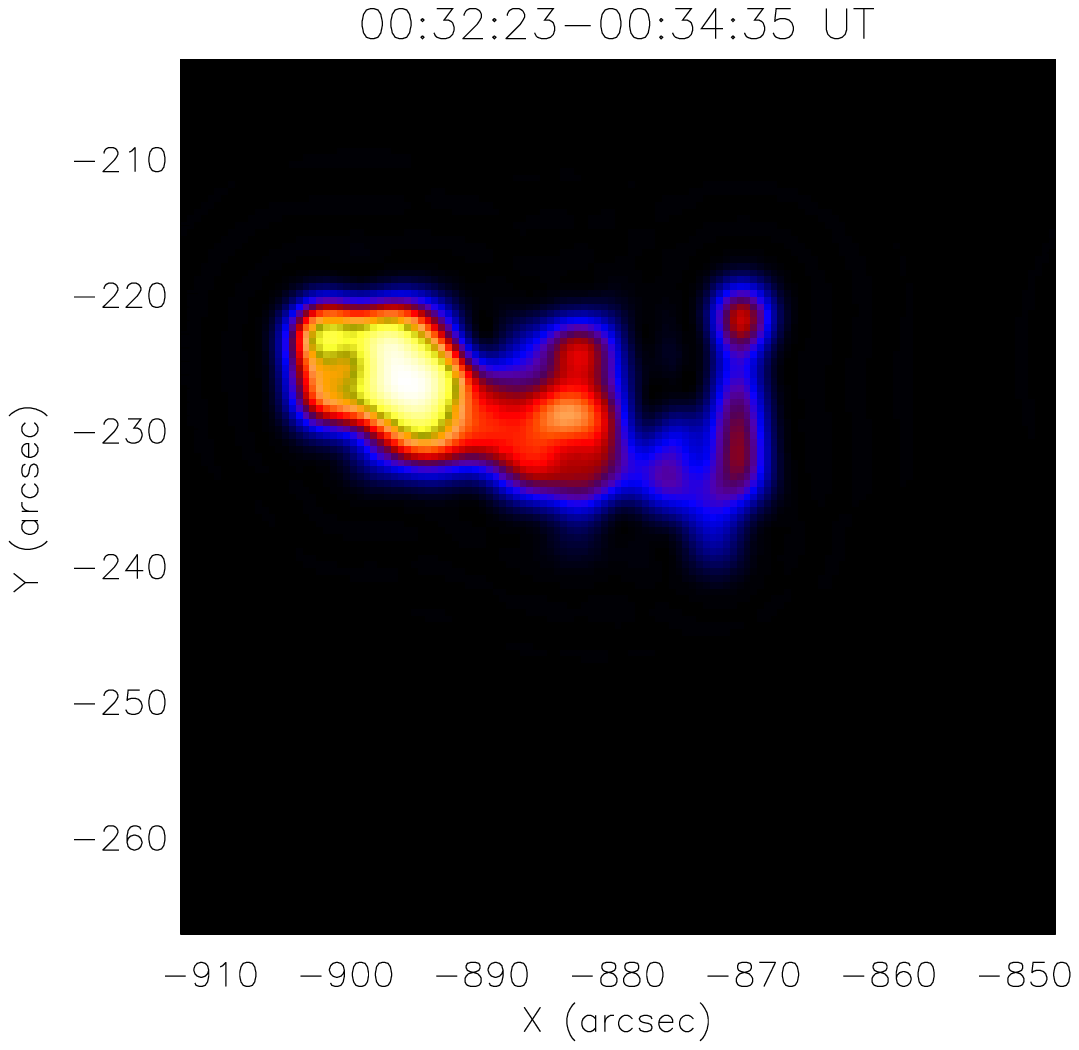}
 \includegraphics[width=0.2\textwidth]{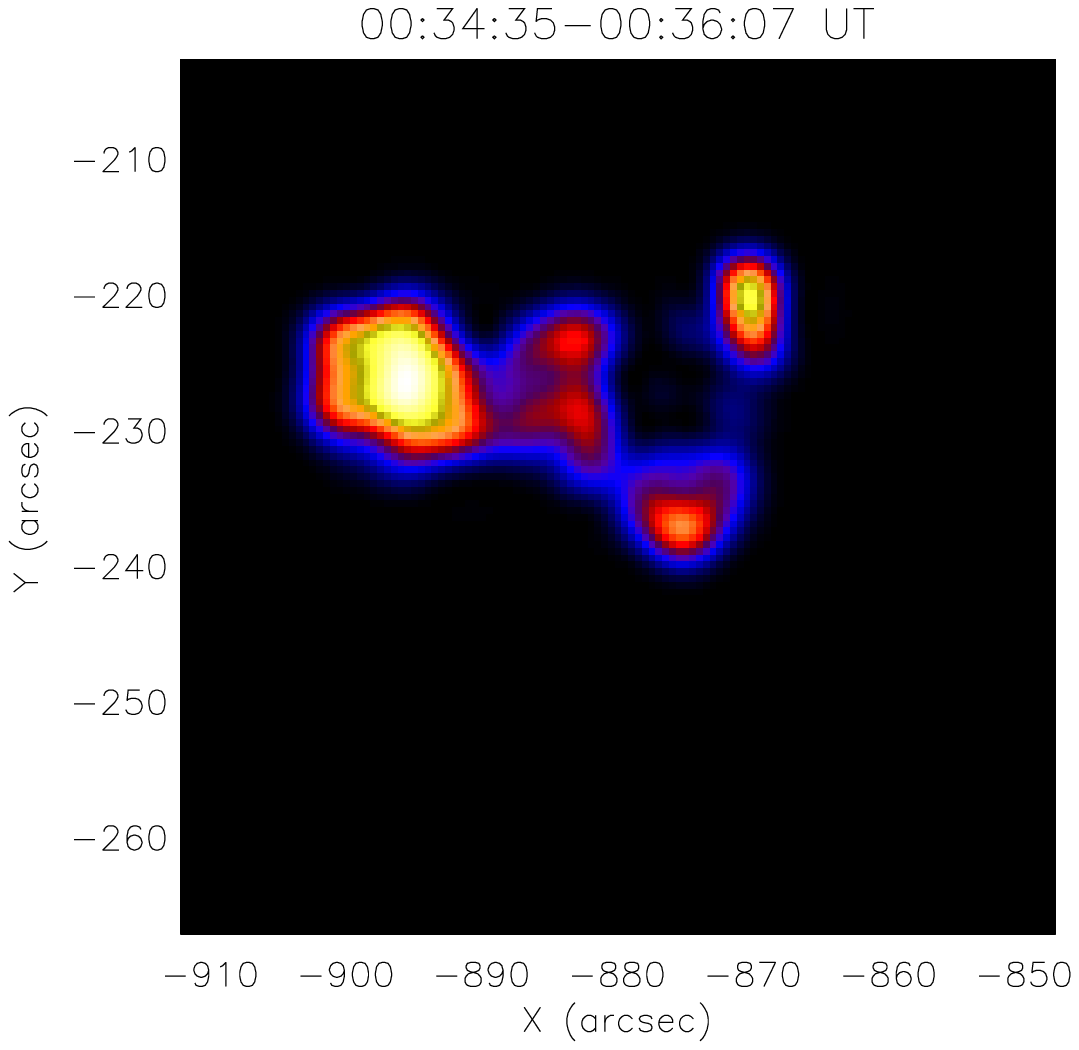}
 \includegraphics[width=0.2\textwidth]{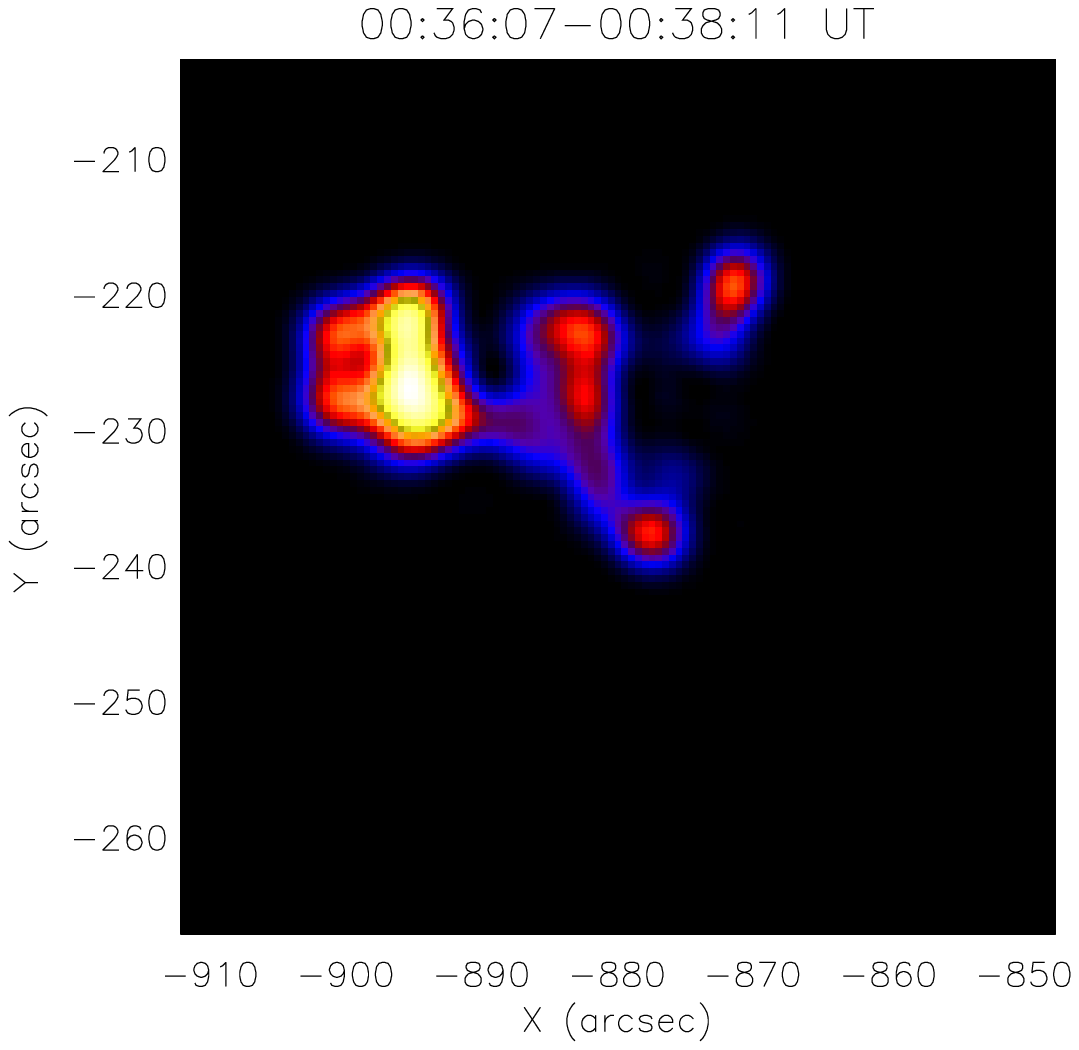}
 \caption{Images reconstructed by 5-CS for the July 23, 2002 flaring event through different time intervals (36-41 keV energy channel). Visibilities collected by {\em{RHESSI}} collimators from 2 to 9 have been used for the reconstructions.}
 \label{fig:rhessi_23-Jul-2002_time_analysis}
\end{figure*}

We first considered five flaring events at specific energy channels and time intervals and compared the reconstructions provided by 5-CS with the ones obtained by two standard visibility-based methods, namely CLEAN \citep{hogbom1974aperture} and uv$\_$ smooth \citep{massone2009hard}. In particular we have considered flaring events characterized by rather different morphologies like double foot-points (20 February 2002), loops (15 April 2002; 13 May 2013), extended sources (31 August 2004), and extended plus compact sources (2 December 2003). 
The results in Figure \ref{fig:rhessi_eval_mixt} show that sparsity promotion and the use of a continuous wavelet formulation reduce the artifacts and provide a higher spatial resolution. For all experiments we used {\em{RHESSI}} detectors from 2 to 9 and a 3-scale decomposition for the wavelet-based deconvolution methods. Figures \ref{fig:rhessi_23-Jul-2002_energy_analysis} and \ref{fig:rhessi_23-Jul-2002_time_analysis} focus on datasets acquired by {\em{RHESSI}} during the July 23 2002 event. In particular, Figure \ref{fig:rhessi_23-Jul-2002_energy_analysis} reproduces the same analysis performed by \citet{emetal03} using CLEAN and clearly points out how 5-CS better preserves the sources' morphology along the energy increase, reduces the artifacts in between the different sources and maintains the image reliability particularly at high energies. On the other hand, Figure \ref{fig:rhessi_23-Jul-2002_time_analysis} shows the time evolution of the flaring emission at a fixed energy channel and probably seems to reject the presence of emission along a curved locus joining the northern and souther sources, in contrast to what argued by \citet{massone2009hard}, but accordingly to the results in \citep{emetal03}. 

%
%

\graphicspath{ {./experiments/rhessi_evaluation/} }

\graphicspath{ {./experiments/rhessi_evaluation/23-Jul-2002_energy_analysis_2-9det/} }

\graphicspath{ {./experiments/rhessi_evaluation/23-Jul-2002_time_analysis_2-9det/} }

\subsection{Computational performance}

The experiments in this paper have been performed by means of a second generation 4 core Intel i7-2600 (3.40 GHz) CPU using IDL 8.4 on Ubuntu 16.04. The computational time complexity of 5-CS is near $\mathcal{O}(KN^2\log N)$ where $N$ is the image size and $K$ is the number of iterations required by FISTA for solving the optimization problem. We did not include the number of employed scales $j_0$ for FIWT on the time complexity analysis, since $j_0$ is usually a small number.

\begin{figure*}[!t]
  \centering
  \subfloat[]{
  \includegraphics[width=0.5\textwidth]{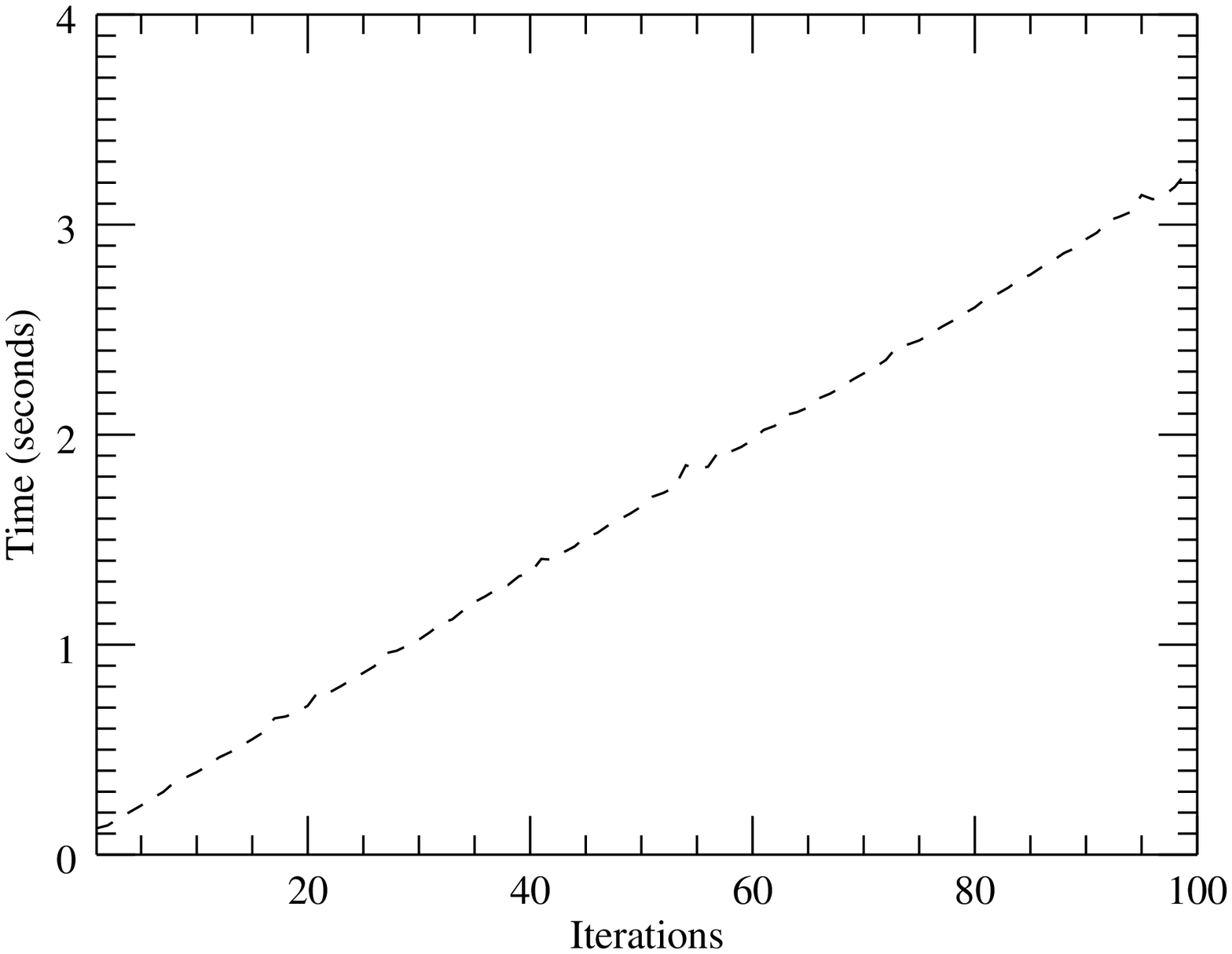}}
  \subfloat[]{
  \includegraphics[width=0.5\textwidth]{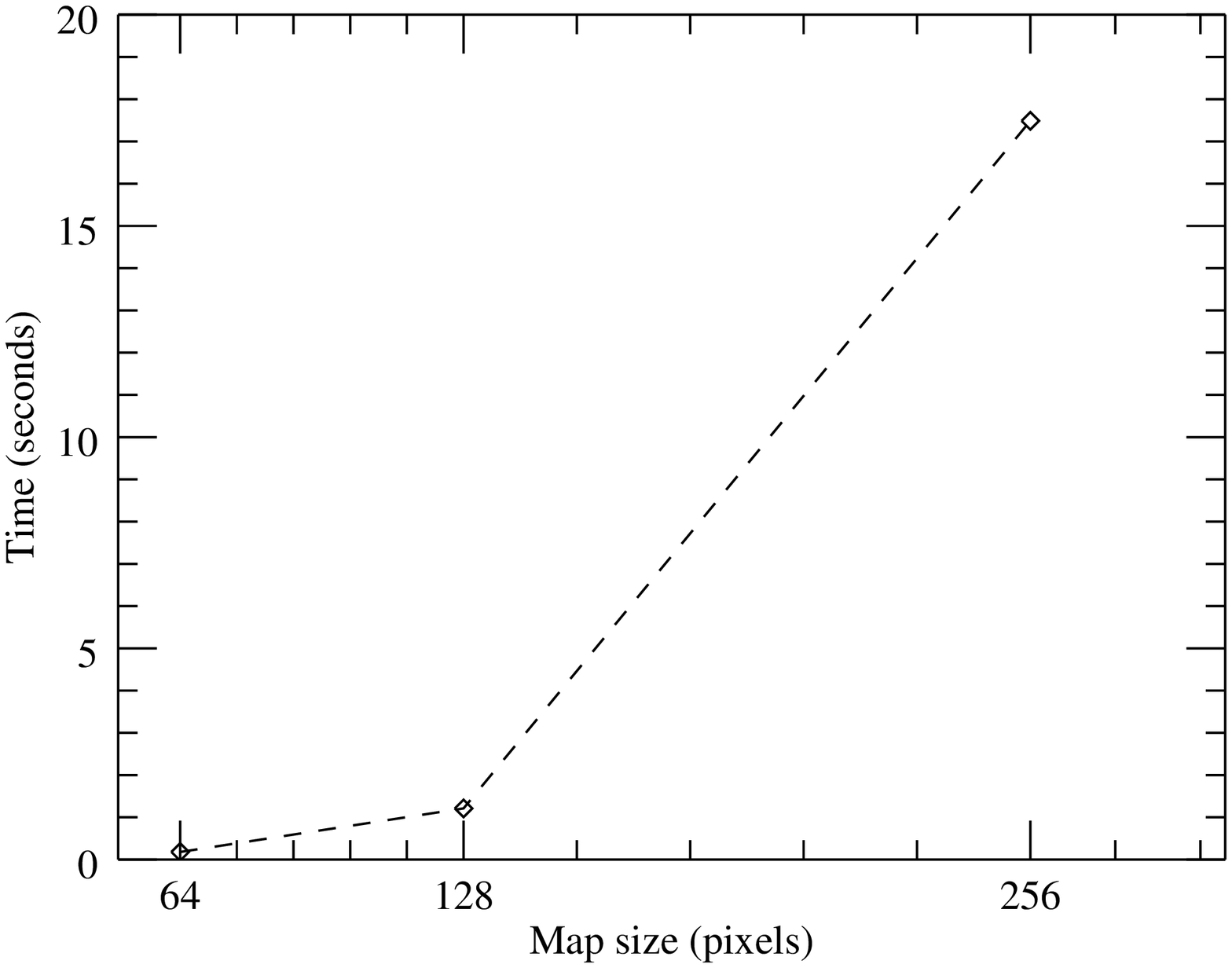}}
  \caption{Computational performance of 5-CS. The computational time is compared versus (a) the number of FISTA iterations and (b) the size of the reconstructed image. The test is made using the synthetic {\em{STIX}} visibilities used in Figure 4(a).}
  \label{fig:speed}
\end{figure*}

Figure \ref{fig:speed} shows the computational performance of 5-CS under different configurations in the case of the first {\em{STIX}} simulation (Figure 4(a)). In Figure \ref{fig:speed} (a) we can observe the linear relation between the employed time and the number of iterations $K$ for the reconstruction of an image map of size $N=128$ and considering $j_0=3$. On the other hand, Figure \ref{fig:speed} (b) shows how the computational time increases when 5-CS recovers images with higher and higher size, where in this case, the number of iterations and considered scales are fixed at $K=30$ and $j_0=3$, respectively. Since most reconstructions are performed with an image size of $128$ pixels and our method usually solves the reconstruction problem with less than 100 iterations, we can conclude that the average time required by 5-CS for reconstructing a single standard image is between 1.5 and 2.5 seconds.

\graphicspath{ {./experiments/} }

\section{Conclusions}
\label{sec:conclusion}
Hard X-ray solar space telescopes typically provide their experimental measurements in the form of visibilities, i.e., sparse samples of the Fourier transform of the incoming flux. Therefore producing images in this setting requires the application of reconstruction methods that realize the inversion of the Fourier transform from limited data. Several methods have been utilized so far but none of them explicitly exploited compressed sensing, i.e. the use of a sparsity-enhancing penalty term in regularization. Here we introduced a wavelet-based deconvolution method promoting sparsity for hard X-ray image reconstruction from visibilities. The main aspects of this method are that
\begin{itemize}
\item It relies on a continuous isotropic wavelet transform, coherently to the fact that X-ray sources are either isotropic or characterized by a slow change of shape.
\item It avoids the use of a computationally demanding catalogue-based compressed sensing.
\item It realizes regularization by means of optimization of a minimum problem where the penalty term promotes sparsity.
\item It realizes numerical optimization by means of a fast iterative algorithm.
\end{itemize}
The applications against both synthetic {\em{STIX}} and experimental {\em{RHESSI}} visibilities show the reliability of the method in terms of both the spatial resolution achieved and the reduction of spurious artifacts. Finally, the computational burden required by the method is low and competitive with respect to possible big data applications. The implementation of the algorithm within Solar SoftWare, which is under construction, will allow the systematic use of this approach against {\em{RHESSI}} observations and future application against {\em{STIX}} measurements.

\begin{acknowledgements}

We would like to acknowledge Federico Benvenuto for useful discussions. This research work has been supported by the ASI/INAF grant "Solar Orbiter ILWS - Supporto scientifico per la realizzazione degli strumenti METIS e SWA/DPU nelle fasi B2-C1"

\end{acknowledgements}

\bibliographystyle{aa}
\bibliography{references}

\end{document}